\documentclass[fleqn,usenatbib]{mnras}
\usepackage{newtxtext,newtxmath}
\usepackage[T1]{fontenc}
\usepackage{graphicx}
\usepackage{amsmath}
\usepackage{anyfontsize} % Extra package that prevents a fontsize error
\usepackage{soul}     % For strikethrough
\usepackage{xcolor}   % For colored text
\usepackage{graphicx}
\usepackage{float}
\usepackage{pgffor}
\usepackage{placeins}

\DeclareRobustCommand{\VAN}[3]{#2}
\let\VANthebibliography\thebibliography
\def\thebibliography{\DeclareRobustCommand{\VAN}[3]{##3}\VANthebibliography}

\title[Advanced Data Analysis of a Dutch SQM Network]{Beyond the Clouds: Advanced Data Analysis of a Dutch Sky Quality Meter Network}

\author[F. R. Shah et al.]{
  Farhan R. Shah,$^{1}$\thanks{Current address: Universität Konstanz, Fachbereich Biologie, AG Stöckl, Universitätsstr. 10, 78464 Konstanz, Germany}
  Reynier Peletier,$^{1}$\thanks{E-mail: peletier@astro.rug.nl}
  Jake Noel-Storr,$^{2}$
  Dirk van der Geest,$^{1}$
  Theo Jurriens,$^{3}$
  \newauthor
  Andreas Hänel,$^{4}$
  Tobias Hoffmann,$^{5,6}$
  Lisa Cordes,$^{5}$
  Robin Will,$^{5}$
  Athleen Selma Rietze,$^{5}$
  Matti Gehlen,$^{5}$
  \newauthor
  Hans Kjeldsen,$^{7}$
  Cristina Nazzari,$^{8}$
  Björn Poppe,$^{5}$
  \\
  $^{1}$Kapteyn Astronomical Institute, University of Groningen, PO Box 800, 9700 AV Groningen, The Netherlands\\
  $^{2}$Access Astronomy\\
  $^{3}$Science LinX, Faculty of Science and Engineering, University of Groningen, PO Box 704, 9700 AK Groningen, The Netherlands\\
  $^{4}$Dark Sky Germany\\
  $^{5}$Division for Medical Radiation Physics and Space Environment, Carl von Ossietzky Universität Oldenburg, Germany\\
  $^{6}$European Space Agency, ESA/ESOC, Darmstadt, Germany\\
  $^{7}$Aarhus University, Aarhus, Denmark\\
  $^{8}$Common Wadden Sea Secretariat\\
}

\date{Accepted XXX. Received YYY; in original form ZZZ}

\pubyear{\the\year{}}

\begin{document}
\label{firstpage}
\pagerange{\pageref{firstpage}--\pageref{lastpage}}
\maketitle

\begin{abstract}
Light pollution is an increasing environmental concern, impacting both ecological systems and human health. This report presents an analysis of light pollution data from the \textit{Was het donker} SQM network from 2020 until 2023, with a focus on indirect light pollution, commonly known as skyglow. By integrating measurements from Sky Quality Meter (SQM) stations in the network and cloud cover data from EUMETSAT, we conducted a comprehensive analysis of night sky brightness across a region encompassing northern Netherlands and the western part of the German Wadden Coast. Yearly changes in brightness for 27 locations were ranked and plotted, revealing that in the darkest areas, light pollution is increasing at a rate of 2.78 to 6.68 percent per year. A trend emerged showing that brighter areas experienced lower variability in brightness, while darker zones exhibited higher variability. This is due to the dominance of artificial light sources, such as street lighting, in brighter areas, which reduces the influence of natural light sources like the Moon, stars, and cloud backscatter. Seasonal patterns and the effects of the Milky Way were also investigated. Density plots were employed to visualize these changes in night sky brightness, helping to identify specific sources of light pollution, such as greenhouse lighting and streetlight turn-off times. These findings emphasize the need for systematic monitoring of light pollution and offer valuable insights that can guide public awareness initiatives and inform light pollution mitigation strategies.
\end{abstract}


%%%%%%%%%%%%%%%%%%%%%%%%%%%%%%%%%%%%%%%%%%%%%%%%%%

%%%%%%%%%%%%%%%%% BODY OF PAPER %%%%%%%%%%%%%%%%%%

\section{Introduction}
Recent research highlights a substantial rise globally in artificial light at night \citep{kyba2017, sanchez21}. Reportedly, this increase is amplified by the introduction of LED lights, which emit light at shorter wavelengths not detectable by current satellite sensors \citep{kyba2017, sanchez21}. This is not only impeding our view of the stars at night, but also harming organisms in the ecosystem. Health effects of light pollution on humans \citep{melatonin} and animals such as birds \citep{birds}, frogs \citep{ecological}, and sea turtles \citep{seaturtle} have been a focus of research. Furthermore, unfocused light emissions wastes a considerable amount of energy. For instance, Turkey saw an 80 percent increase in energy released to space from 2012 to 2019 while the population increased only by 9.1 percent in that period \citep{turkeytemporal, turkpop}. To create awareness for these problems, it is important that governments and organizations invest in accurate and precise light pollution measurement systems/networks and formulate policies based on their monitoring.

The components of light pollution include glare, sky glow, light trespass, and clutter \citep{darksky2024}. Here, sky glow is a type of indirect light pollution which reduces the contrast of the night sky and consists of light being scattered back to the ground \citep{types}. In this paper, we focus on the measurement of indirect light pollution (sky glow). A device that is widely used for this purpose is a photometer called \textit{Sky Quality Meter} (SQM) produced by Unihedron \citep{sqm}. It has become a widely-used tool in light pollution studies, with numerous applications in ecological, environmental, and astronomical research \citep{reviewpap}.

Regional networks for monitoring artificial light at night (ALAN) have provided critical insights into light pollution trends and variability. For instance, \citet{padua} discussed a Sky Quality Meter (SQM) network in Veneto, Italy. They investigated long-term trends and localized variations in night sky brightness (NSB). Their work revealed bimodal NSB distributions in urban areas driven by cloud effects and identified seasonal variations in darker locations, attributing it to the Milky Way and snow cover. Complementing these efforts, \citet{eumet} combined SQM data with satellite observations to confirm the relationship between the SQM data and cloud cover. Another study utilized 26 SQM stations across Austria and identified significant seasonal trends, with darker skies in summer and brighter skies in winter due to cloud cover and snow reflection \citep{posch2018systematic}. Additionally, recent research by \citet{kocifaj2023} introduced a novel method to estimate atmospheric aerosol parameters, such as aerosol optical depth (AOD), directly from daytime sky images. Their results demonstrate that higher aerosol concentrations amplified sky glow due to enhanced scattering of artificial light, further reinforcing the role atmospheric conditions play in modulating NSB trends.

Building on these studies, recent analyses further clarified the drivers of observed NSB changes. \citet{puschnignew} integrated atmospheric modeling to separate anthropogenic light trends from atmospheric variability, finding that albedo and vegetation had significant impacts on NSB, with average annual increases in ALAN of approximately 1.7 to 3.7 percent, dependent on urbanization levels. Similarly, \citet{italygroupnew} conducted an in-depth investigation of SQM sensor aging and atmospheric influences, identifying significant sensor responsivity decay and highlighting aerosol optical depth (AOD) as a major factor influencing long-term NSB trends, particularly evident in sites affected by substantial light pollution, where atmospheric variations explained more than 75 percent of NSB changes.

\subsection{Keep it Dark (KID)}
'KID – Keep it Dark" was a project of the Interreg North Sea Region Programme 2021-2027. The project was led by the Dutch \textit{Rijksuniversiteit Groningen}. The project ran from October 1, 2022 until March 31, 2024. Collaborating partners included the University of Oldenburg (Germany) and Aarhus University (Denmark). The primary objective of KID was to develop a robust measurement system for assessing the brightness of the night sky in the Wadden Sea area and to establish guidelines for accurate monitoring of sky brightness \citep{KID}. These guidelines aimed to assist various organizations, such as nature conservationists and jurisdictions, in tracking the evolution of light pollution on a local scale.

\subsection{Sky Quality Meter}
\begin{figure}
	\includegraphics[width=\columnwidth]{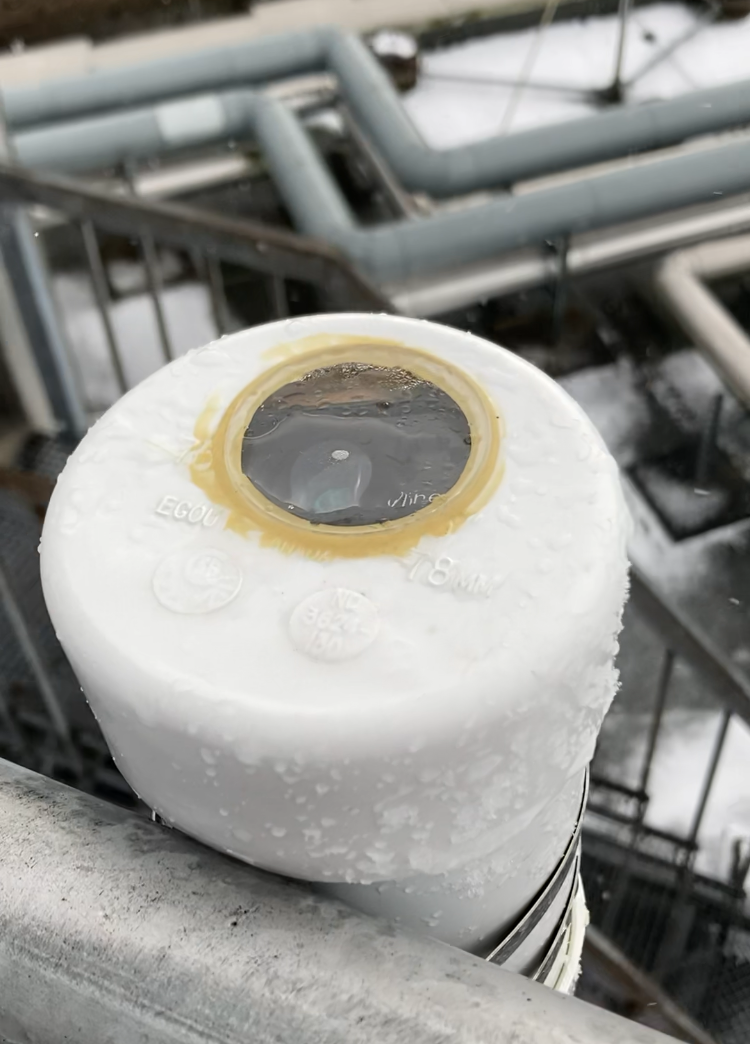}
    \caption{Photograph showing an SQM in Oldenburg (Germany) installed in its protective housing. Drops of rain are visible on top.}
    \label{fig:sqm}
\end{figure}

The Sky Quality Meter (SQM), originally developed by Unihedron, is an easy-to-use and affordable device for measuring light pollution. Fig.~\ref{fig:sqm} shows a typical SQM. This one-dimensional instrument uses a single channel, typically observing at zenith, employs a solid-state light-to-frequency detector (TAOS TSL237S) and operates in the green, with a wide spectral band \citep{hanel, colorworld}. The SQM reports brightness values in the unit mag/arcsec\textsuperscript{2}, a logarithmic scale but it can be converted to a linear luminance scale—expressed in candela per square meter (cd/m$^2$)—using the relation:
\begin{equation}
L\ (\mathrm{cd/m}^2) = L_0 \times 10^{-0.4\, m},
\end{equation}

where $m$ is the measured brightness in mag/arcsec$^2$ and $L_0$ is the luminance corresponding to 0 mag/arcsec$^2$. Note that the value of $L_0$ depends on the photometric calibration and the reference system chosen. A value frequently used in literature is \(1.08 \times 10^5\) \citep{kyba15,hanel} and is also provided in Unihedron's SQM Operator's Manual, but \citet{barazero} questioned its precise origin and range of validity. At artificially lit locations, the sky becomes redder with increasing cloud cover, and in that condition, this conversion would likely overestimate the luminance. Others have used different values, such as \(0.90 \times 10^5\) by \citet{kyba13} and \citet{grauer} states it to be \(1.475 \times 10^5\) in the AB System and \(1.08 \times 10^4\) in the Vega system. The absolute radiometric calibration of the SQM is described by \citet{baraabs} where its photometric zero-points can be inferred. In this paper, we do not carry out any conversions.

The SQM-L version, which we primarily discuss, incorporates a lens that restricts the device's field of view to 20° (FWHM) \citep{sqmlfwhm}, thereby enhancing consistency in value measurements, particularly in the presence of direct lighting.

Over extended periods, SQMs may experience an ageing effect due to loss of sensitivity, resulting in darkening. This effect has been quantified in a study utilizing twilight as a calibrator \citep{twilight}. They calculated an average darkening effect of 0.0440±0.0003 mag/arcsec\textsuperscript{2} per year across three SQMs. This means a 4.1 percent darker reading after one year. In this paper, we prefer not to correct our data for darkening, but we take it into account in the discussion. Despite limitations, the SQM remains a cost effective device for monitoring light pollution and gathering valuable data. In this paper, we analyse SQM data in different ways and present its use cases.

\section{Data acquisition and processing}
This study utilized data from two sources. The main source was the SQM network operating in parts of the Netherlands, Germany, and Denmark with data from 67 stations accessible through \texttt{washetdonker.nl} (as of August 2024). The other source included cloud cover data obtained from the European Organisation for the Exploitation of Meteorological Satellites (EUMETSAT) \textit{Cloud Mask - MSG - 0} product, which served as a reference to filter out parts of the SQM data where clouds were present.

\subsection{'Was het donker'} \label{washetdonkersection}
\begin{figure}
	\includegraphics[width=\columnwidth]{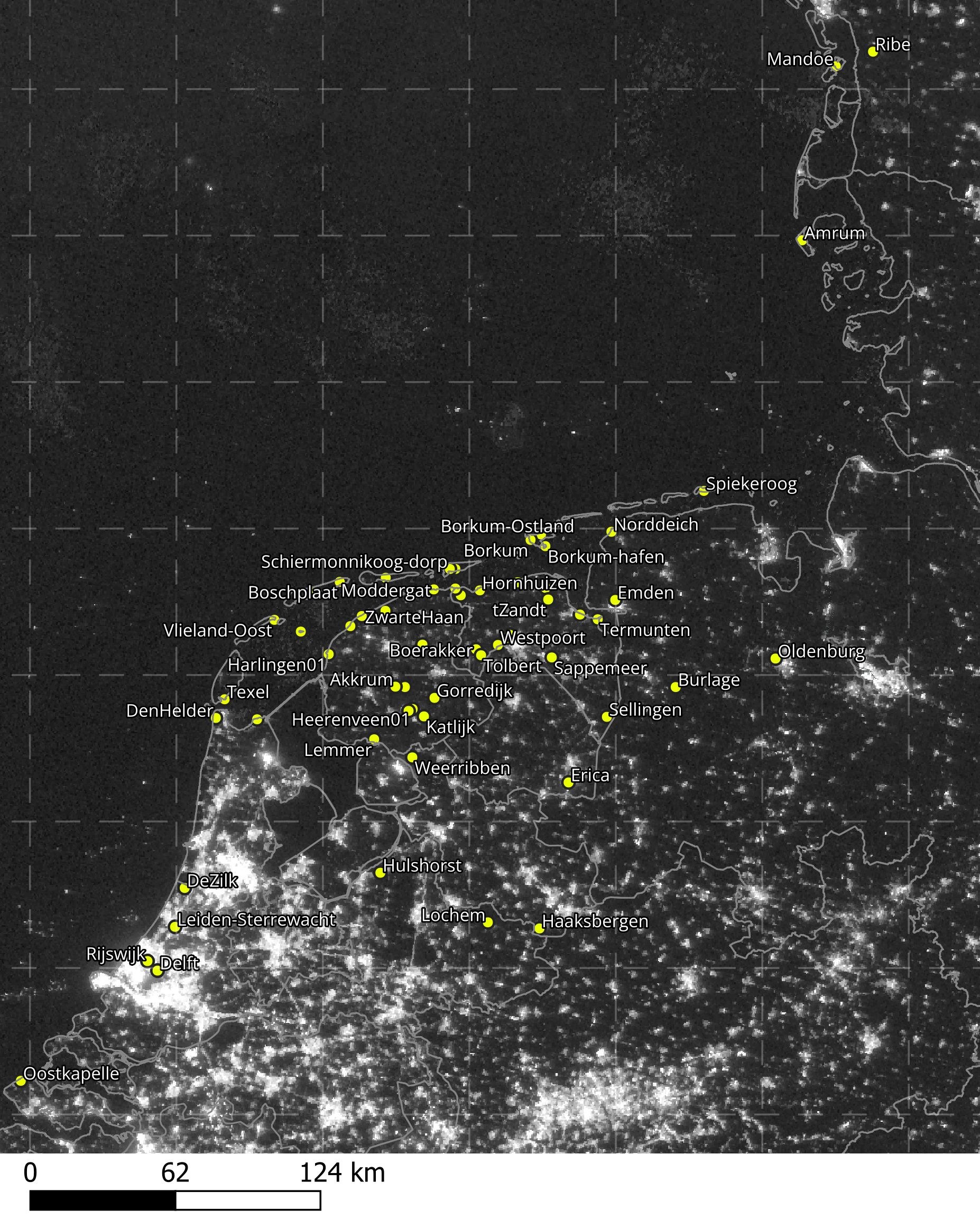}
    \caption{The \textit{Was het donker} network as of March 2024. The background image is sourced from NASA's Black Marble product (VNP46A2) from September 2023.}
    \label{fig:washetdonkerfigure}
\end{figure}

\texttt{Washetdonker.nl} is an initiative of the Kapteyn Astronomical Institute of the Faculty of Science and Engineering in the University of Groningen. It is a network of 67 Unihedron SQM-LUs (as of August 2024) working in combination with Raspberry Pis to make available data in real-time to the website every night. From dusk to dawn, the SQMs record the sky brightness at 42 second intervals. A high level overview of creating such a system is found in literature \citep{pribadi2019iot}. Fig.~\ref{fig:washetdonkerfigure} overlays the locations of the observational sites onto a nighttime satellite image, providing spatial context for the data collection points.

The data from the network is organized on the website in the form of .dat files, structured by directories based on location, year, and month. A python script was used to extract all the data through web scraping \citep{farhancodes}. After initial data cleaning, we obtained a substantial set of usable data. This included the coordinates, date, time, and night sky brightness (NSB). We then analysed it in two major ways: using "jellyfish" plots, and trend analysis. Similar work was done between 2009-2019 (NachtMeetnet) where 15 SQMs were installed at various locations in The Netherlands and annually calibrated \citep{dutchreport}. These results have not been presented in the refereed literature. Our work expands by working on a wider dataset in terms of number of locations, cloud filtering based on satellite data, providing more visualisations, and a different way of analyzing the data.

\subsection{Trend Analysis}

Trend analysis involves examining data over time to identify patterns, tendencies, or changes in a particular variable or set of variables. We aimed to discern and understand the direction and magnitude of changes, whether they are increasing, decreasing, or remaining stable, over the detection period.

The brightness of the night sky is dependent on many variables \citep{bara2019monitoring}. In our trend analysis, we identified these variables and excluded data segments where they exerted influence. The primary factors considered were solar, lunar, and cloud effects. Our aim was to obtain the most pristine dataset feasible, isolating temporal variations solely attributable to artificial lighting.

\subsubsection{The Sun and the Moon}
In astronomy, when the Sun is between 12° to 18° below the horizon, it is known as Astronomical Twilight. Below that, its scattered light is negligible and the sky is essentially astronomically dark. The Moon, however, is far less luminous than the Sun and its effect on sky brightness vanishes at a much smaller depression angle. There is no single universally adopted “lunar twilight” angle analogous to the Sun’s 18°, because the Moon’s impact depends strongly on its phase (illumination) and sky conditions. Nonetheless, scientific studies and observatory practices do identify rough guidelines for when moonlight ceases to brighten the night sky. A full Moon’s scattered light becomes negligible around 5° below the horizon, and less illuminated phases need even less (on the order of 1–3° or no depression at all for thin crescents) \citep{schaaf}. In observatory practice, a specific angular threshold for the Moon is sometimes used, but it varies by institution: the Liverpool Telescope (LJMU)\footnote{\url{telescope.livjm.ac.uk/PropInst/Phase2/?sf=SkyBrightness}} uses 3° as the cutoff for “dark sky” conditions for the full moon, while ESO considers the Moon’s effect gone once it’s just set (especially for smaller phases) \citep{patat}. Researchers have also used up to 10° in ultra-conservative cases, \citep{grauer} primarily to account for edge cases (very bright full Moon just set with slight haze, etc.). In general astronomy usage, however, such a large depression is rarely necessary. In summary, there isn’t a universal single number like 18° for the Moon, however, around 3–5° below horizon, even a full Moon’s residual glow is under roughly 0.1 mag/arcsec\textsuperscript{2} brightness increase – essentially negligible \citep{schaaf}.

We used PyEphem to filter our data based on time and location \citep{farhancodes}. It is a Python library for performing astronomy computations, such as calculating the positions of celestial objects. In our analysis, no data points were considered where the Sun and the Moon were less than 18° and 3° below the horizon, respectively. For the Moon, we used the 3° threshold used by the Liverpool Telescope group. To minimize data exclusion, a more conservative angle was not chosen.

\subsubsection{Clouds}
\begin{figure}
	\includegraphics[width=\columnwidth]{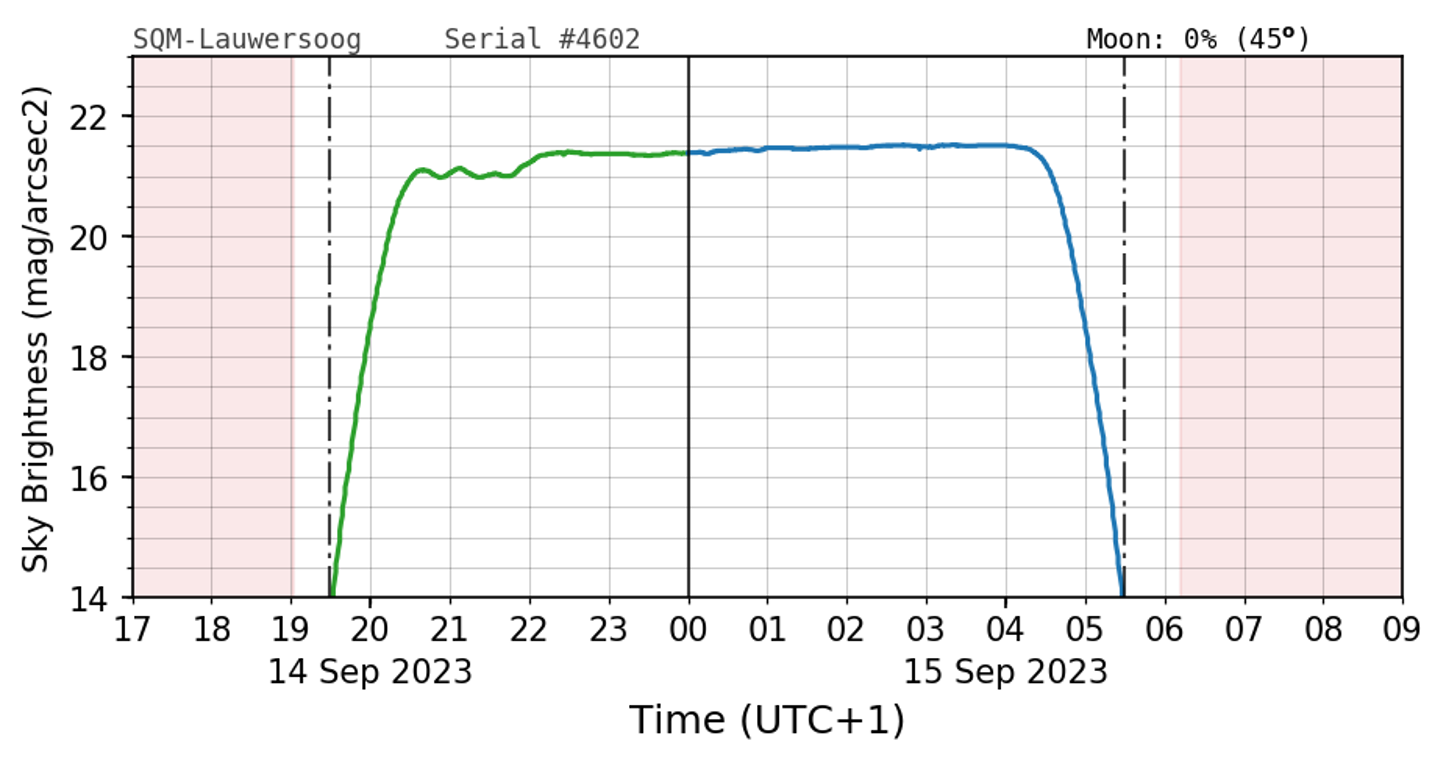}
    \caption{A typical graph from \texttt{washetdonker.nl} showcasing nightly data from a SQM.}
    \label{fig:14_sep}
\end{figure}

Filtering out cloudy data poses a greater challenge relative to that of the Sun and Moon. In our study, we explored two methods for this purpose. The first involves a statistical approach. Fig.~\ref{fig:14_sep} illustrates a graph depicting night measurements leading to the morning of September 15, 2023, for the dark sky park at Lauwersoog. The shaded pink regions denote times when the moon is above the horizon, though for this particular night, lunar illumination is at zero percent. The dot-dash lines indicates the times of sunset and sunrise. Noticeable fluctuations in the brightness direction occur between 20:30 and 22:00 (UTC+1), attributed to presence of clouds in the sky. This phenomenon is more pronounced in brighter locations, where clouds scatter the light back to the ground \citep{kyba2011cloud, cloud2, cloud3}. Leveraging this understanding, we filtered data based on hourly standard deviation values of the hourly sampled data, however, these values can differ from location to location. Receiver-Operating Characteristic (ROC) curves aided in determining the optimal thresholds. In our study, the optimal hourly standard deviation thresholds determined via ROC analysis (described in Appendix Section~\ref{rocsection}) are below 0.081 for Oldenburg and below 0.063 for Lochem. These values are slightly lower than those reported by \citet{eumet}, who derived thresholds spanning approximately 0.096–0.141 for sites with average magnitudes in a similar range. These discrepancies are likely attributable to differences in the temporal sampling resolution of the data considered (hourly versus 9–10‐minute intervals). Despite these methodological differences, the fact that our thresholds are of comparable order of magnitude reinforces the validity of using SQM standard deviation as an effective cloud indicator. Nevertheless, for the purpose of conducting trend analysis, we have relied on satellite data from EUMETSAT.

\begin{figure}
	\includegraphics[width=\columnwidth]{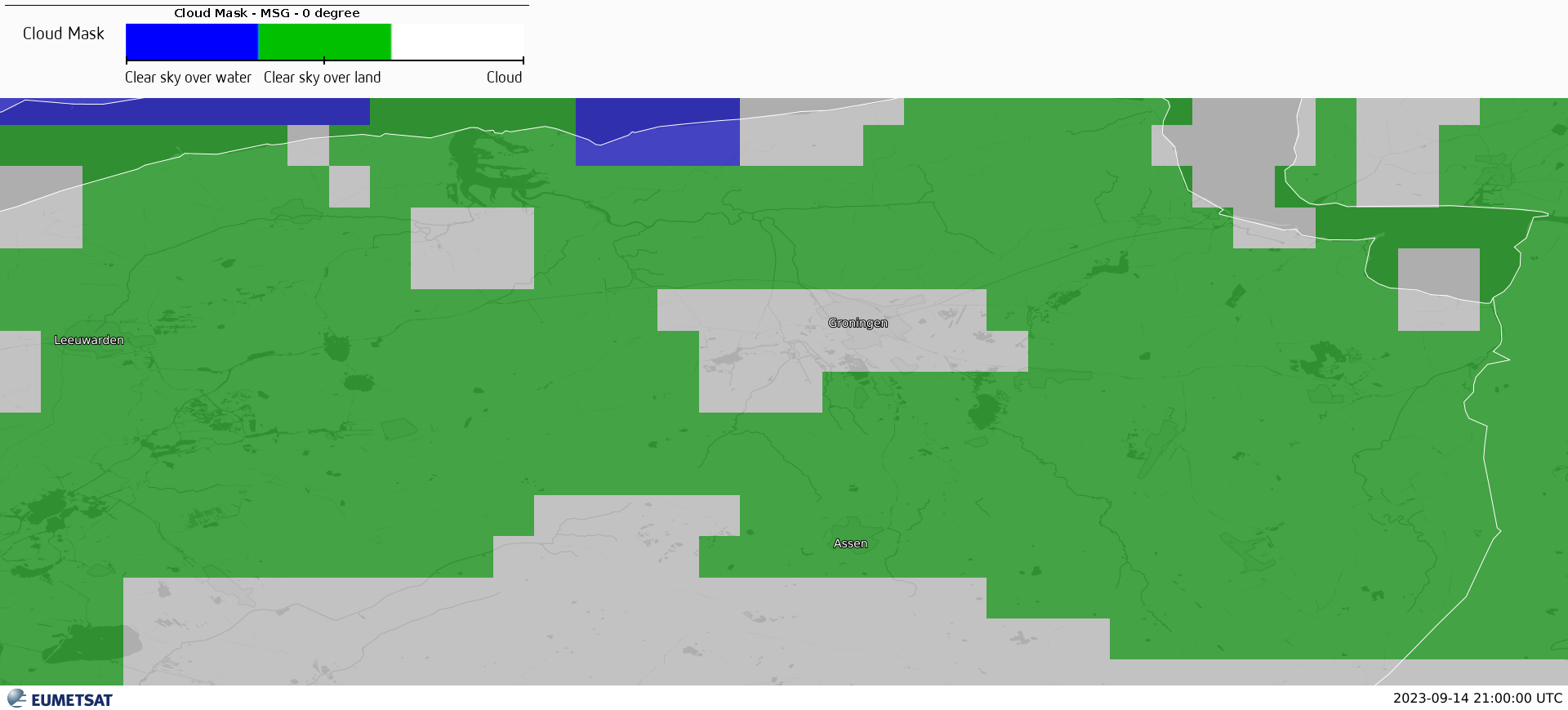}
    \caption{EUMETSAT Cloud Mask - MSG - 0 degree. (Groningen region).}
    \label{fig:gron}
\end{figure}

EUMETSAT is the European operational satellite agency for monitoring weather, climate and the environment from space. Their Cloud Mask (CLM) product (Cloud Mask - MSG - 0 degree) which describes the scene type (either 'clear' or 'cloudy') on a pixel level. Each pixel is classified as one of the following four types: clear sky over water, clear sky over land, cloud, or not processed (off Earth disc). Whether a pixel is cloudy or not is determined by analyzing spectral signatures in satellite imagery, identifying characteristic patterns associated with clouds such as reflectance and thermal radiation. Sophisticated algorithms then apply thresholds and texture analysis to classify pixels as cloudy or cloud-free, enhancing weather forecasting and climate monitoring capabilities. This is explained in the document for the algorithm's theoretical basis \citep{EUMETSAT2007}. Fig.~\ref{fig:gron} shows what a typical map would look like. The pixel size is 0.041 degrees which means that for the latitude of The Netherlands, a pixel represents an area of approximately 11 km\textsuperscript{2}. The data files for this product are publicly available for every 15 minute interval. We performed a detailed temporal analysis of a 3×3 pixel region surrounding each SQM location, iterating through all intervals from January 2020 to November 2023. Selecting nine pixels ensured greater coverage (at maximum cloud ceiling altitude) than that of the 20-degree field of view of the SQMs. Hours during which clouds were detected in any of the nine pixels were classified as cloudy sky hours and subsequently excluded from the SQM dataset.

\subsubsection{Data availability and sampling patterns}\label{sec:sampling_patterns}

A key consideration in long-term trend analysis is the completeness and distribution of usable data. Although our Sky Quality Meters (SQMs) record continuously at a 42-second cadence, only a small fraction of this data is usable after applying filters to isolate clear, dark-sky conditions. To estimate the impact of filtering, we ran a simulation (using the Open-Meteo weather API \citep{Zippenfenig_Open-Meteo}) for randomly selected locations in northern Netherlands, northern Germany, and Denmark. While not based on our network data, this simulation reflected typical conditions in the region and provides a representative estimate of data retention.
From 2020 to 2023, out of 1,051,920 potential hourly measurements:
\begin{itemize}
    \item 72.4\% were removed due to sunlight (solar altitude $> -18^\circ$),
    \item 53.8\% of the remaining hours were removed due to moonlight,
    \item 93.9\% of the remaining moon-free hours were removed due to cloud cover.
\end{itemize}

This resulted in approximately 8,207 usable hours, or just 0.8\% of the original dataset. Importantly, data availability is strongly seasonal: spring (February–April) yielded 4,697 usable hours, while summer (May–July) contributed only 88 hours due to limited nighttime. Winter and autumn provided moderate coverage.

For our trend analysis, there is coverage for all weeks for which there are dark, no cloudy data. We discuss the implications of this seasonal sampling structure further in Section~\ref{sec:discussion_limitations}.

\subsubsection{Sensitivity to sampling and data gaps}
\label{sec:sampling}

Two sources dominate the formal trend error:
(i) statistical scatter of the weekly medians and (ii) the irregular pattern of missing weeks due to data gaps or filtering.

To quantify the second term we generated $N_{\rm MC}=10\,000$ bootstrap replicates per location, re‑drawing the \emph{presence/absence} mask of each calendar week with replacement while keeping the original brightness values fixed. For every replicate we fitted a straight line with slope $m_{j}$ through the weekly medians and defined

\begin{equation}
\sigma_{\rm gap} \;=\; \sqrt{\frac{1}{N_{\rm MC}-1}\,
                 \sum_{j=1}^{N_{\rm MC}}\bigl(m_{j}-\langle m\rangle\bigr)^{2}} .
\end{equation}

The total uncertainty that we quote in this paper is

\begin{equation}
\sigma_{\rm tot} \;=\;
\sqrt{\sigma_{\rm stat}^{2} \;+\; \sigma_{\rm gap}^{2}},
\label{eq:sigma_tot}
\end{equation}

where $\sigma_{\rm stat}$ is the standard error of the slope returned by ordinary least–squares.

\begin{figure}
	\includegraphics[width=\columnwidth]{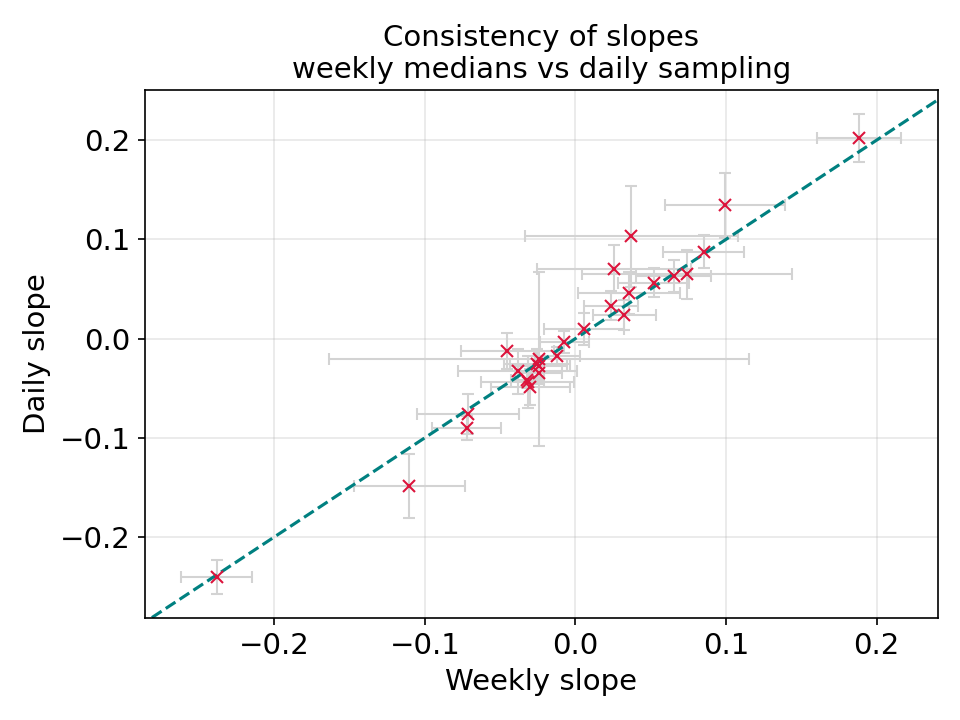}
    \caption{Correlation plot for slopes from weekly and daily samplings.}
    \label{fig:corr1}
\end{figure}

\begin{figure}
	\includegraphics[width=\columnwidth]{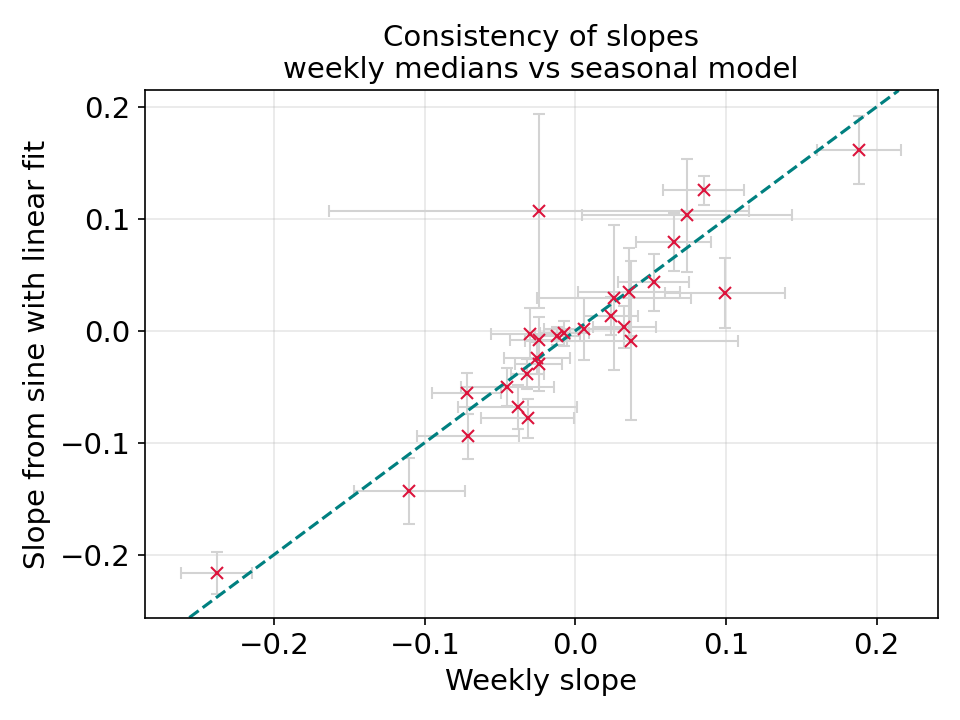}
    \caption{Correlation plot for linear weekly fits and sinusoidal fits.}
    \label{fig:corr2}
\end{figure}

\paragraph*{Weekly medians versus daily fits.}
To verify that our choice of a 1‑week cadence does not bias the results we repeated the entire procedure using nightly median magnitudes
($\approx$1‑night cadence). The two sets of slopes are almost identical (Pearson $r\!=\!0.98$; see Fig. \ref{fig:corr1}). For every site the absolute difference, $\lvert m_{\rm daily}-m_{\rm weekly}\rvert$, remains below the combined error $\sigma_{\rm tot}$ defined in Eq.~(\ref{eq:sigma_tot}); no re‑weighting is therefore required. We also compared the linear fits with the sinusoidal fits for all locations and the correlation plot is shown in Fig. \ref{fig:corr2}. A concise comparison of linear and sine fits, together with the sampling statistics (weeks observed / missing), is given in Appendix Section~\ref{seasonal_appendix}.

\section{Results}
The examination of key findings encompasses various analytical approaches. For each location in the network, we created "jellyfish" plots with accompanying histograms, developed time-magnitude plots for trend analysis, explored "roller coaster" plots (Appendix Section~\ref{rolsection}), ranked locations, and conducted receiver operating characteristic (ROC) analysis (Appendix Section~\ref{rocsection}). Detailed explanations of these approaches will be provided in subsequent sections. All mentioned times henceforth are in UTC, unless specified.

\subsection{Trend Plots}
This section presents magnitude values as a function of time. The filtering outlined in Section~\ref{washetdonkersection} is applied to the data, wherein information affected by cloudy nights, the Moon, and the Sun is excluded. A methodological approach was employed to ensure the robustness of the analysis. Initially, SQM locations that were actively recording data at the time of retrieval and covered a significant span of time were chosen. Then, for all locations, a weekly median was computed for nightly hours devoid of clouds, moonlight, and sunlight. Weekly medians were selected as they smooth out short-term variations and are easier to interpret in long-term plots. Subsequently, data points that deviated more than two standard deviations from this median were identified as outliers. These outliers could be due to a variety of local factors, including but not limited to, severe thunderstorms, light from construction activities, or firework displays. Such data points were filtered out and the refined dataset was then subjected to a linear regression analysis to discern any major trends over time.

\begin{figure}
	\includegraphics[width=\columnwidth]{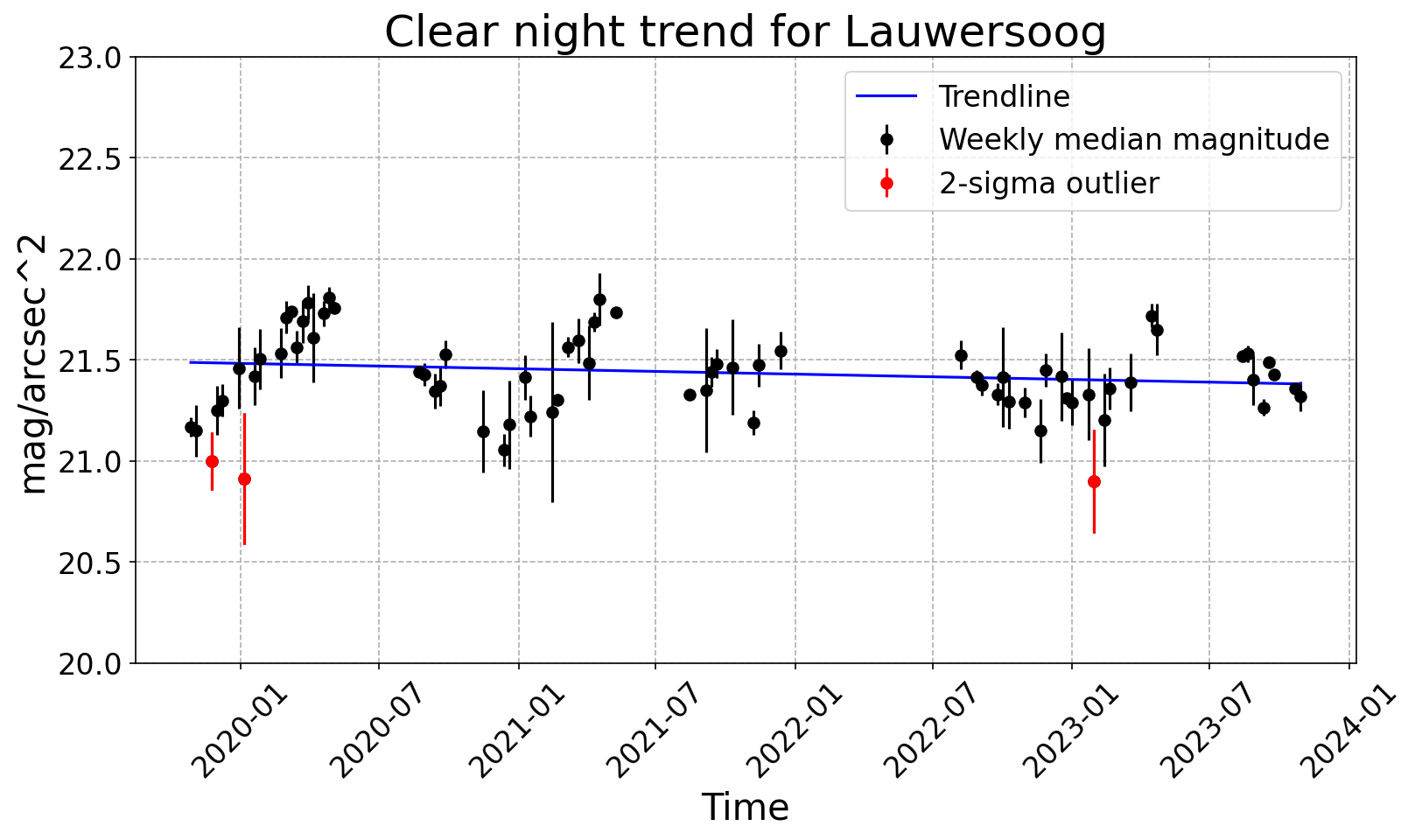}
    \caption{Trend of clear nights for the Lauwersoog Dark Sky Park.}
    \label{fig:lauwersoogtrend}
\end{figure}

The location of Lauwersoog in Fig.~\ref{fig:lauwersoogtrend} provides an example. Situated in the northern part of The Netherlands, the SQM is located in an established Dark Sky Park. By cross-checking weekly-median
trends against daily-sample regressions and by propagating an additional gap-imputation error derived from 10,000
Monte-Carlo realisations per site, we show that the net slope uncertainties increase by at most 30 percent, yet none of
the scientific conclusions change. For example, the weekly-median regression, augmented with our Monte Carlo gap imputation, yields a slope of $m_{\rm week} = -0.027 \pm 0.017\;\mathrm{mag\,arcsec^{-2}\,yr^{-1}}$ (within $1\sigma$ of zero). The annual-sine fit gives $m_{\rm sine} = -0.005\;\mathrm{mag\,arcsec^{-2}\,yr^{-1}}$. No statistically significant brightening can yet be claimed.

\begin{figure}
	\includegraphics[width=\columnwidth]{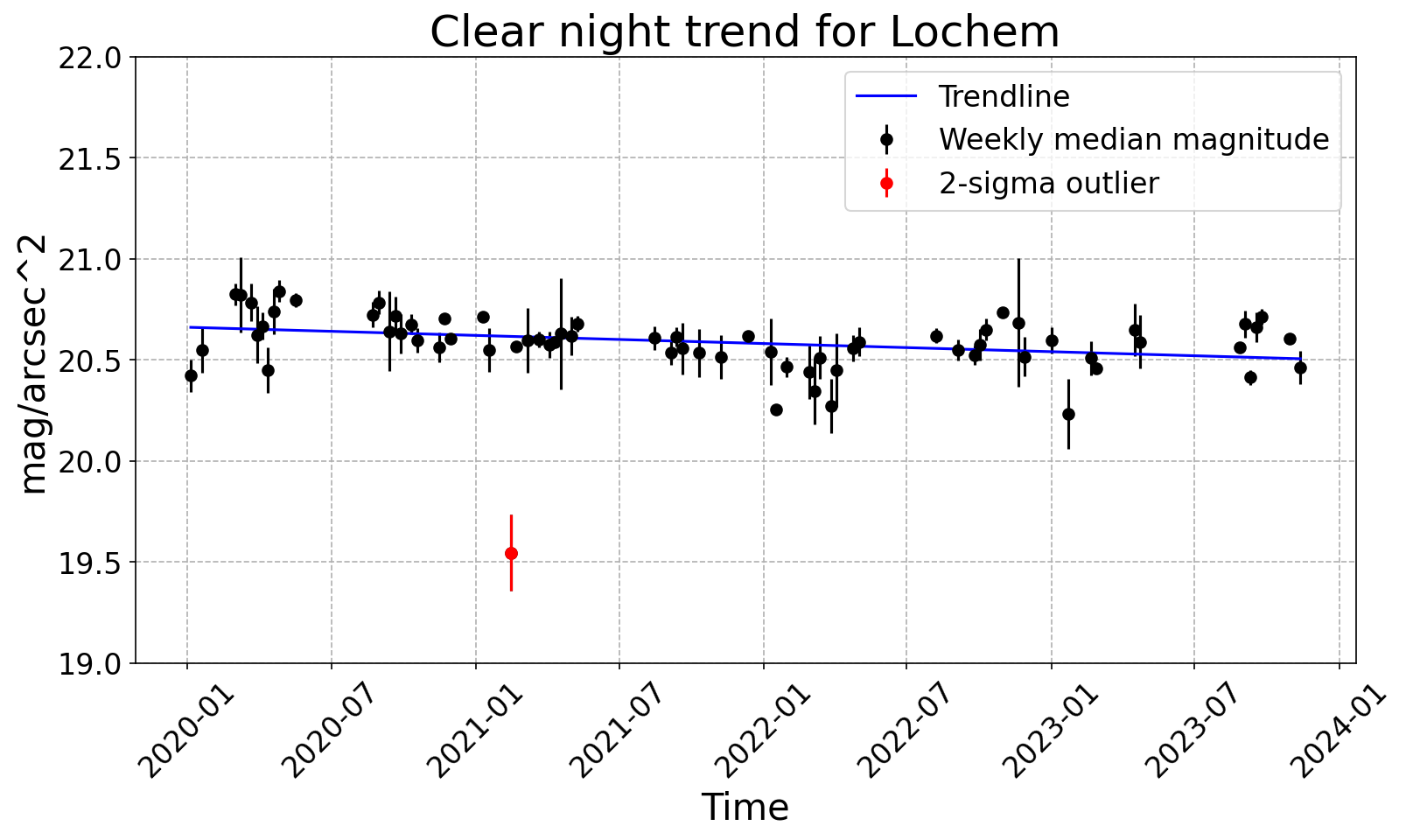}
    \caption{Trend of clear nights for Lochem.}
    \label{fig:lochemtrend}
\end{figure}

We continue our analysis to another location, Lochem, which is a city in the province of Gelderland in the eastern part of The Netherlands. Fig.~\ref{fig:lochemtrend} shows a gradual change at a rate of $m_{\rm sine} = -0.0403\;\mathrm{mag\,arcsec^{-2}\,yr^{-1}}$. According to the Central Bureau of Statistics dashboard “Inwoners per gemeente,” Lochem’s population was 33,729 in 2020 and 34,289 on in 2024, indicating a minimal increase of just 560 residents over four years. \citep{cbs} Given its status as a smaller municipality, significant urban development or demographic changes are not anticipated and thus the observed changes in sky brightness may not be attributed to it.

Examining the data from De Held in the city of Groningen, reveals a significant trend. This location stands out among all analysed areas with the fastest increase in sky brightness, recorded at a rate of -0.22~mag/arcsec\textsuperscript{2} per year. The rapid brightening of the sky over the years could be attributed to ongoing construction activities in the vicinity. This is particularly evident from some data points with high standard deviations, indicating large changes in brightness levels in the week. It's important to note that the SQM used for measurements is situated in a private garden, potentially introducing variability and affecting accuracy due to factors such as stray light and other environmental influences. However, despite potential limitations, the analysis shows a significant upward trend in brightness attributed to local factors such as construction projects. It is noteworthy that this phenomenon cannot be generalized to the entire city of Groningen, as evidenced by a comparison with another SQM located in Zernike campus (Appendix Section~\ref{appendixtrend}), which does not exhibit a similar magnitude of change in brightness over the same period.

\begin{figure}
	\includegraphics[width=\columnwidth]{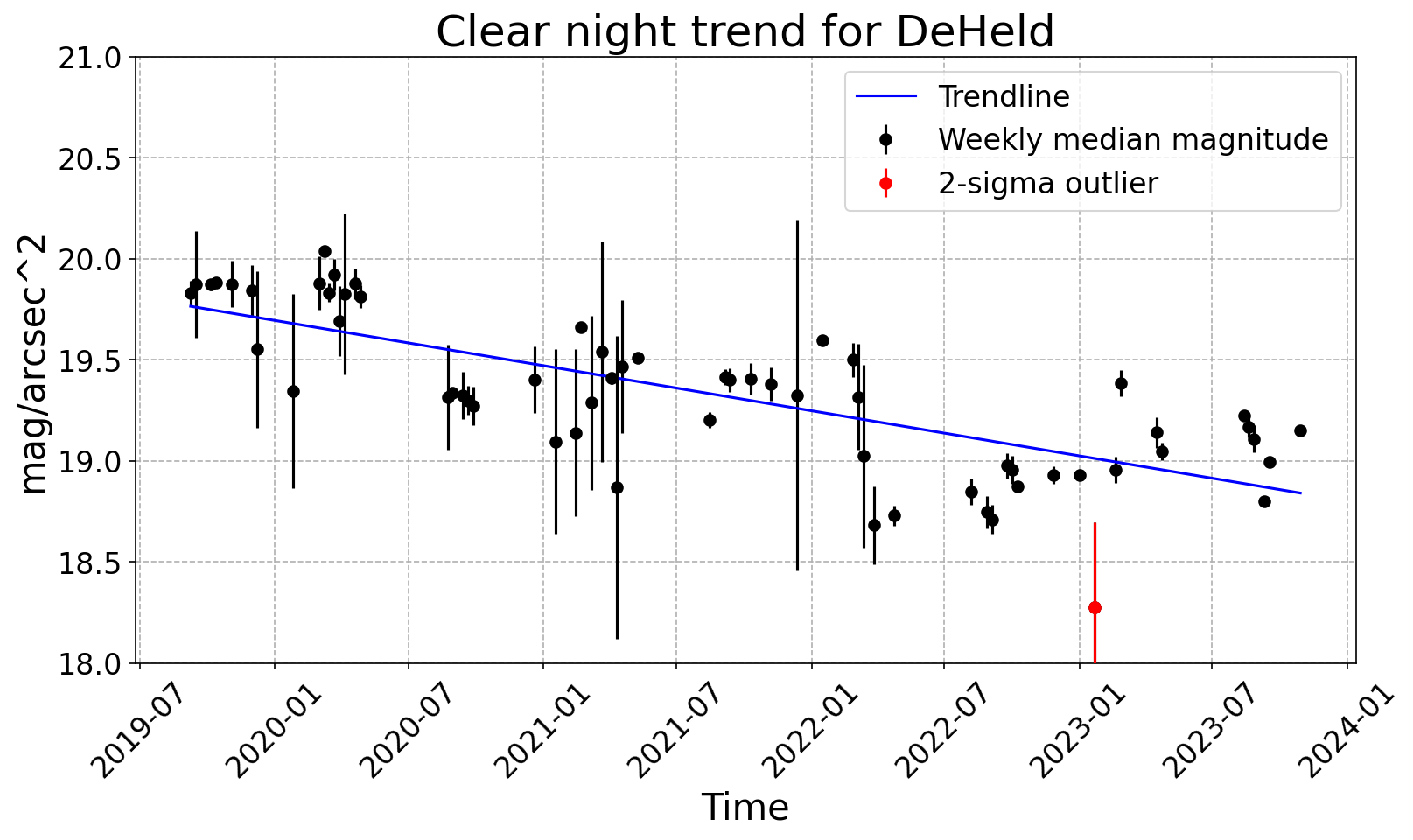}
    \caption{Trend of clear nights for De Held, Groningen.}
    \label{fig:deheldtrend}
\end{figure}

The analysis of sky brightness across multiple locations yields valuable insights into temporal variations and underlying factors influencing light levels. From Lauwersoog to Lochem to De Held in Groningen, each locale presents unique characteristics contributing to observed trends. Table~\ref{tab:overallslopes} summarizes the change in magnitude per year, standard error, and percent change per year for all locations. Here, the standard error of the slope was calculated using the residuals from a linear least squares regression. Specifically, after fitting a linear model to the weekly median magnitudes over time (in years), the standard error of the slope $\sigma_m$, was calculated using the formula:
\[
\sigma_m = \sqrt{\frac{\chi^2_\nu}{\sum (x_i - \bar{x})^2}}
\]
where $\chi^2_\nu$ is the reduced chi-squared (sum of squared residuals divided by degrees of freedom), $x_i$ is time in years, and $\bar{x}$ is the mean of $x_i$. This accounts for both the scatter in the data and the temporal distribution of observations. The time-magnitude plots generated for other locations can be found in Appendix Section~\ref{appendixtrend}.

\begin{table*}
 \caption{Results of trend analysis. A negative change indicates the night sky is getting brighter. These are 27 locations from the \textit{Was het donker} network. The remaining were either newly established, non-operational at the time, or lacked the continuous long-term data required for this specific analysis. Starting NSBs represent the brightness levels during January 2020. The unit for starting NSB, change, and uncertainties is mag/arcsec\textsuperscript{2}. Latitude and longitude are in degrees North and East, respectively.}
 \label{tab:overallslopes}
 \begin{tabular}{llrrrrrr}
  \hline
  Location & Lat, Lon & Starting NSB & Change per year & $\sigma_{\rm stat}$ & $\sigma_{\rm gap}$ & $\sigma_{\rm tot}$ & Percent Change per year \\
  \hline
    Hornhuizen       & 53.403, 6.352 & 21.770 & -0.070 & 0.023 & 0.018 & 0.029 & -6.68\% \\
    Vlieland-Oost    & 53.295, 5.092 & 21.627 & -0.041 & 0.026 & 0.023 & 0.035 & -3.89\% \\
    Sellingen        & 52.938, 7.131 & 21.611 & -0.015 & 0.019 & 0.021 & 0.028 & -1.41\% \\
    Moddergat        & 53.406, 6.069 & 21.518 & -0.073 & 0.040 & 0.020 & 0.045 & -6.94\% \\
    Ostland          & 53.607, 6.727 & 21.513 & -0.029 & 0.031 & 0.017 & 0.035 & -2.67\% \\
    Lauwersoog       & 53.385, 6.235 & 21.483 & -0.027 & 0.015 & 0.008 & 0.017 & -2.48\% \\
    Nes              & 53.449, 5.775 & 21.433 &  0.081 & 0.025 & 0.026 & 0.036 &  7.20\% \\
    Hippolytushoef   & 52.929, 4.986 & 21.219 & -0.010 & 0.022 & 0.014 & 0.026 & -0.96\% \\
    Roodeschool      & 53.412, 6.755 & 21.038 & -0.013 & 0.034 & 0.020 & 0.039 & -1.19\% \\
    Boerakker        & 53.187, 6.329 & 20.988 & -0.124 & 0.037 & 0.029 & 0.047 & -12.05\% \\
    Texel            & 53.003, 4.787 & 20.970 & -0.001 & 0.027 & 0.028 & 0.039 & -0.08\% \\
    Borkum           & 53.587, 6.663 & 20.969 &  0.036 & 0.051 & 0.065 & 0.082 &  3.22\% \\
    Oostkapelle      & 51.572, 3.537 & 20.821 &  0.007 & 0.016 & 0.011 & 0.020 &  0.62\% \\
    tZandt           & 53.369, 6.771 & 20.810 & -0.009 & 0.140 & 0.086 & 0.164 & -0.80\% \\
    Gorredijk        & 53.008, 6.074 & 20.809 &  0.020 & 0.034 & 0.039 & 0.052 &  1.86\% \\
    Weerribben       & 52.788, 5.938 & 20.764 &  0.091 & 0.027 & 0.013 & 0.030 &  8.03\% \\
    Lochem           & 52.172, 6.401 & 20.662 & -0.040 & 0.011 & 0.013 & 0.017 & -3.78\% \\
    Haaksbergen      & 52.149, 6.718 & 20.387 & -0.027 & 0.016 & 0.025 & 0.029 & -2.47\% \\
    Heerenveen-St.   & 52.960, 5.915 & 20.151 & -0.005 & 0.021 & 0.019 & 0.028 & -0.48\% \\
    Tolbert          & 53.165, 6.359 & 20.117 &  0.083 & 0.071 & 0.071 & 0.100 &  7.34\% \\
    Heerenveen       & 52.967, 5.940 & 20.084 & -0.085 & 0.031 & 0.017 & 0.035 & -8.09\% \\
    Oldenburg        & 53.152, 8.165 & 19.920 &  0.125 & 0.040 & 0.031 & 0.051 & 10.87\% \\
    Zernike          & 53.240, 6.536 & 19.882 &  0.004 & 0.070 & 0.050 & 0.086 &  0.34\% \\
    DeHeld           & 53.228, 6.512 & 19.695 & -0.223 & 0.024 & 0.018 & 0.030 & -22.83\% \\
    DeZilk           & 52.301, 4.542 & 19.514 &  0.015 & 0.020 & 0.017 & 0.024 &  1.38\% \\
    Leiden           & 52.155, 4.483 & 18.910 &  0.049 & 0.024 & 0.026 & 0.035 &  4.44\% \\
    Rijswijk         & 52.026, 4.314 & 18.160 &  0.178 & 0.028 & 0.031 & 0.042 & 15.11\% \\
  \hline
 \end{tabular}
\end{table*}

\subsubsection{Overall trend}
\begin{figure}
	\includegraphics[width=\columnwidth]{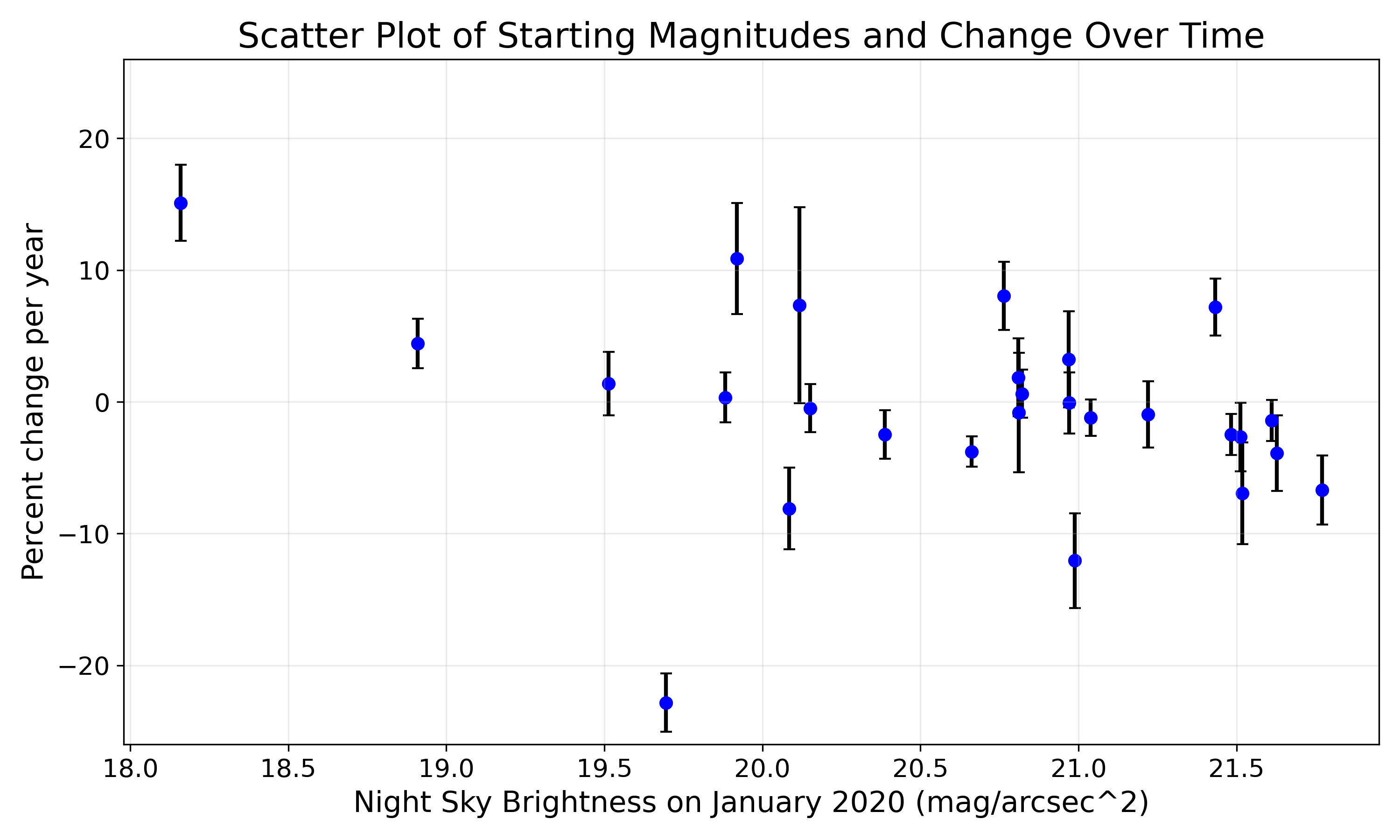}
    \caption{Percent change per year (over the measured period) relative to the starting magnitudes, where negative and positive changes indicate increasing and decreasing brightness, respectively.}
    \label{overallscatter}
\end{figure}

We further examined the results by plotting the starting magnitudes for each location against the percent change per year (for the measured period). The starting magnitudes are characterized as the sky brightness levels in January 2020, which is the earliest date when measurements were available for all the chosen locations. We considered dark locations to be those with their starting night sky brightnesses to be above 21~mag/arcsec\textsuperscript{2}. For the dark locations, the standard deviation of percent change was 4.9 and the median absolute deviation from zero was 2.7. On the other hand, for bright locations, the values were 8.8 and 3.5 respectively. All the darkest locations (except Nes) seem to be getting brighter. Considering the SQMs exhibit a darkening effect over time, the actual change is even larger. The network is mostly present in dark locations in the north of The Netherlands, but the few stations there are in brighter places exhibited greater variability when it comes to changes over time in the measured period. This data is summarized in Fig.~\ref{overallscatter}.

\subsubsection{Seasonal trend (effect of the Milky Way)}
\begin{figure}
	\includegraphics[width=\columnwidth]{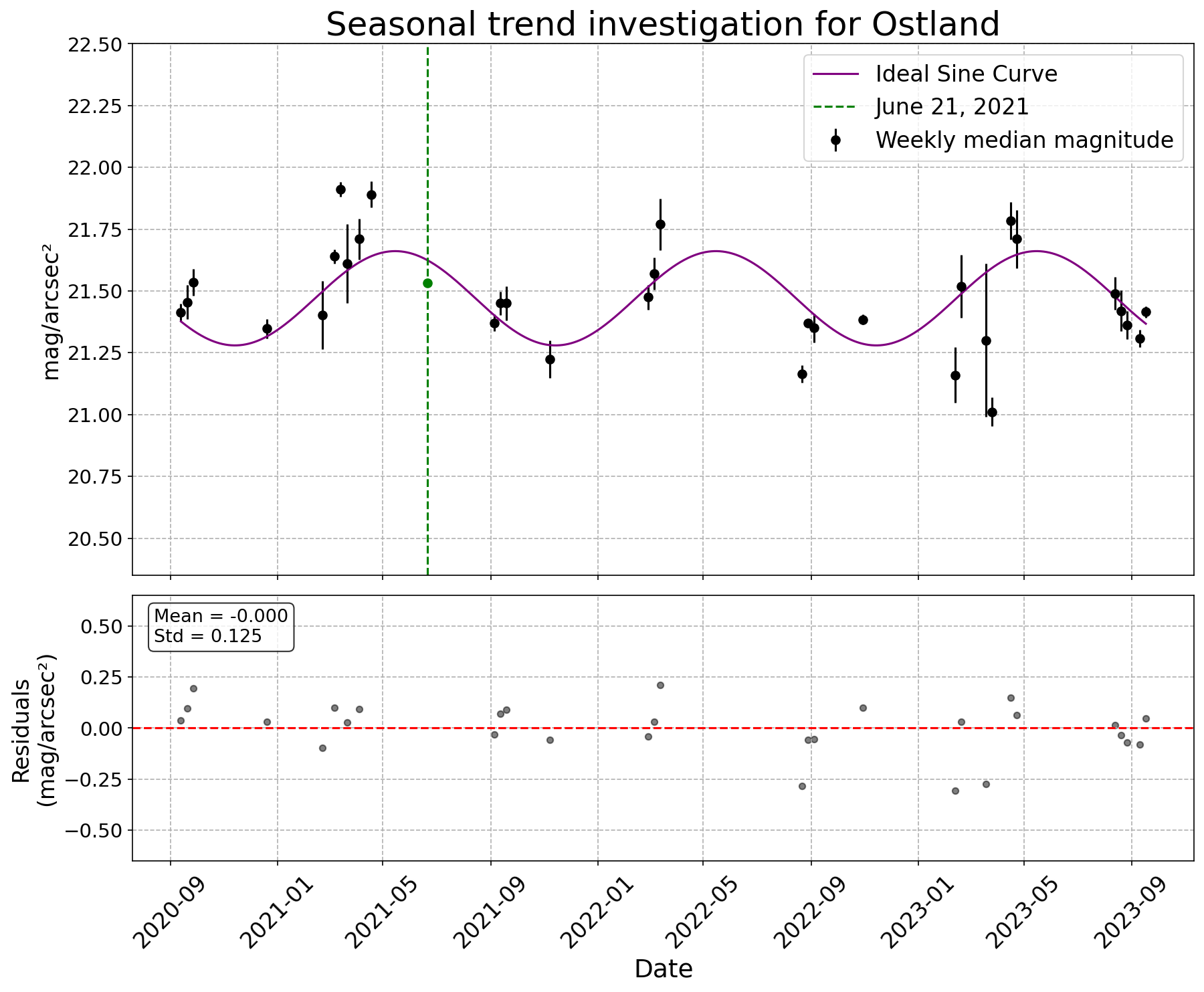}
    \caption{Fitting a sinusoidal curve to the Borkum-Ostland trend plot with a summer crest and winter trough. Fitted Amplitude = 0.191}
    \label{milkyost}
\end{figure}

\begin{figure}
	\includegraphics[width=\columnwidth]{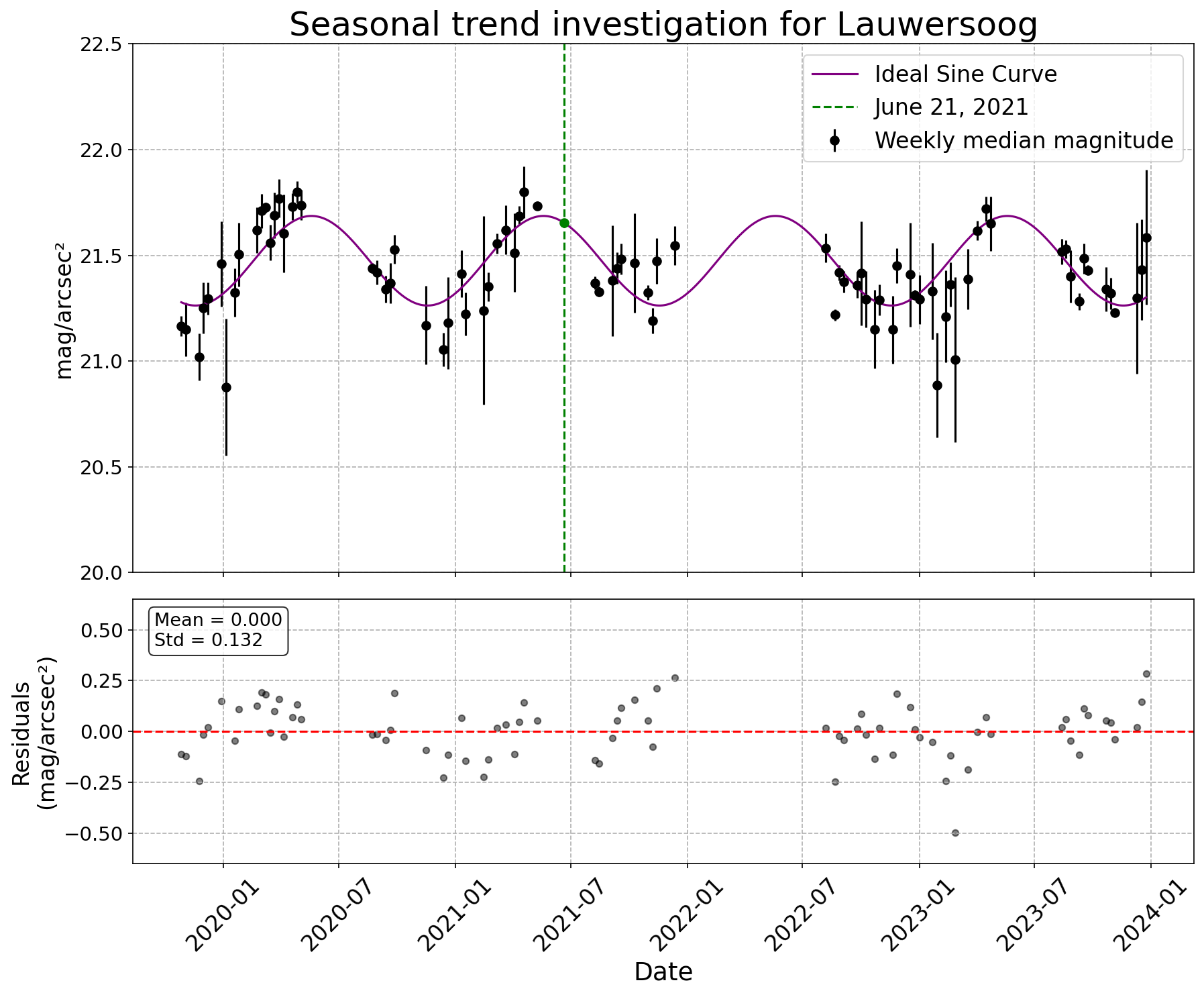}
    \caption{Fitting a sinusoidal curve to the Lauwersoog trend plot with a summer crest and winter trough. Fitted Amplitude = 0.212}
    \label{milkylau}
\end{figure}

\begin{figure}
	\includegraphics[width=\columnwidth]{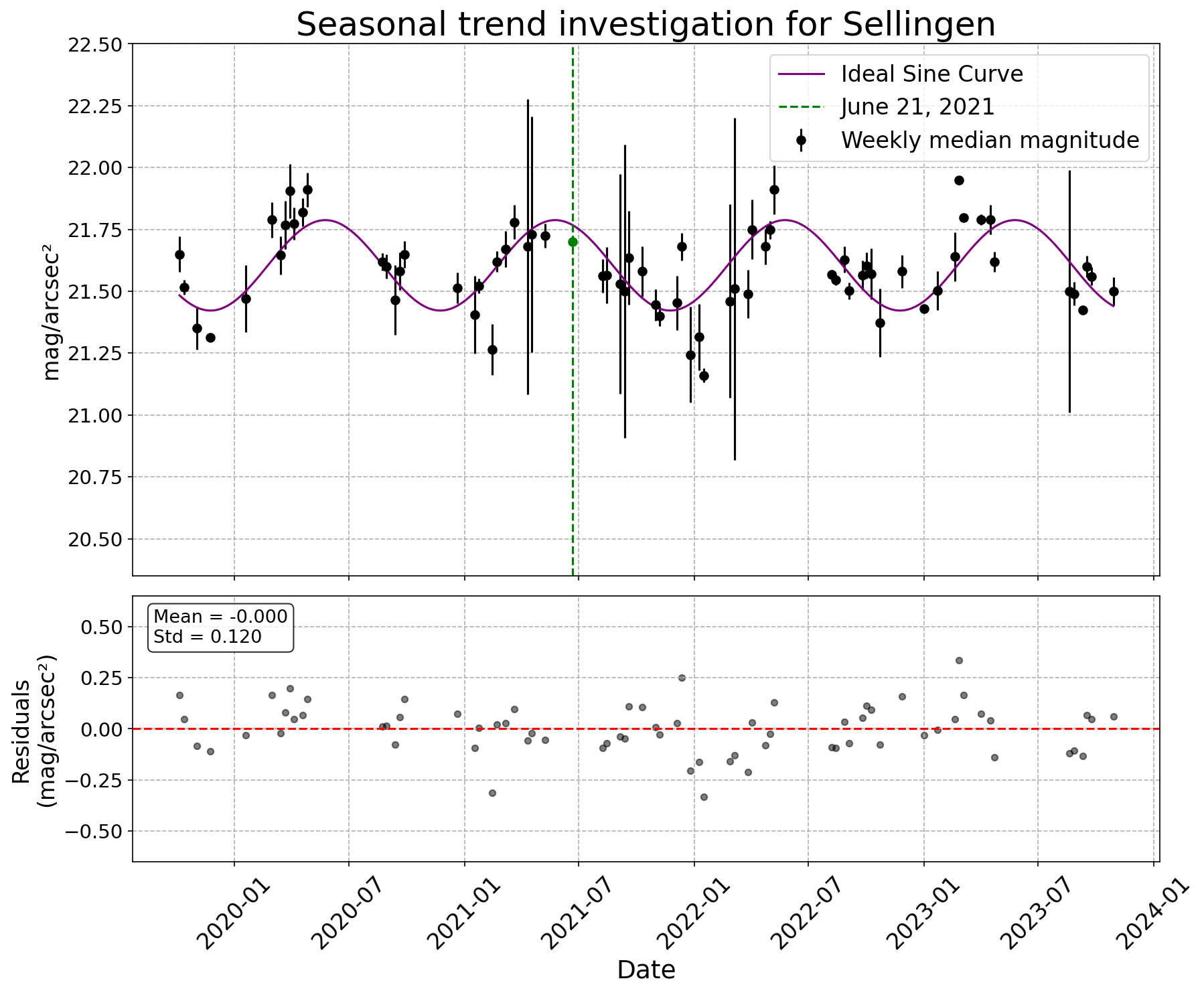}
    \caption{Fitting a sinusoidal curve to the Sellingen trend plot with a summer crest and winter trough. Fitted Amplitude = 0.183}
    \label{milkysel}
\end{figure}

An important yet overlooked factor in NSB variability is the contribution of natural celestial sources, particularly the Milky Way. As noted by \citet{padua}, the Milky Way is close to the northern horizon in spring when their dark site "Passo Valles" has a maximum value of NSB. This pattern aligns with the fact that in the northern hemisphere, the Milky Way becomes prominent in the zenith region as summer approaches. We also observe a seasonal pattern in some of our darkest locations: Borkum-Ostland, Lauwersoog, and Sellingen. Fig.~\ref{milkyost}, Fig.~\ref{milkylau}, and Fig.~\ref{milkysel} show the plots fitted with a sinusoidal curve of a one year frequency, peaking in the summer. Appendix Section \ref{seasonal_appendix} shows 6 other locations. The lower halves plot the residuals. This method is only useful in extremely dark sites where the Milky Way or galactic core significantly impacts the sky brightness, which is why these three locations were chosen for this analysis. In all figures, June 21, 2021 is shown as a reference point for summer.

\subsection{"Jellyfish" Plots} \label{jellyfishplotsresultssection}

Density plots (or "jellyfish" plots) are a common way of presenting night sky brightness data \citep{posch2018systematic}. They essentially involve plotting the entire dataset on a single graph. They resemble the tendrils of a jellyfish, hence the name. We note that there was no weather (or Sun/Moon) exclusion carried out for these plots. The brightness values are plotted with respect to the hours of the night. These plots, in a visually intuitive manner, illustrate the predominant magnitude of the NSB observed at any specific location.

\begin{figure}
	\includegraphics[width=\columnwidth]{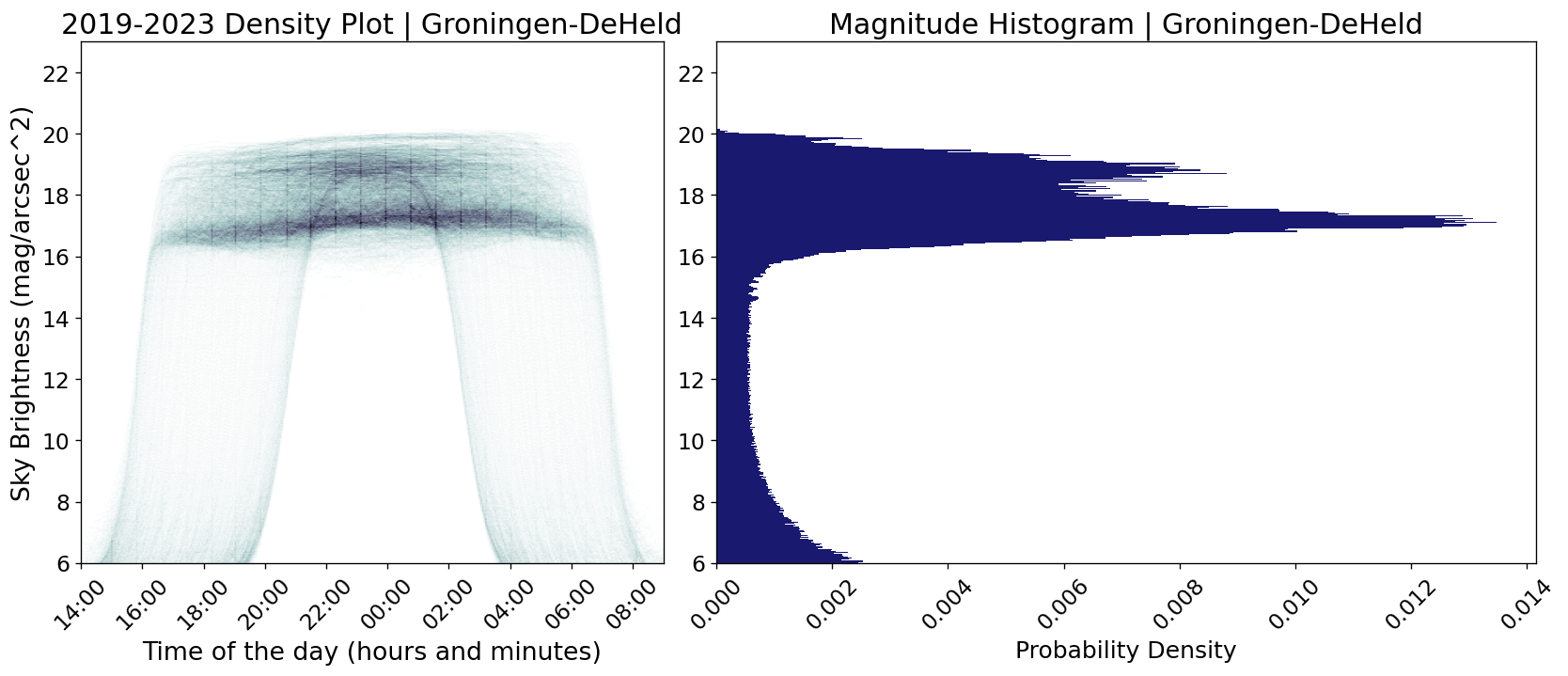}
  \caption{"Jellyfish" (left) and density plot (right) representing a bright area (Groningen-DeHeld).}
  \label{fig:sfig00}
\end{figure}

\begin{figure}
	\includegraphics[width=\columnwidth]{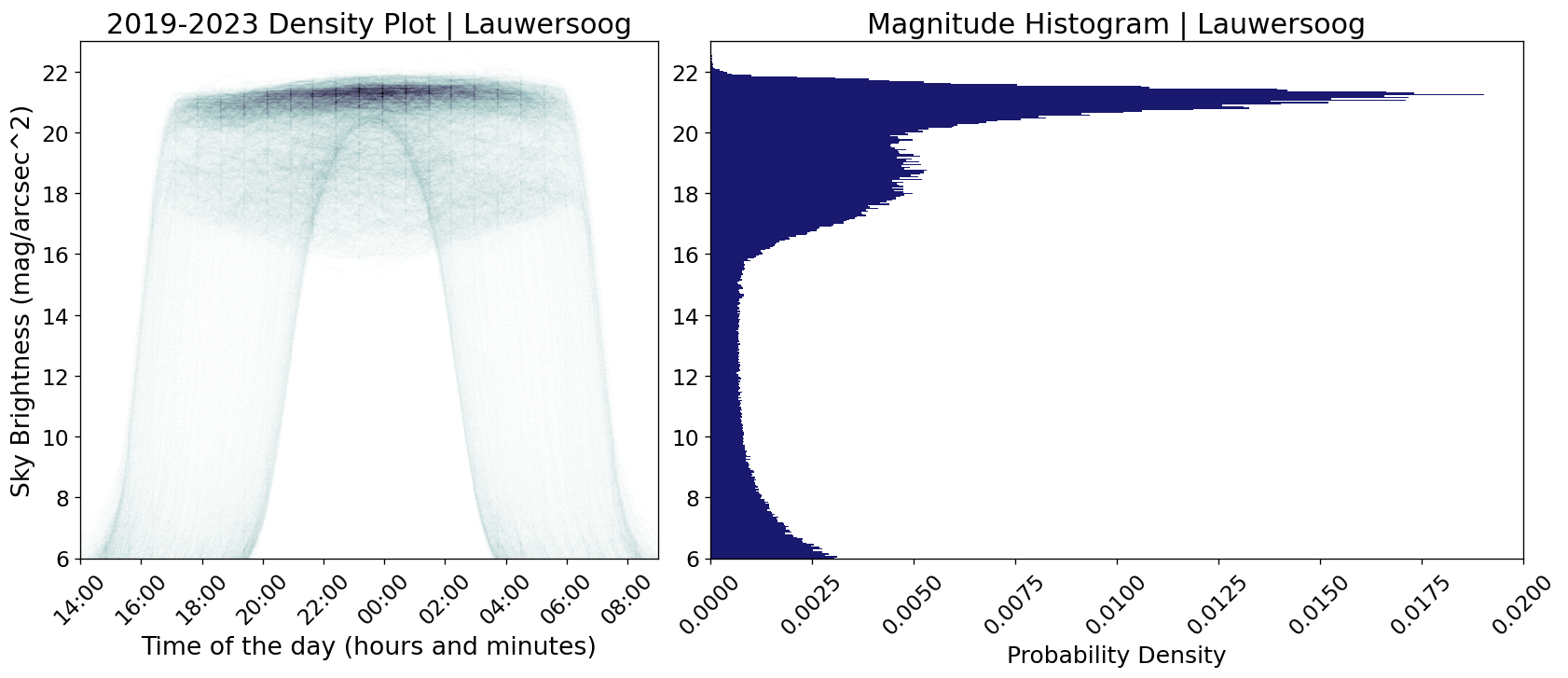}
  \caption{"Jellyfish" (left) and density plot (right) representing a dark area (Lauwersoog).}
  \label{fig:sfig01}
\end{figure}

Fig.~\ref{fig:sfig00} and Fig.~\ref{fig:sfig01} show the density plots for two locations (Groningen~DeHeld and Lauwersoog) in our data set. Each graph is divided into two subplots. The left halves depict "jellyfish" plots with time of day and magnitudes. The right halves provide a more quantitative representation of the same data, showcasing a histogram of the night sky brightness values based on relative frequencies.

Overall, bright locations such as cities exhibit two major peaks. The lower magnitude peak indicates cloudy nights, while the higher magnitude one is indicative of clear nights. Taking the example of two bright locations, \textit{Groningen-DeHeld} and \textit{Groningen-ZernikeCampus}, both display dual peaks. De Held has one at 17.16±0.56 mag/arcsec\textsuperscript{2} and the other at 18.81±0.72 mag/arcsec\textsuperscript{2}. Zernike Campus has one at 16.87±0.70 mag/arcsec\textsuperscript{2} and the other at 19.67±0.52 mag/arcsec\textsuperscript{2}. For both locations, the lower magnitudes occur more often. This is attributed to the region experiencing more cloudy nights throughout the year than clear ones. Clouds, with their light scattering effect, contribute to an increase in the brightness of the night sky. Although both De Held and Zernike Campus are in the same city, De Held stands out as notably bright during clear nights due to ongoing construction work.

Conversely, darker locations tend to exhibit a singular peak. Lauwersoog (a dark sky park) reaches its peak at 21.14±0.51 mag/arcsec\textsuperscript{2}. A noticeable bump at 18.78±1.74 mag/arcsec\textsuperscript{2} is observed, attributed to factors such as the influence of the moon, fog, and snow. There are some instances of high magnitude nights above the peak, indicating overcast nights as clouds in darker regions obstruct the light from stars, making the night sky even darker.

\begin{figure}
	\includegraphics[width=\columnwidth]{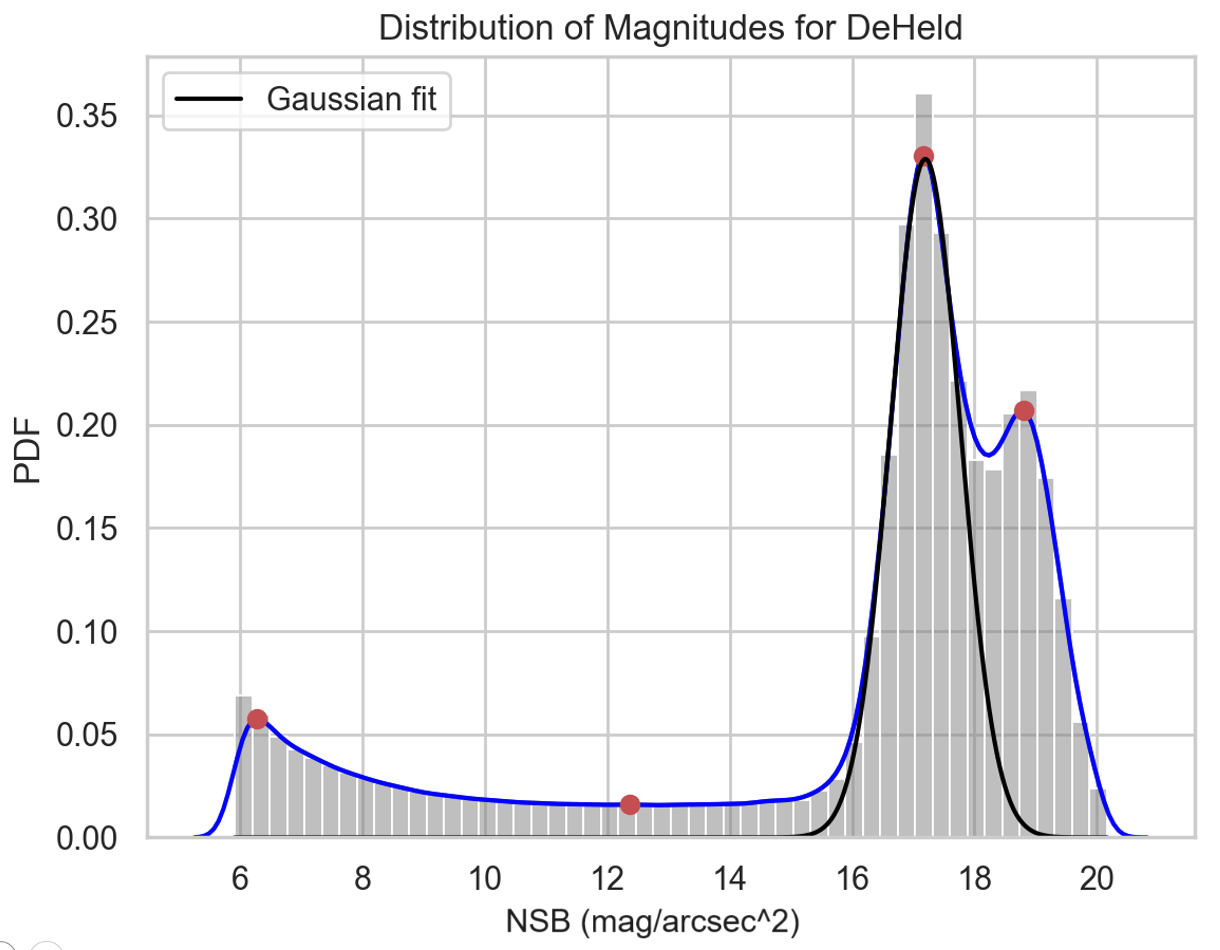}
  \caption{Example from Groningen-DeHeld illustrates how we fit a Gaussian to the brighter peak.}
  \label{deheldpeaks}
\end{figure}

\begin{figure}
	\includegraphics[width=\columnwidth]{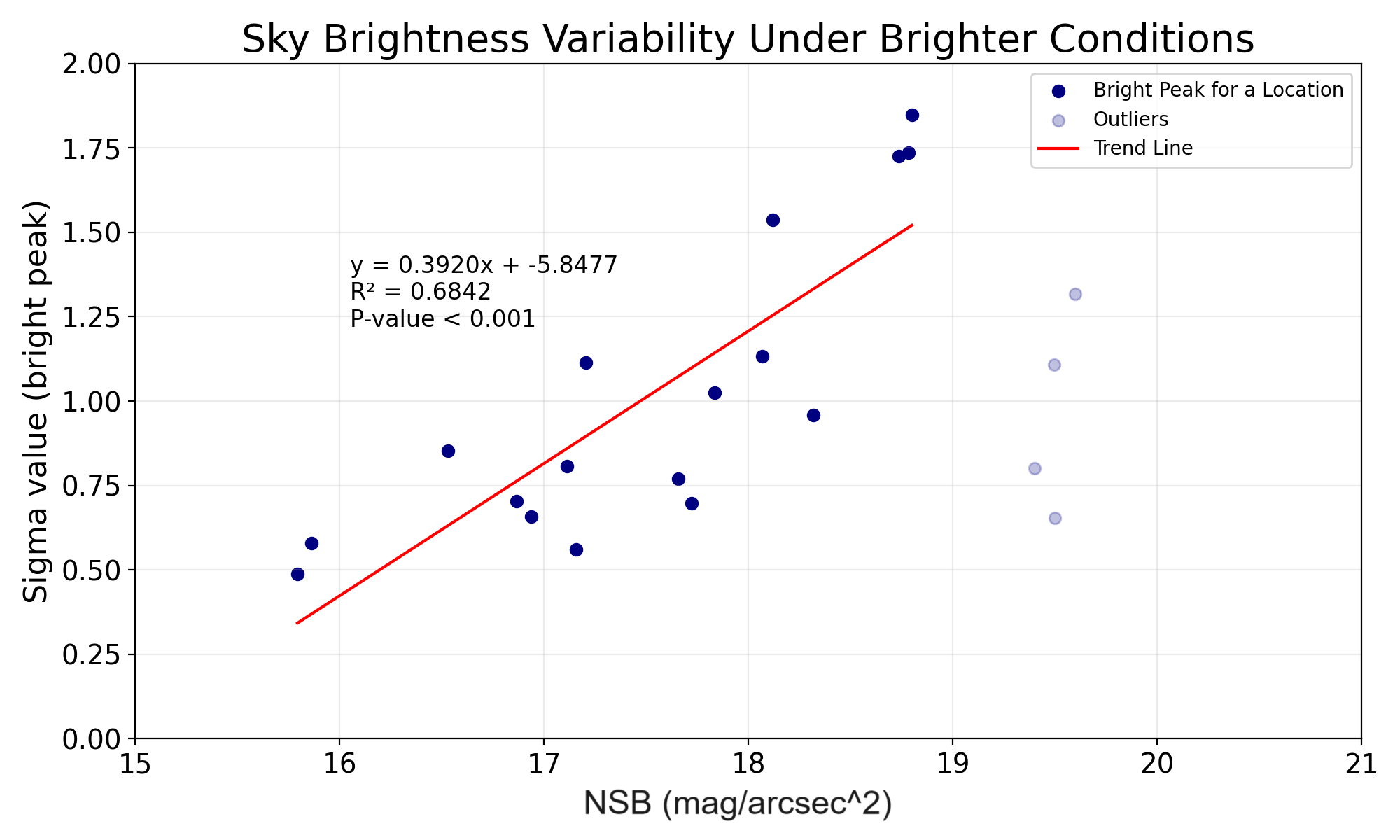}
    \caption{Final plot of all bright peaks vs. their widths.}
    \label{cloudpeak}
\end{figure}

The "jellyfish" plots generated for other locations can be found in appendix Section~\ref{appendixjelly}. The corresponding histograms for every location were examined further and the results are shown in Table~\ref{tab:peaks}. We sorted the locations into two categories, those having two clear peaks, and those with a single peak. For the two peak group, we looked at the magnitude where the brighter peak laid, fit a Gaussian to it, and calculated the standard deviation value for that peak. The example of Groningen-DeHeld is shown in Fig.~\ref{deheldpeaks}. The standard deviation values represent the width of the peak and were plotted against the magnitudes. We observed a linear trend (excluding higher magnitudes) as shown in Fig.~\ref{cloudpeak}. As locations became brighter, the variability of sky brightness during cloudy conditions decreased, and vice versa. This phenomenon can be explained by considering that as locations become brighter, factors such as the moon and back-scatter from clouds become less significant. We note that this trend breaks after the bright peak crosses 19 mag/arcsec\textsuperscript{2} as visible in Fig.~\ref{cloudpeak}. Artificial light emerges as the primary factor in brightening the sky, and overshadows other factors.

\begin{table*}
 \caption{Night sky brightness values in mag/arcsec\textsuperscript{2} for dark and bright (cloud) peaks with their corresponding standard deviations for various locations. - indicates the absence of the corresponding peak.}
 \label{tab:peaks}
 \begin{tabular}{rrrrr}
  \hline
  Location&Dark peak&Sigma of dark peak&Cloud peak&Sigma of cloud peak \\
  \hline
        Hornhuizen&21.377&0.457&18.799&1.849 \\
        Vlieland-Oost&21.259&0.494&-&- \\
        Nes&21.237&0.640&19.597&1.318 \\
        Sellingen&21.172&0.612&-&- \\
        Ostland&21.194&0.462&18.120&1.538 \\
        Lauwersoog&21.141&0.512&18.784&1.737 \\
        Moddergat&21.017&0.484&18.736&1.726 \\
        Roodeschool&20.777&0.609&19.497&0.654 \\
        Borkum&20.750&0.627&18.068&1.133 \\
        Texel&20.662&0.604&19.495&1.108 \\
        Weerribben&20.623&0.691&-&- \\
        Oostkapelle&20.574&0.625&19.398&0.802 \\
        Gorredijk&20.571&0.622&18.317&0.958 \\
        Lochem&20.402&0.434&17.658&0.770 \\
        Hippolytushoef&20.190&0.773&-&- \\
        Haaksbergen&20.098&0.475&17.114&0.807 \\
        Tolbert&20.080&0.676&17.833&1.025 \\
        Oldenburg&19.952&0.609&17.204&1.115 \\
        Heerenveen St.&19.908&0.533&16.939&0.658 \\
        tZandt&19.735&0.870&-&- \\
        Zernike&19.674&0.518&16.866&0.703 \\
        Heerenveen&19.519&0.582&16.529&0.854 \\
        DeZilk&19.425&0.438&17.722&0.697 \\
        Boerakker&19.352&0.811&-&- \\
        Erica&18.945&0.757&14.638&1.744 \\
        Leiden&18.854&0.433&15.864&0.580 \\
        DeHeld&18.805&0.716&17.157&0.560 \\
        Rijswijk&18.451&0.505&15.793&0.489 \\
  \hline
 \end{tabular}
\end{table*}

\subsubsection{Special case: Erica}
\begin{figure}
	\includegraphics[width=\columnwidth]{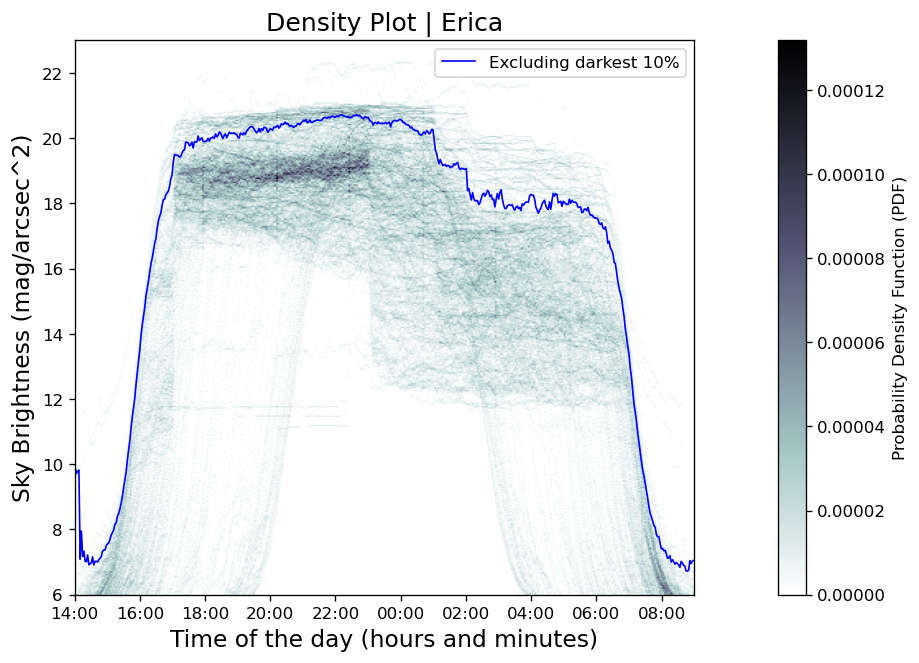}
    \caption{"Jellyfish" plot for Erica (NL) station (whole year).}
    \label{ericaspecial}
\end{figure}

\begin{figure}
	\includegraphics[width=\columnwidth]{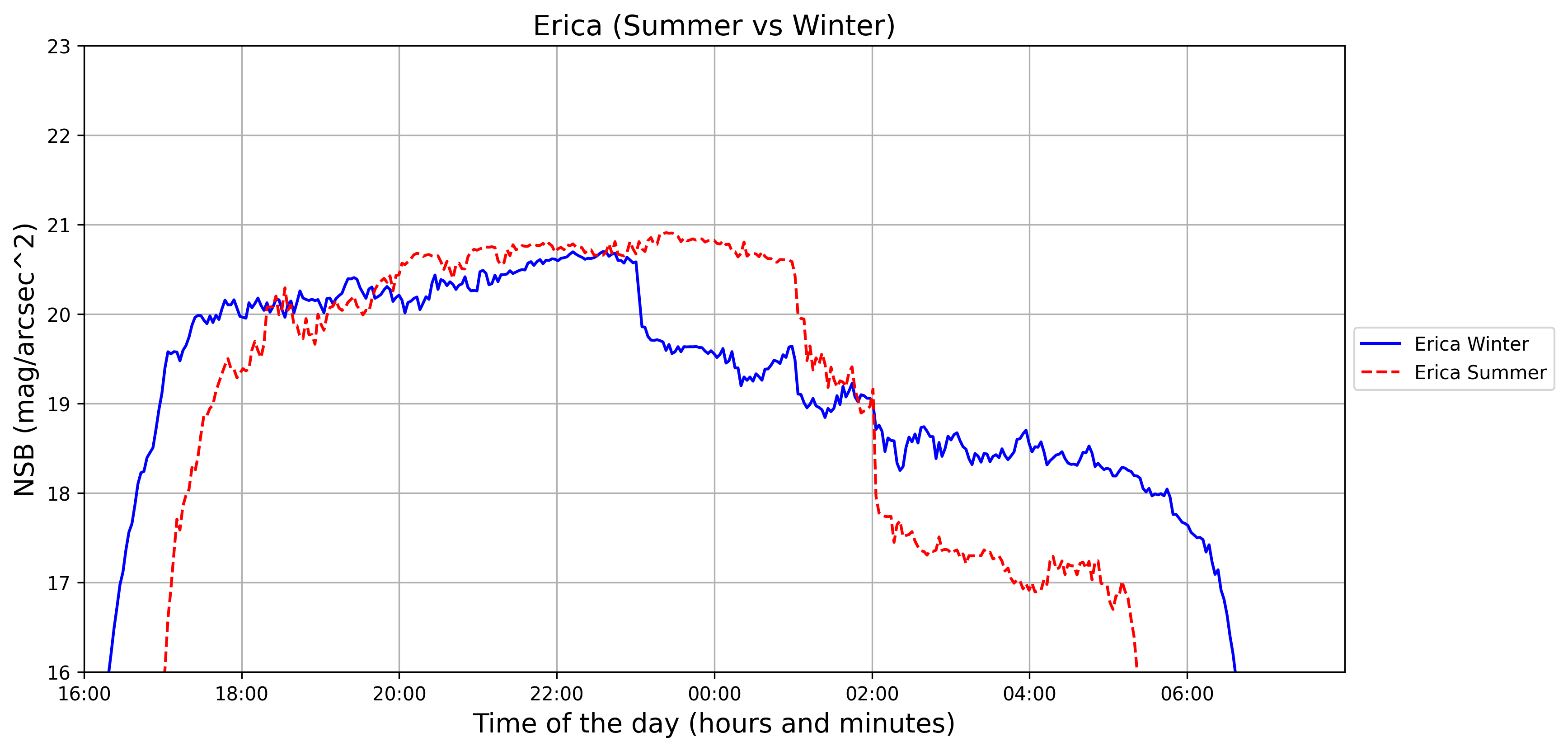}
    \caption{Difference between summer and winter at Erica (NL) station.}
    \label{closererica}
\end{figure}

Erica is a town in north-eastern Netherlands. The SQM present in that location is placed close to a greenhouse. The greenhouses, after around midnight, are allowed to turn on their lights. This produces a significant increase in night-sky brightness at a specific time of the night and it is easily visualized in the corresponding "jellyfish" plot shown in Fig.~\ref{ericaspecial}. Fig.~\ref{closererica} displays a blue line representing winter and a red line representing summer. The time on the x-axis is UTC. A noticeable observation is that the significant surge in night sky brightness occurs at distinct local times (offset from UTC by +1 hour in winter and +2 hours in summer) for each season. During winter, this increase happens at midnight, with a further spike at 2 a.m. However, in summer, the spikes are at 3 a.m. and 4 a.m. This variance may be attributed to the later sunset during summer months, likely reducing the necessity to activate greenhouse lighting as early as during winter.

\section{Discussion}
SQMs offer a cost-effective means of measuring nighttime brightness, yet their one-dimensional nature poses a notable limitation. Without visual data, pinpointing the specific causes behind variations in brightness is challenging, necessitating reliance on third-party data for comprehensive analysis. Despite this drawback, the affordability and demonstrated capabilities of SQMs render them invaluable tools for monitoring light pollution.

Our results underline the importance of insights derived from SQM data, particularly through the generation of informative visualizations like "jellyfish" plots. "Jellyfish" plots can serve as powerful tools for public awareness campaigns and aiding authorities in monitoring efforts. For example, the significant increase in brightness after midnight in the Erica region (NL), attributed to greenhouse lighting, highlights the practical utility of SQMs in identifying sources of light pollution. Our use of jellyfish plots for visualizing NSB variability aligns with the methodologies employed by \citet{posch2018systematic}, who used similar representations to highlight the divergence between urban and rural sites. They observed bimodal NSB distributions in urban areas, similar to ours, and commented how the width of the brightness peaks changed depending on various effects. We delve deeper into this relationship and observe a linear trend up to a NSB of 18.8 mag/arcsec\textsuperscript{2}.

\subsection{Limitations of Trend Analyses and Future Directions}\label{sec:discussion_limitations}

\begin{figure}
	\includegraphics[width=\columnwidth]{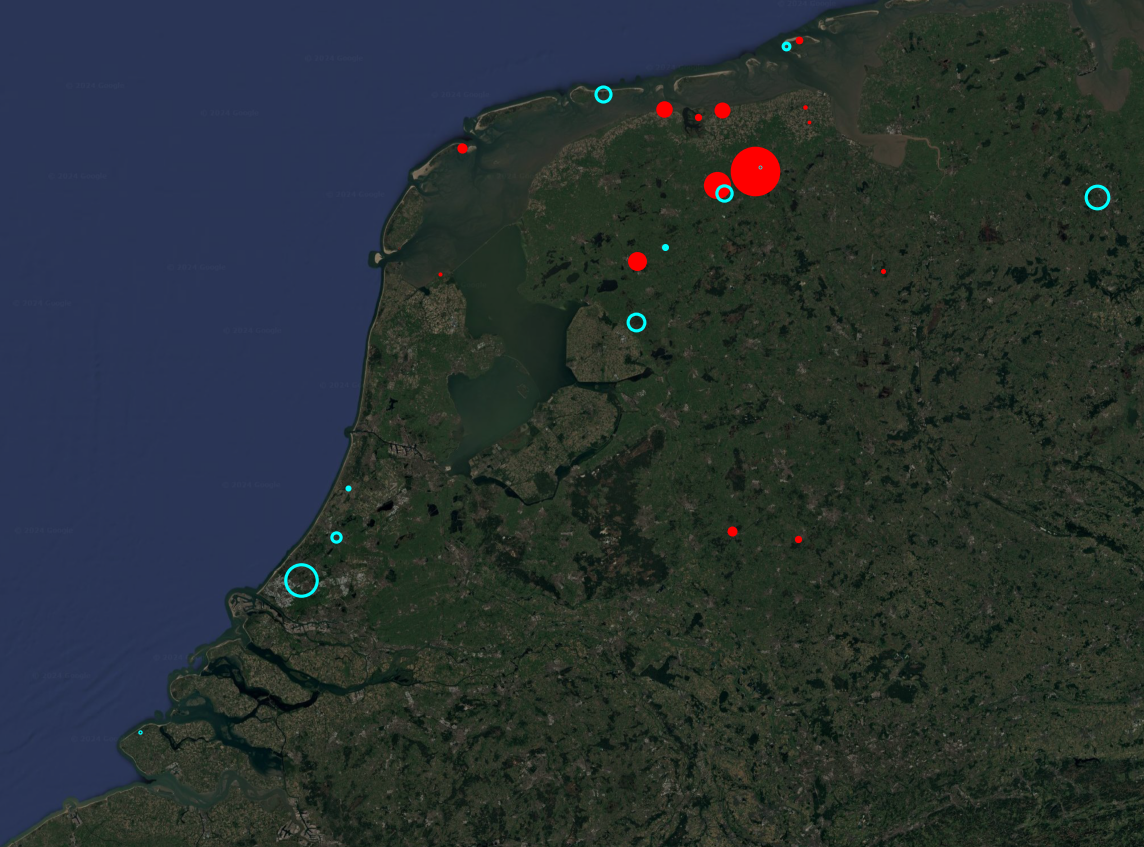}
    \caption{Changes for each location plotted on a map of The Netherlands. The bigger the circle, the larger the change was. Filled red indicates an increase in brightness while unfilled cyan indicates a decrease. Some bright sites, notably Leiden, decreased in brightness.}
    \label{cartopy}
\end{figure}

Trend analysis, depicted in Fig.~\ref{cartopy}, plots changes in brightness on a map, illustrating regional variations. However, without additional contextual data, understanding the underlying drivers of these trends remains elusive. Continued deployment of SQMs, coupled with the integration of All Sky Cameras into the network, will provide a more nuanced understanding of localized variations in light pollution. 

\paragraph*{Sampling bias and its impact.} Our analysis of data availability (Section~\ref{sec:sampling_patterns}) show that data gaps are not random but follow a seasonal pattern. The severe under-representation of summer months and the clustering of data in spring can bias the linear trend fits. While our bootstrap method, which re-samples the presence/absence mask of calendar weeks, accounts for the specific pattern of missing data, it still implicitly assumes that the underlying reasons for missingness are random. To see whether this sampling bias systematically influences our results, we can examine the residuals of the linear fits, as shown in Figs.~\ref{milkyost}, \ref{milkylau}, and~\ref{milkysel}.

\paragraph*{Seasonal effects in residuals.} For seasonal trends in our study, we observed brightness peaks during winter for dark locations like Lauwersoog, Borkum-Ostland, and Sellingen parallel those reported by \citet{padua} who attributed such patterns to snow cover and the Milky Way. In the Netherlands during this period, snow cover can, however, be neglected. In our Milky Way investigations, we fit a sine curve to our darkest locations. \citet{wallner2023reliability} previously used a triangular model on dark sky areas in Austria with fluctuating residuals, indicating a good capture of trends. For us however, the results are varying. Although the sinusoidal trend is present, there appear to be additional effects influencing the residuals. The residual patterns across our sites show varying degrees of structure, with some locations displaying more systematic deviations than others, but generally without strong seasonal periodicity that would indicate systematic sampling bias. Appendix Section \ref{seasonal_appendix} shows 6 other locations. Nevertheless, in all three locations, visually there seems to be a gradual increase of darkness of 0.5 magnitudes from Winter to Spring and subsequent drop. Since our data cleaning steps already filtered out data with clouds and the Moon, we attribute this seasonal fluctuation to the geometric position and seasonal visibility of the Milky Way.

\paragraph*{Long-term atmospheric variability.} Our filtering removed the three dominant atmospheric contributors (Sun, Moon, and clouds). Other residual meteorological signals could arise from small year‑to‑year shifts in aerosol optical depth (AOD) or tropospheric transparency. A check of the Copernicus CAMS \citep{Cop} re‑analysis for the Netherlands (2020--2023) shows quarterly total AOD values (at 550nm) centered on a mean of $0.132$, with deviations ranging from $-0.060$ to $+0.077$ (maximum absolute deviation $\pm0.077$), and no significant monotonic trend (slope~$=0.00022\ \mathrm{quarter}^{-1}$, $p=0.93$). Assuming Rayleigh‑like scattering, (i.e.\ weighting the aerosol optical‑depth fluctuation by the standard Rayleigh phase function at zenith) and converting optical‑depth changes to magnitudes via the Beer–Lambert law \citep{hansen1974light,BohrenHuffman1983}, a maximum $\Delta\mathrm{AOD}=0.077$ corresponds to $\Delta m \lesssim 0.010\,\mathrm{mag\,arcsec^{-2}}$. A fluctuation of this magnitude would change the zenith sky luminance by $\lesssim0.010\,\mathrm{mag\,arcsec}^{-2}$--- less than our total error floor (Section~\ref{sec:sampling}), which does not include aerosol variability. With the present four‑year baseline we therefore treat the atmosphere as effectively static and interpret our slopes only as \emph{apparent} trends in artificial light at night (ALAN), irrespective of the detailed physical cause. Longer data streams or co‑located aerosol measurements will be needed to decouple ALAN from climate variables---beyond the scope of the current work but a worthwhile goal for future studies.

\paragraph*{Future direction: towards physically motivated models.} A more robust approach for future work would be to move beyond a purely statistical treatment and employ a physically motivated model. Such a model could explicitly account for seasonal brightness variations (e.g., the Milky Way's contribution, albedo changes from vegetation) by fitting a sinusoidal or other periodic function simultaneously with a linear trend. This would allow for a cleaner separation between long-term anthropogenic trends and predictable, periodic fluctuations. Furthermore, integrating ancillary data on AOD and other atmospheric parameters would help to decouple the true ALAN trend from atmospheric variability. This approach would be less susceptible to the sampling biases identified in our data and represents a critical next step for refining light pollution trend analysis.

\subsection{Variability in Sky Brightness}

Our analysis indicates that as locations experience increases in brightness, the overall variability of sky brightness during cloudy conditions decreases.

For clear conditions, brighter sites exhibit greater variability in the direction and magnitude of sky brightness changes (Fig.~\ref{overallscatter}). This can be attributed to factors such as urban development, local lighting policies, weather patterns, and public awareness of light pollution mitigation efforts. Urban areas often exhibit higher variability in light pollution trends compared to rural areas. This variability is influenced by the density of artificial light sources, the type of lighting used (e.g., LED vs. traditional bulbs), and the effectiveness of local light pollution control measures. Urban centers experience more pronounced changes in light pollution levels due to the concentrated presence of artificial light sources and their dynamic nature \citep{kyba2017}.

The observed variability in sky brightness aligns with patterns reported by \citet{kyba2011cloud}, \citet{padua} and \citet{eumet}. \citet{kyba2011cloud} demonstrated that clouds amplify NSB in polluted urban areas while darkening skies in pristine regions, a phenomenon we also observe in sites like Groningen, Oldenburg, Leiden. In contrast, rural sites exhibited reduced variability under similar conditions, echoing the observations by \citet{padua} of more stable NSB distributions in minimally polluted regions. Together, these findings suggest a complex interplay between natural and artificial factors influencing NSB trends.

\subsection{Stability and Correction of SQM Measurements}

The stability of SQMs over time is critical for ensuring accurate long-term measurements of light pollution. A known issue with SQMs is the darkening effect caused by housing degradation and sensor aging, resulting in progressively lower readings. This darkening effect has been quantified in previous studies \citep{baradarkening, twilight, ageing}.

To account for this aging effect, various correction methods can be applied. One approach involves using twilight data as a calibrator, which helps adjust SQM readings for the darkening effect \citep{twilight}. Incorporating such corrections would allow for more accurate measurements of night sky brightness over extended periods.

\section{Conclusion}
In conclusion, the extensive use of SQMs across multiple locations has provided valuable insights into the temporal and spatial variations of night sky brightness. The data reveal that in urban areas, such as De Held in Groningen, there is a variety of behaviour, some places get brighter while others get fainter. This trend is influenced by factors like ongoing construction and local lighting policies, highlighting the importance for targeted light pollution mitigation strategies tailored to specific regions. In addition, for the darkest locations, sky brightness had been increasing between 1 to 7 percent per year.

The analysis also indicates a strong correlation between increased brightness and reduced variability during cloudy conditions in darker areas. This suggests that artificial lighting is beginning to dominate natural light sources in these regions. Continued monitoring and expansion of the SQM network, coupled with the integration of All Sky Cameras, are essential for a more comprehensive understanding of light pollution dynamics. These efforts will aid in identifying critical areas of concern and implementing effective interventions to preserve the integrity of our night skies.

Overall, the KID project has successfully demonstrated the importance of systematic light pollution monitoring and provided a framework for future research and public policy initiatives. By increasing awareness and fostering collaboration among stakeholders, we can mitigate the adverse effects of artificial light at night and protect our natural and urban environments for future generations.

\section*{Acknowledgements}
We acknowledge that the work described in this paper was supported by funding from a project of the Interreg North Sea Region Programme 2021-2027, under the name 'KID – Keep it Dark.'

%%%%%%%%%%%%%%%%%%%%%%%%%%%%%%%%%%%%%%%%%%%%%%%%%%

\section*{Data Availability}
The SQM measurements utilized in this study are publicly accessible through the repository at \texttt{washetdonker.nl}. All computational analyses were performed using custom Python scripts, and have been made available in their entirety \citep{farhancodes}. The Cloud Mask product data from EUMETSAT can be obtained via their Product Navigator platform (\texttt{navigator.eumetsat.int}). All datasets and analysis tools used in this study are available to the astronomical community to facilitate reproducibility of our results.



\bibliographystyle{mnras}
\bibliography{references}

%%%%%%%%%%%%%%%%% APPENDICES %%%%%%%%%%%%%%%%%%%%%

\clearpage

\appendix
\FloatBarrier
\section{Darkness rankings}

\begin{figure}
	\includegraphics[width=\columnwidth]{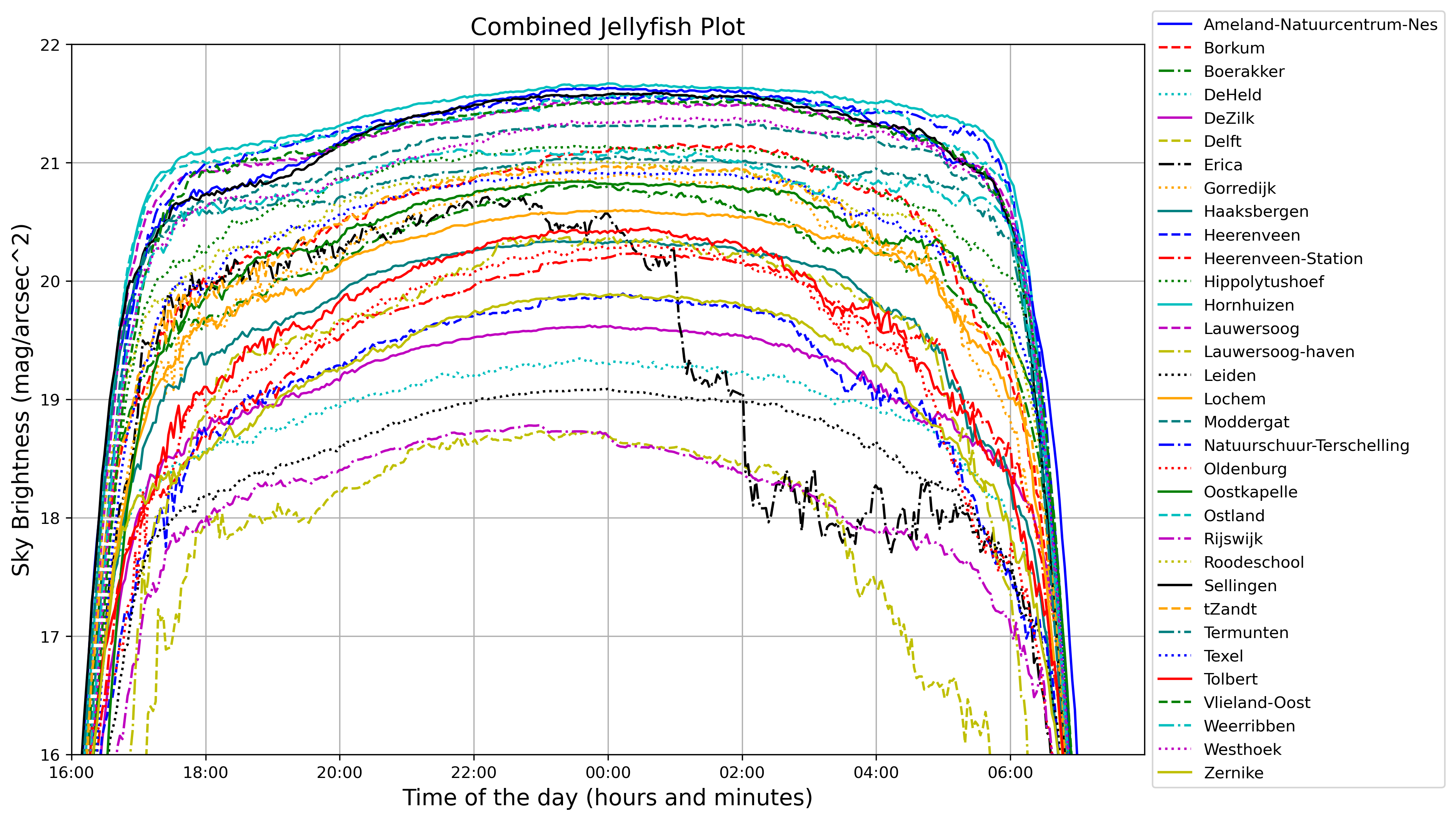}
    \caption{Approximate of clear and moonless nights zone (90th percentile) of each "Jellyfish" combined into one plot.}
    \label{combinedjellyfishplot}
\end{figure}

In Fig.~\ref{combinedjellyfishplot}, we plot the 90th percentile of the night sky brightness data against the time for each location on one graph. This graph essentially combines all the 'jellyfish' plots into one and is done by excluding the darkest 10\% of values. Thus, the lines here approximate the nights of each location when it is clear and moonless. The Figure clearly illustrates the average night sky brightness of the clear sky for each location and is a good way to rank these locations. Most sites typically exhibit a brightness ranging from 18.5 to 21.5 mag/arcsec\textsuperscript{2}, with brightness diminishing as the night progresses, likely due to the people switching off lights.

\section{"Roller coaster" plots}\label{rolsection}

\begin{figure}
	\includegraphics[width=\columnwidth]{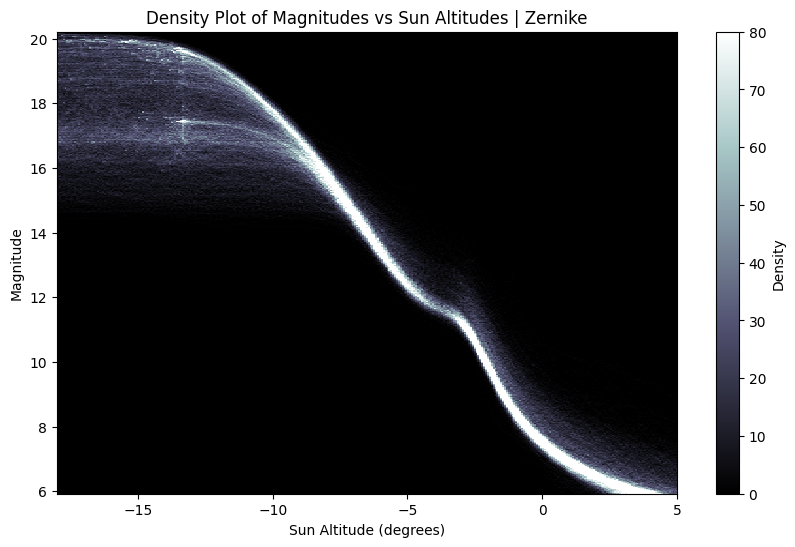}
    \caption{Roller coaster plot of Zernike Campus, Groningen}
    \label{fig:zernikecoaster}
\end{figure}

We examined how SQMs behaved at twilight because we planned to use twilight data to track the ageing effect of these devices over time. In Fig.~\ref{fig:zernikecoaster}, we plotted the altitude of the Sun against the brightness of the sky at Zernike Campus, Groningen. We noticed a change in the slope of the graph when the Sun got to about 3 degrees below the horizon. The shape of the plot resembled a roller coaster, thus we named it "roller coaster" plots. The same condition was observed at other locations.

At first, we thought it was because SQMs are intended to measure brightness between 16 and 23 mag/arcsec\textsuperscript{2}, but this is not the case since their measurement range is stated to be from 7 to 23 mag/arcsec\textsuperscript{2} \citep{sqmle_manual}. The true reason is different. The Python library we used to find the Sun's altitude (PyEphem) adjusts for atmospheric refraction based on the observer's temperature and pressure. If we don't account for this refraction, we don't see the aforementioned change in slope. This is supported by the fact that we get a normal graph when instead of the solar altitude, time is plotted against brightness.

Thus, the reason for the change in slope is because even though the true altitude of the Sun is below the horizon, its apparent altitude is still higher due to atmospheric refraction, which slows down the change in sky brightness.

\section{Receiver Operating Characteristic (ROC) Curve Analyses}\label{rocsection}

\begin{figure}
	\includegraphics[width=\columnwidth]{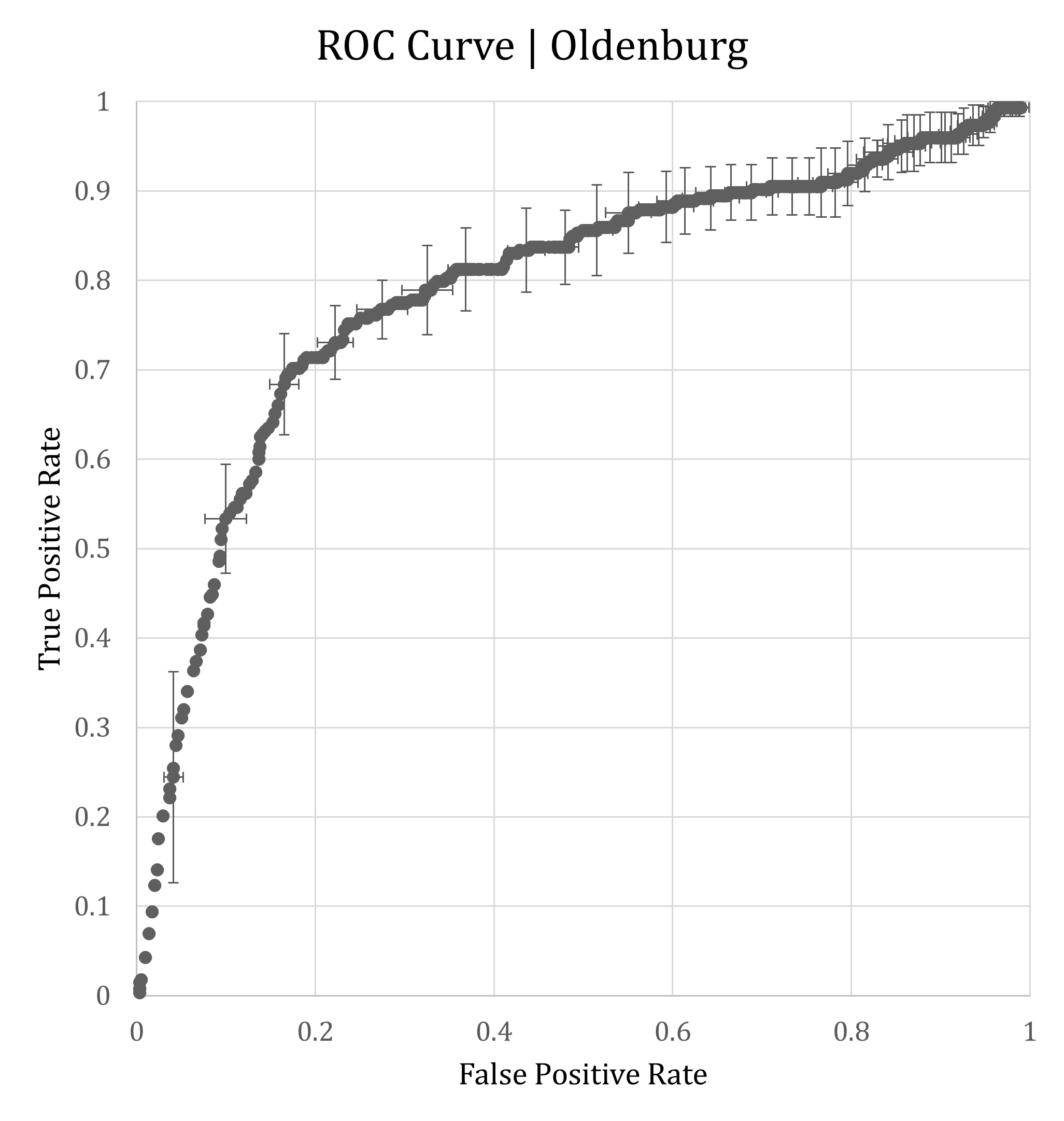}
    \caption{Receiver Operating Characteristic Curve for Oldenburg}
    \label{rocoldenburg}
\end{figure}

We conducted ROC curve analysis, focusing on Oldenburg. In this analysis, we utilized Receiver Operating Characteristic Curves, where the true positive rate was determined by satellite data indicating clear nights. Different standard deviation filters were applied to assess true positives and false positives. These filters are plotted on a graph in Fig.~\ref{rocoldenburg} and revealed that the optimal standard deviation for Oldenburg was less than 0.081, with a true positive rate of 74.47 (± 4.33) percent and a false positive rate of 23.28 (± 2.31) percent. However, this result is biased, as it included 23 percent false positives and discarded 25 percent true positives. To correct this bias, we employ the Rogan-Gladen estimator, \citep{rogan}

\begin{equation}
    P_{\text{corr}} = \frac{P - \text{FPR}}{\text{TPR} - \text{FPR}}
\end{equation}

where P and P\textsubscript{corr} are percentages of clear sky before and after correction. FPR is false positive rate and TPR is true positive rate.

Utilizing this correction, we can estimate the number of clear nights in any given year for Oldenburg accurate to the nearest percent. If we want the FPR cutoff value to be less than 10 percent, the TPR drops to 54 percent and the standard deviation filter drops to less than 0.036. The ROC analysis was also conducted for Lochem, which is a location darker than Oldenburg. The optimum standard deviation value for Lochem is at less than 0.063, which is lower than Oldenburg's 0.081 while retaining similar TRP and FPR values.

\FloatBarrier
\section{"Jellyfish" plots} \label{appendixjelly}
The "Jellyfish" and density plots created for all other locations are included here.

% Reset figure counter for appendix and set format to D1, D2, etc.
\setcounter{figure}{0}
\renewcommand{\thefigure}{D\arabic{figure}}

% Define the directory containing your images
\newcommand{\imagedir}{Jellyfishes/}

% Loop through all 41 images
%\foreach \i in {1,...,41} {%
    \begin{figure}
        \includegraphics[width=\columnwidth]{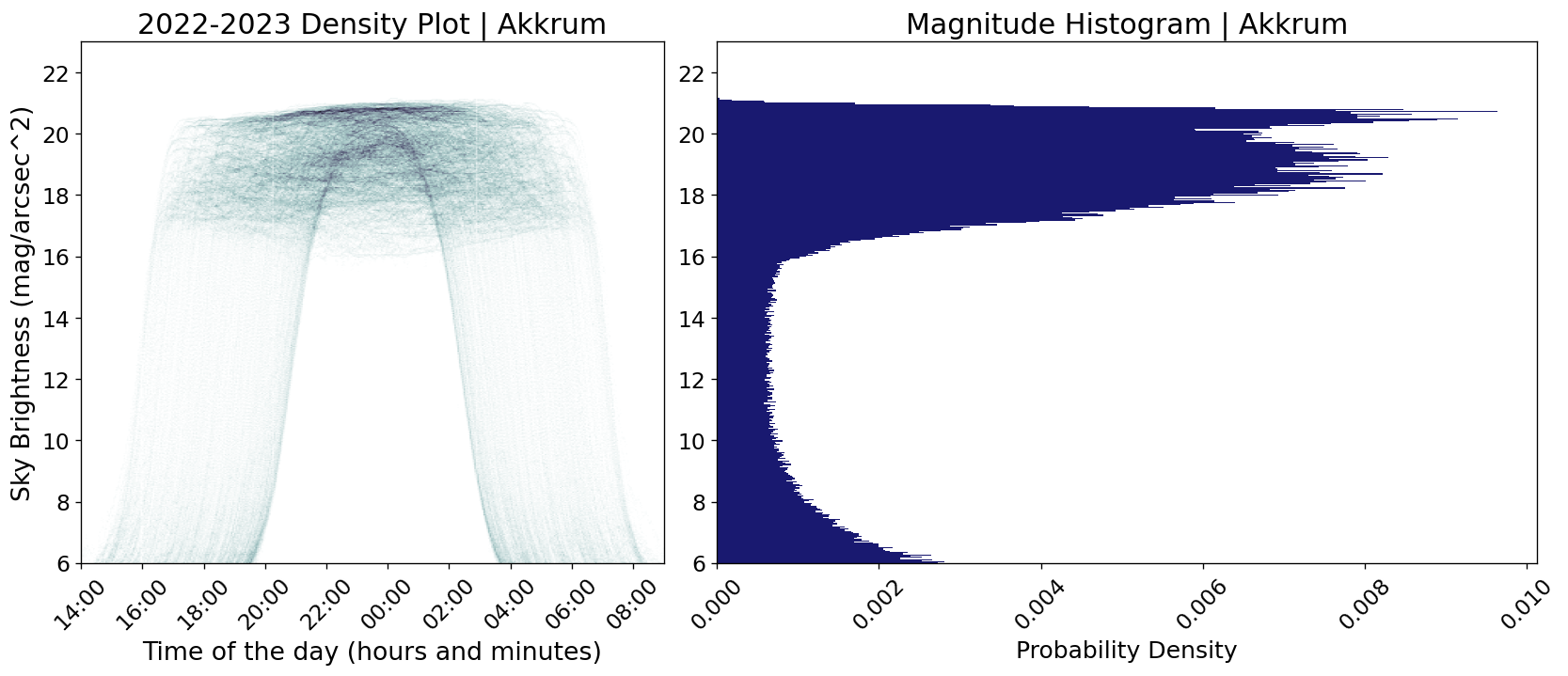}
        \caption{}
        \label{fig:jellyfish1}
    \end{figure}
    \begin{figure}
        \includegraphics[width=\columnwidth]{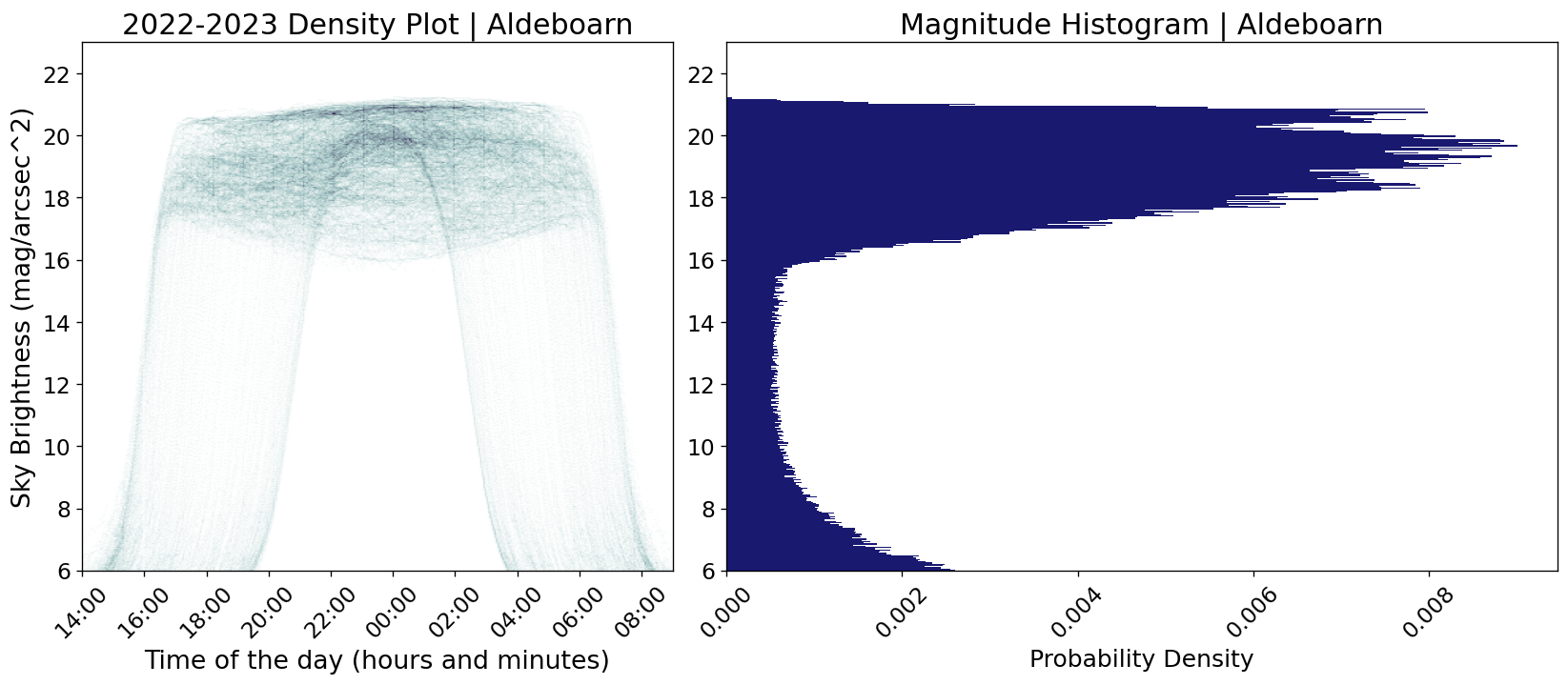}
        \caption{}
        \label{fig:jellyfish2}
    \end{figure}
    \begin{figure}
        \includegraphics[width=\columnwidth]{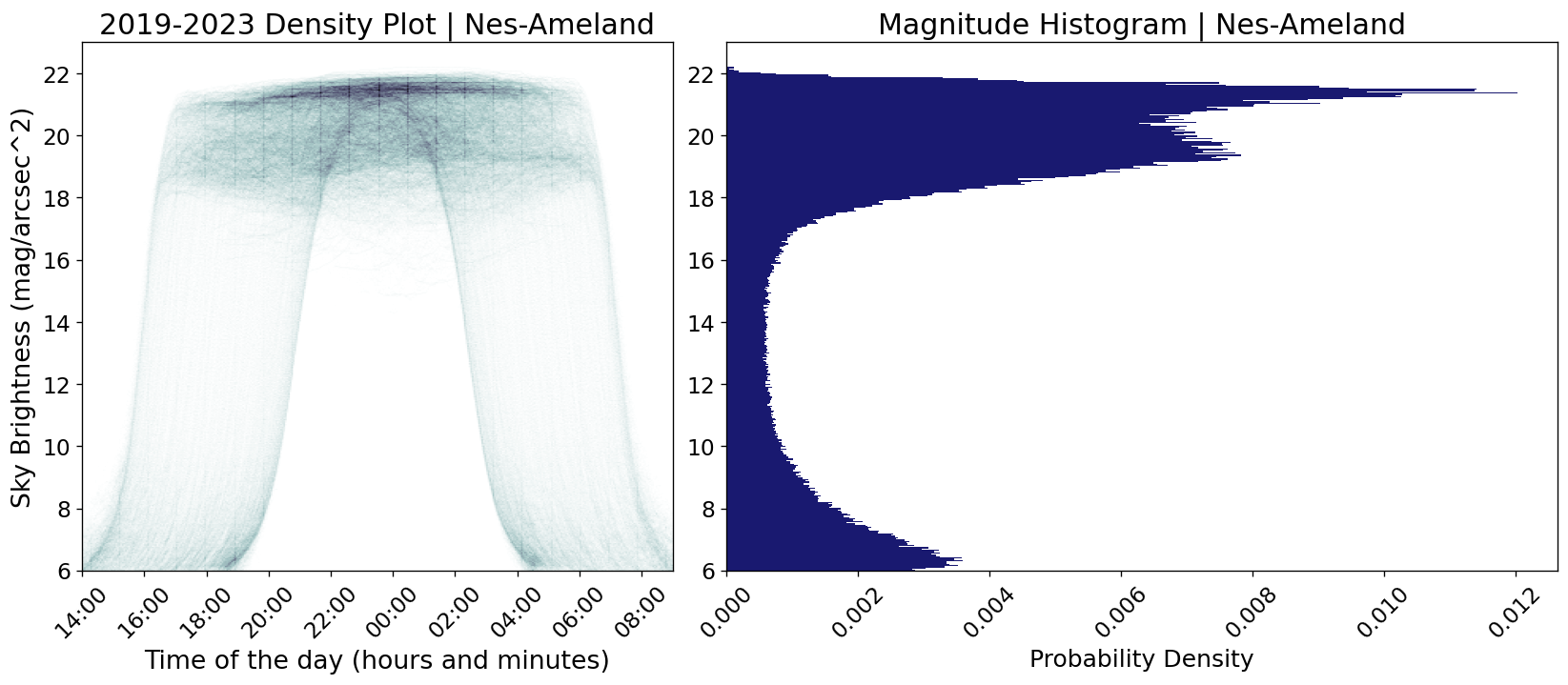}
        \caption{}
        \label{fig:jellyfish3}
    \end{figure}
    \begin{figure}
        \includegraphics[width=\columnwidth]{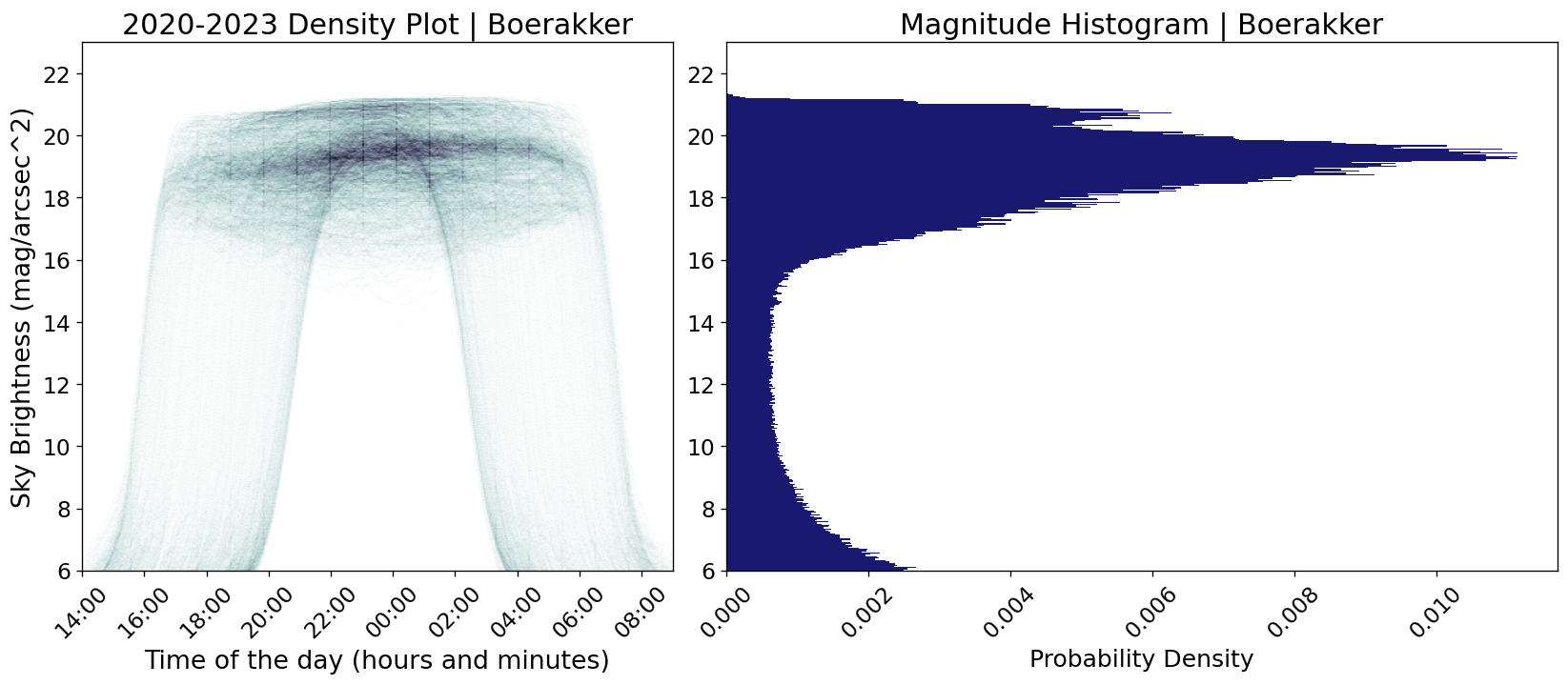}
        \caption{}
        \label{fig:jellyfish4}
    \end{figure}
    \begin{figure}
        \includegraphics[width=\columnwidth]{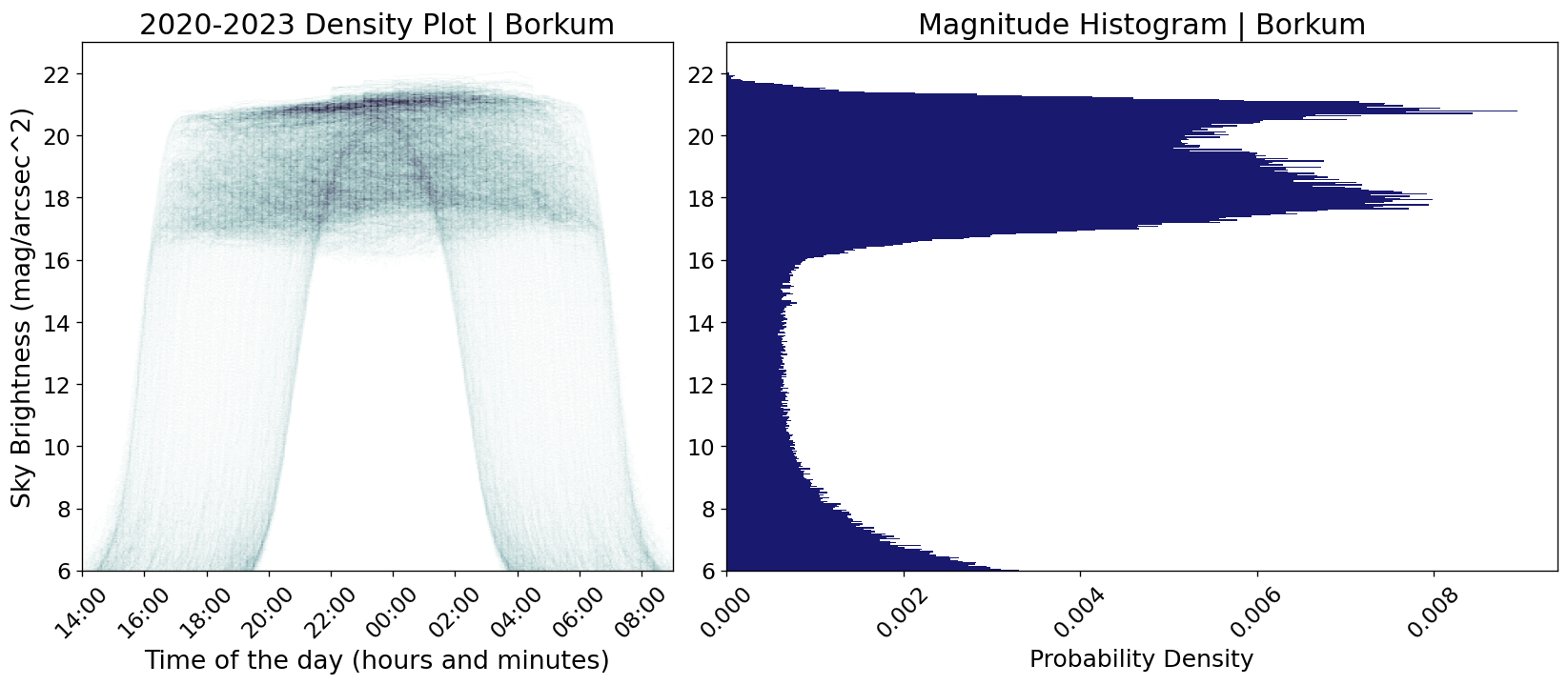}
        \caption{}
        \label{fig:jellyfish5}
    \end{figure}
    \begin{figure}
        \includegraphics[width=\columnwidth]{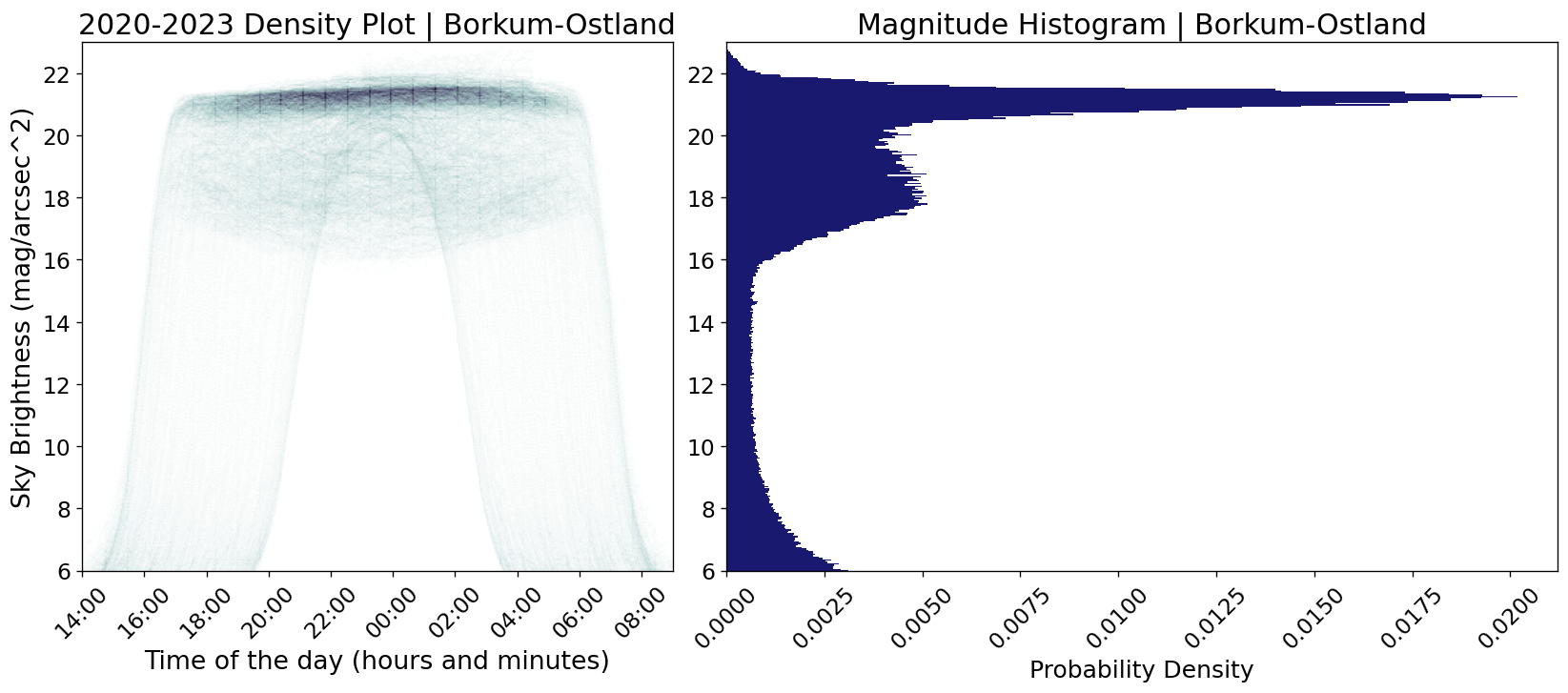}
        \caption{}
        \label{fig:jellyfish6}
    \end{figure}
    \begin{figure}
        \includegraphics[width=\columnwidth]{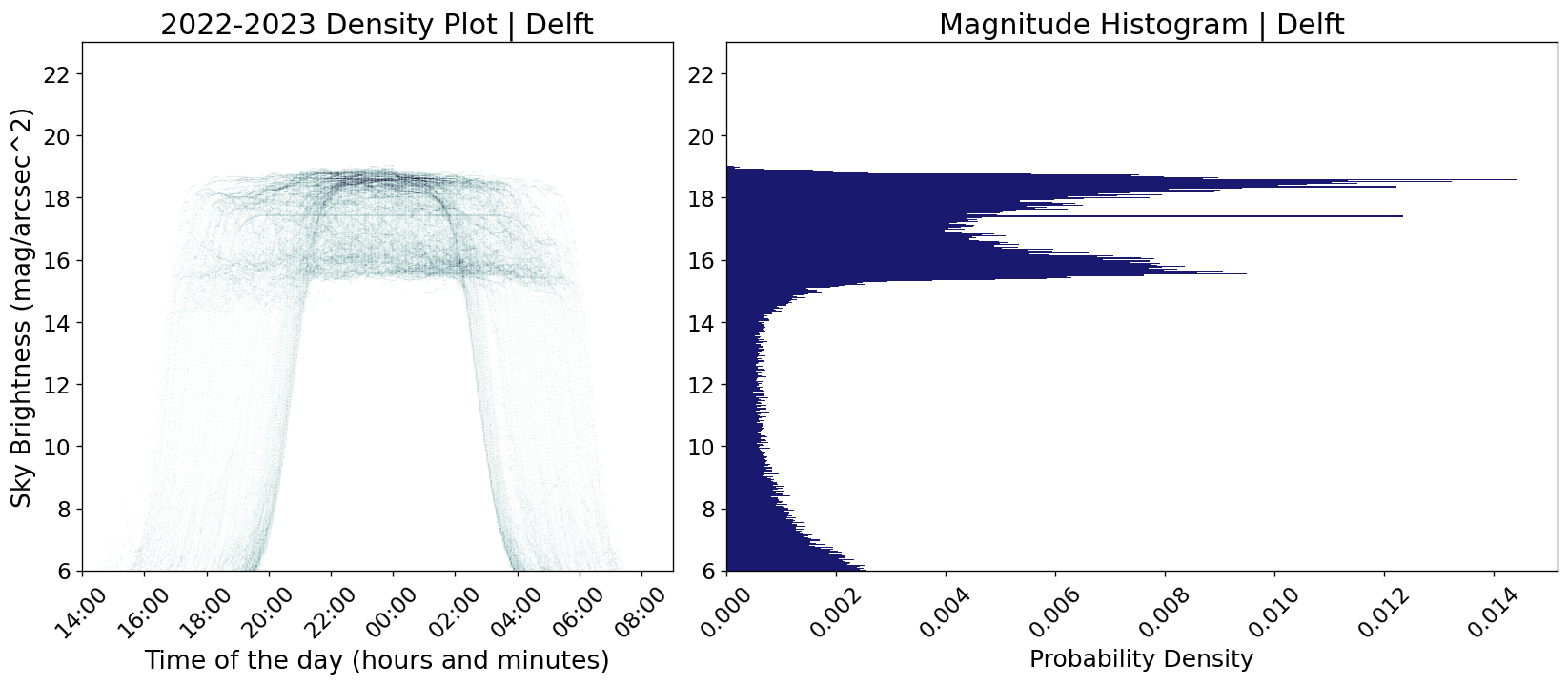}
        \caption{}
        \label{fig:jellyfish7}
    \end{figure}
    \begin{figure}
        \includegraphics[width=\columnwidth]{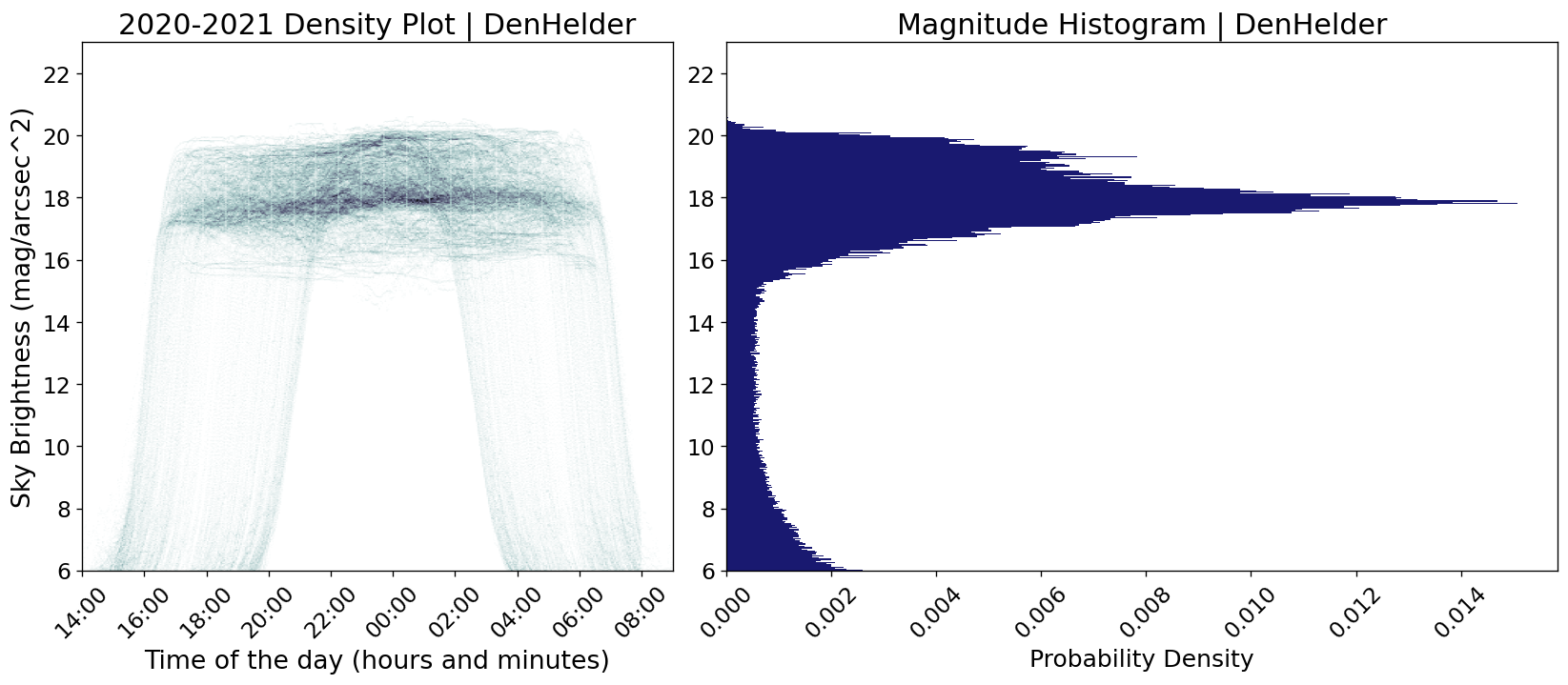}
        \caption{}
        \label{fig:jellyfish8}
    \end{figure}
    \begin{figure}
        \includegraphics[width=\columnwidth]{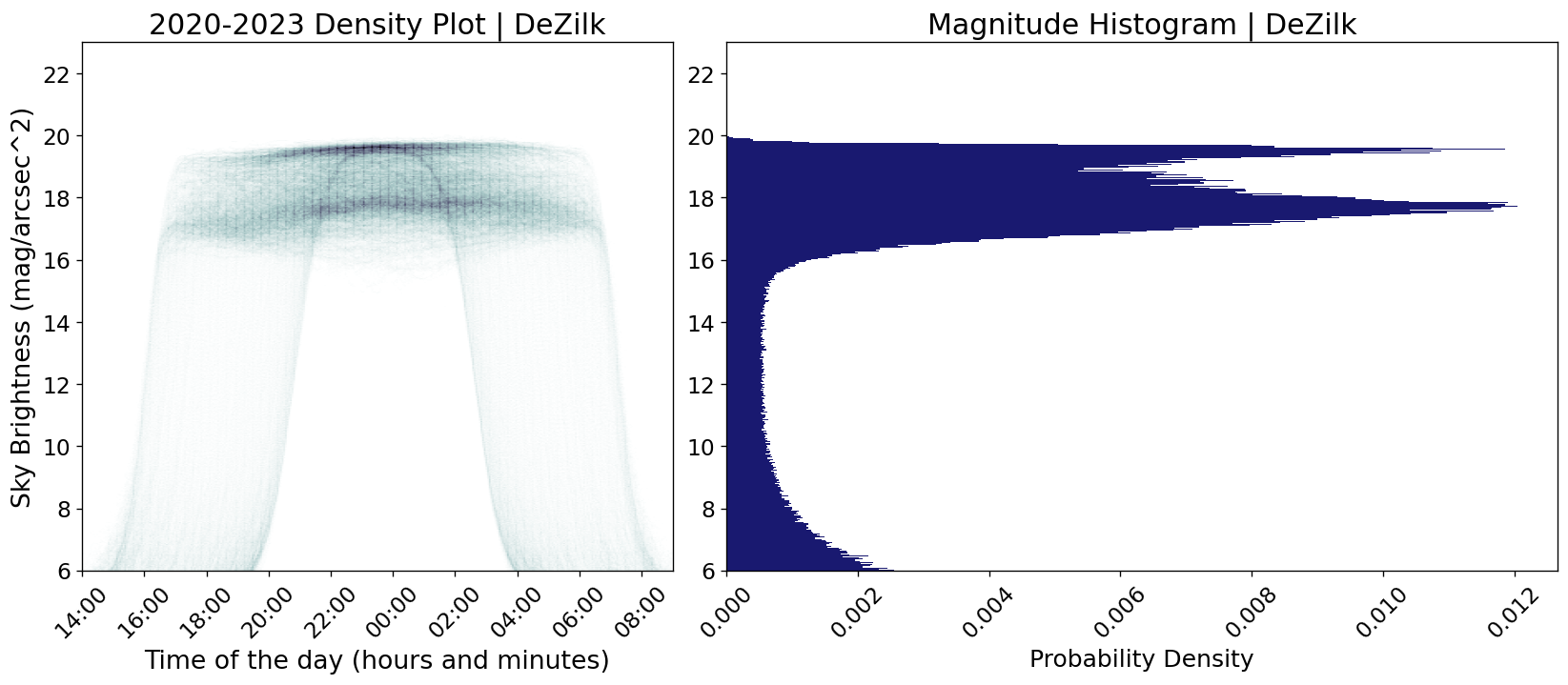}
        \caption{}
        \label{fig:jellyfish9}
    \end{figure}
    \begin{figure}
        \includegraphics[width=\columnwidth]{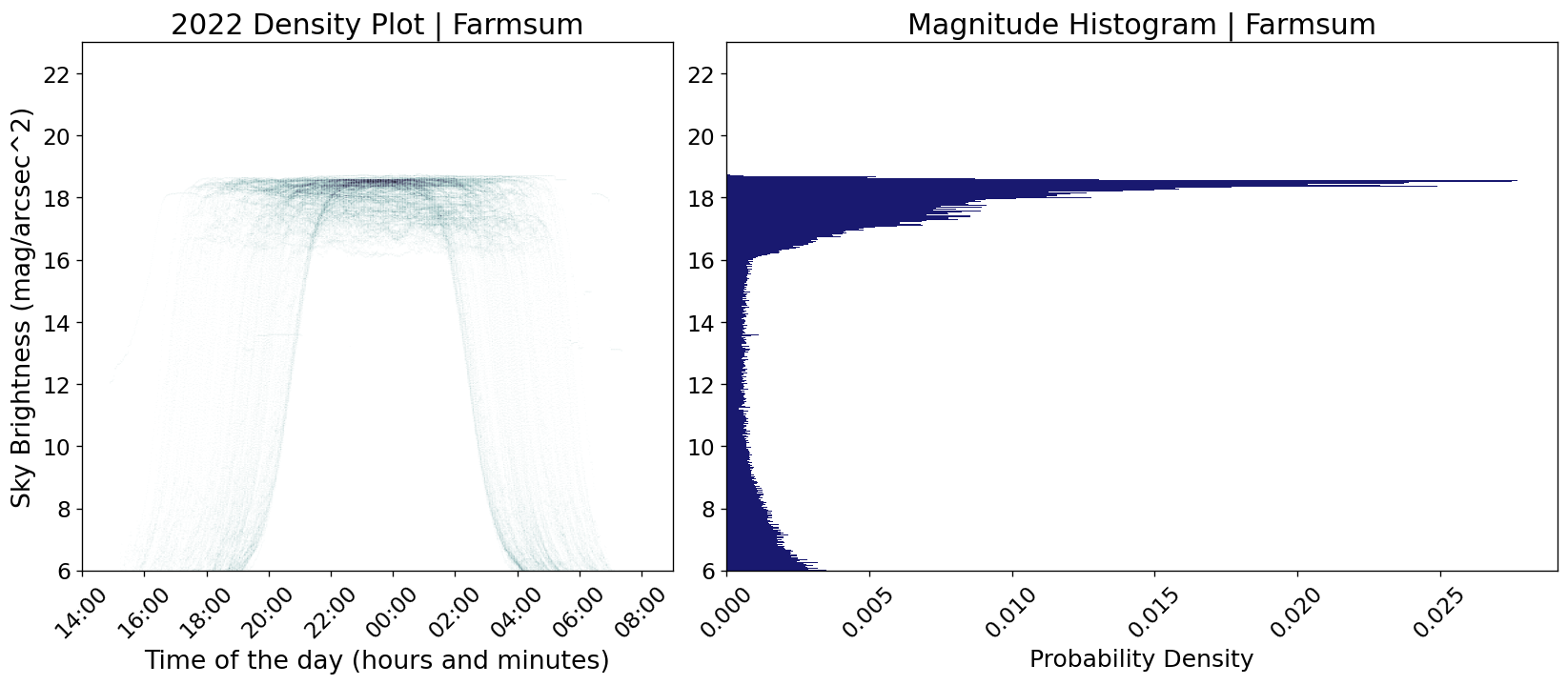}
        \caption{}
        \label{fig:jellyfish10}
    \end{figure}
    \begin{figure}
        \includegraphics[width=\columnwidth]{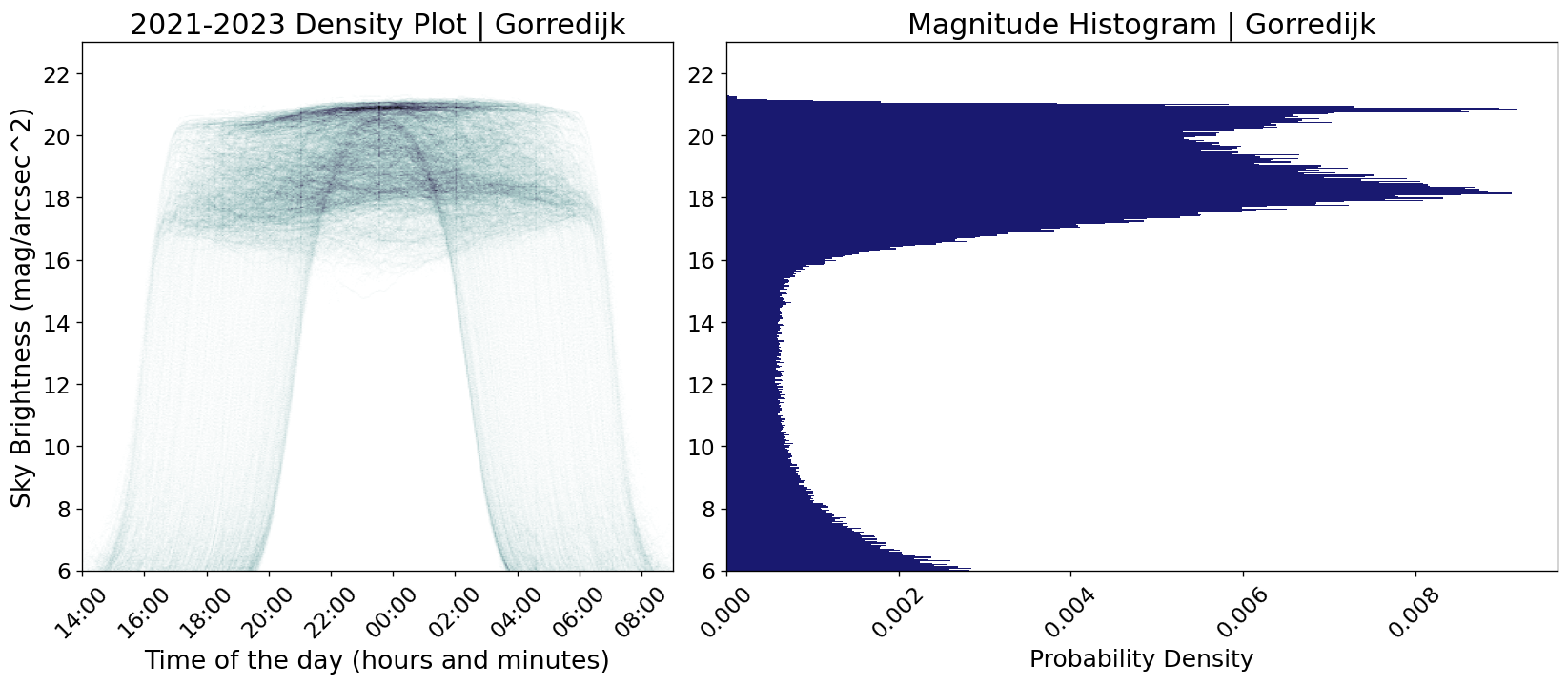}
        \caption{}
        \label{fig:jellyfish11}
    \end{figure}
    \begin{figure}
        \includegraphics[width=\columnwidth]{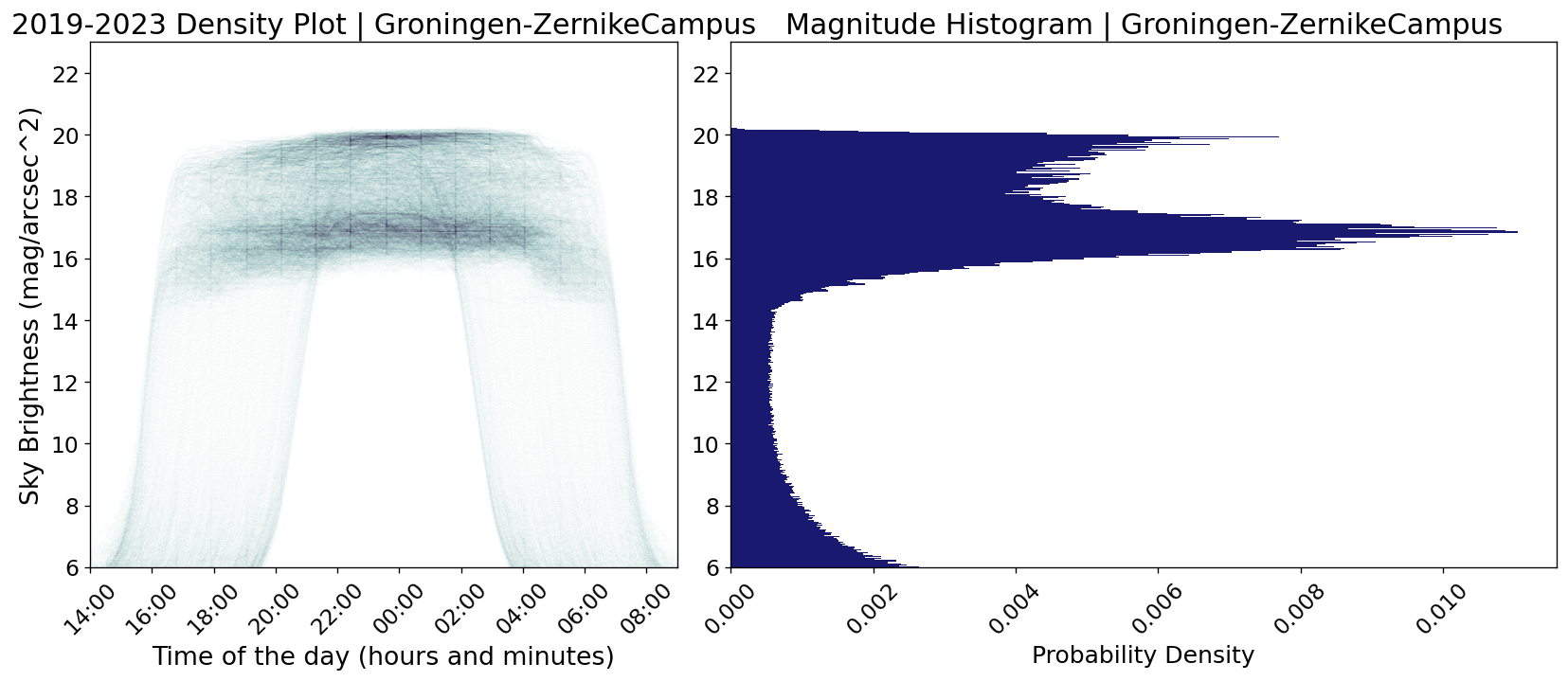}
        \caption{}
        \label{fig:jellyfish12}
    \end{figure}
    \begin{figure}
        \includegraphics[width=\columnwidth]{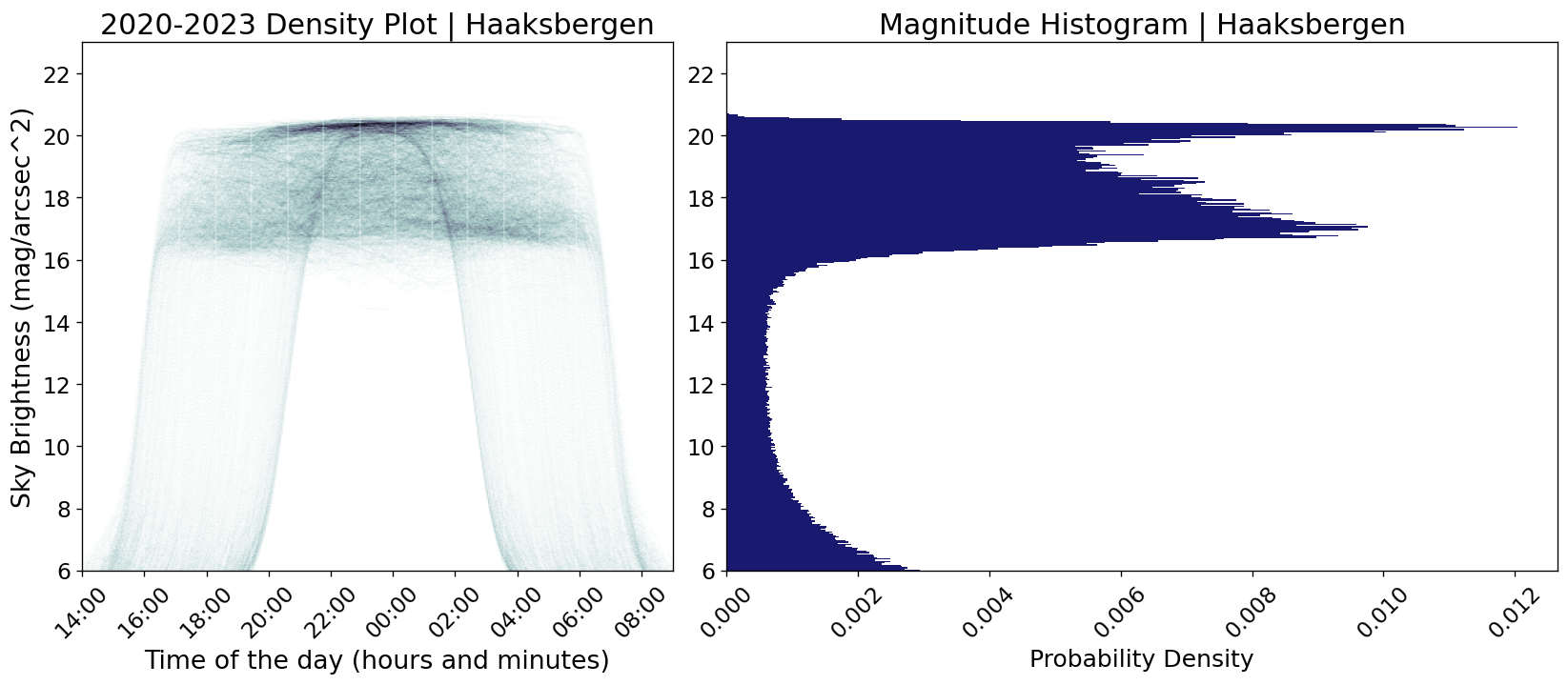}
        \caption{}
        \label{fig:jellyfish13}
    \end{figure}
    \begin{figure}
        \includegraphics[width=\columnwidth]{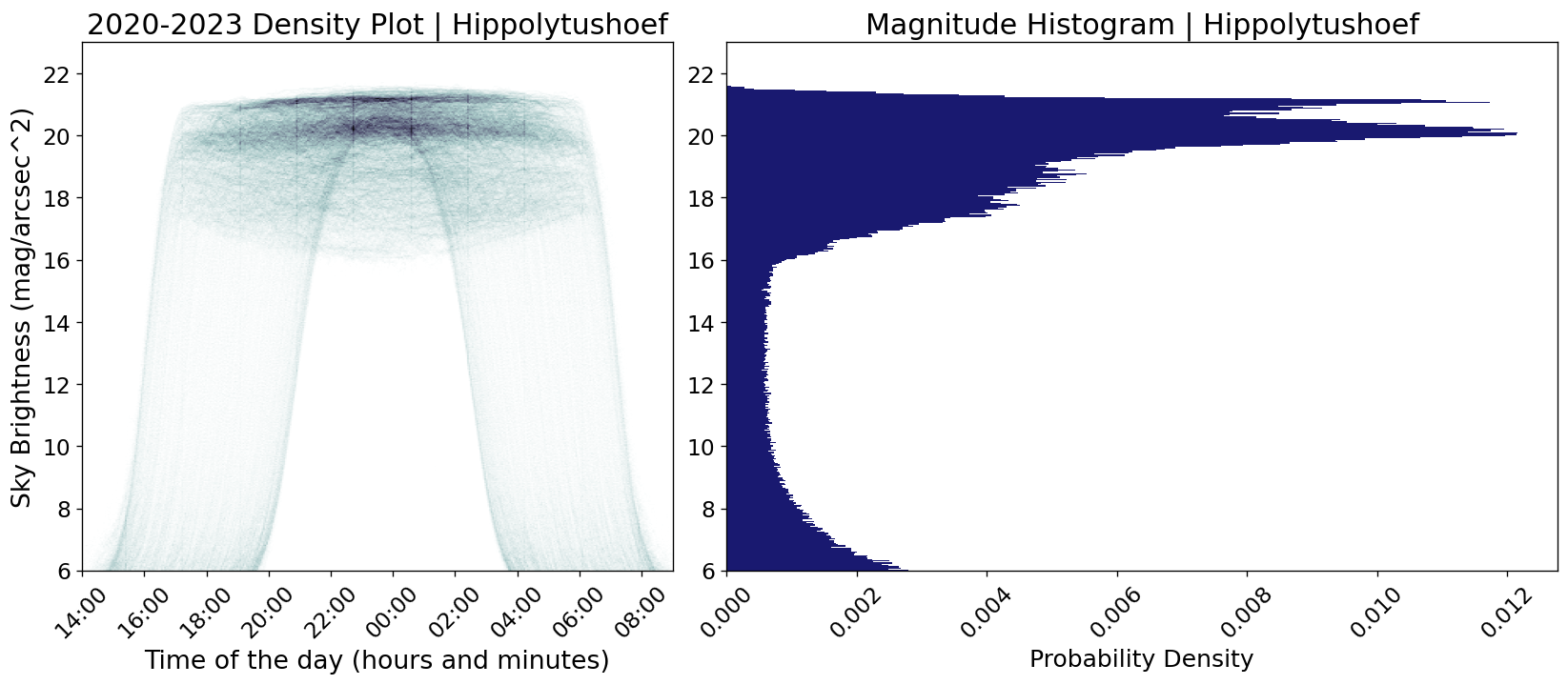}
        \caption{}
        \label{fig:jellyfish14}
    \end{figure}
    \begin{figure}
        \includegraphics[width=\columnwidth]{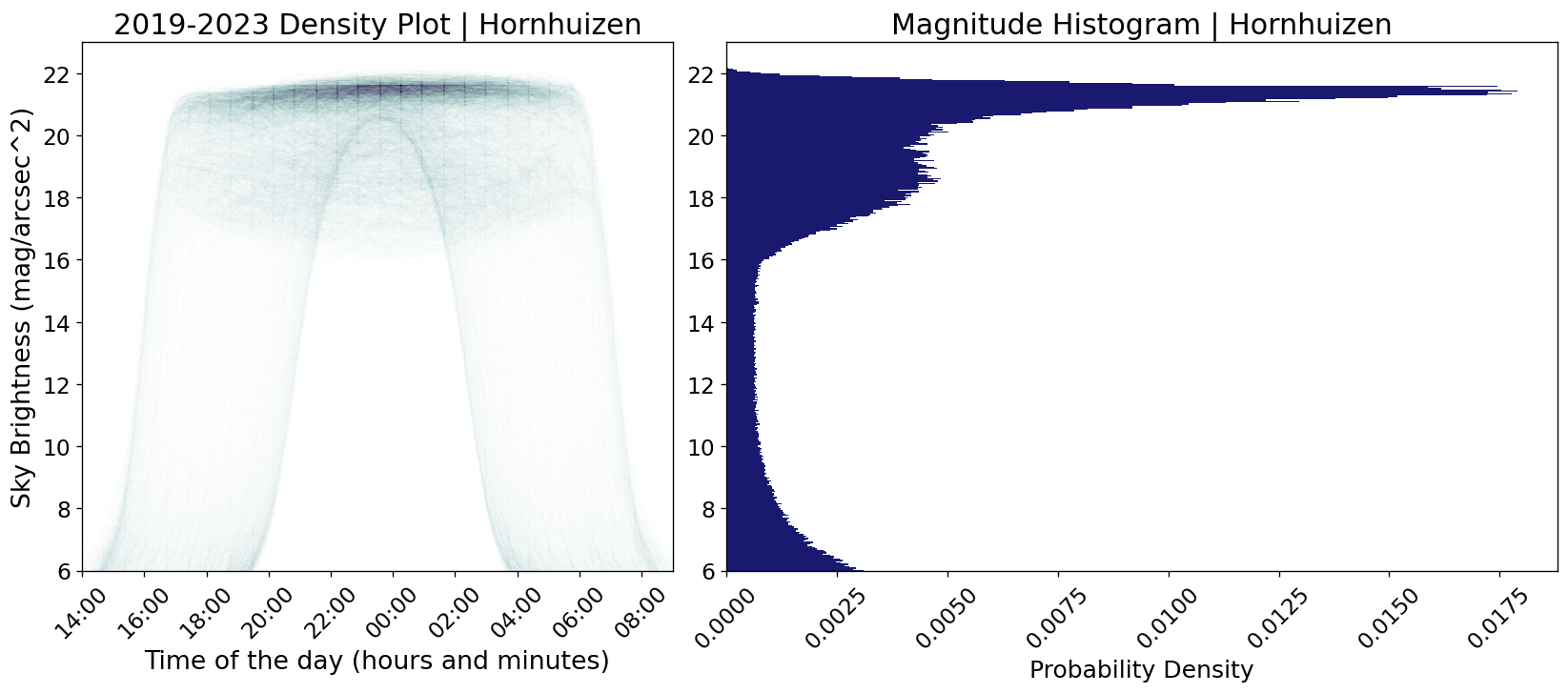}
        \caption{}
        \label{fig:jellyfish15}
    \end{figure}
    \begin{figure}
        \includegraphics[width=\columnwidth]{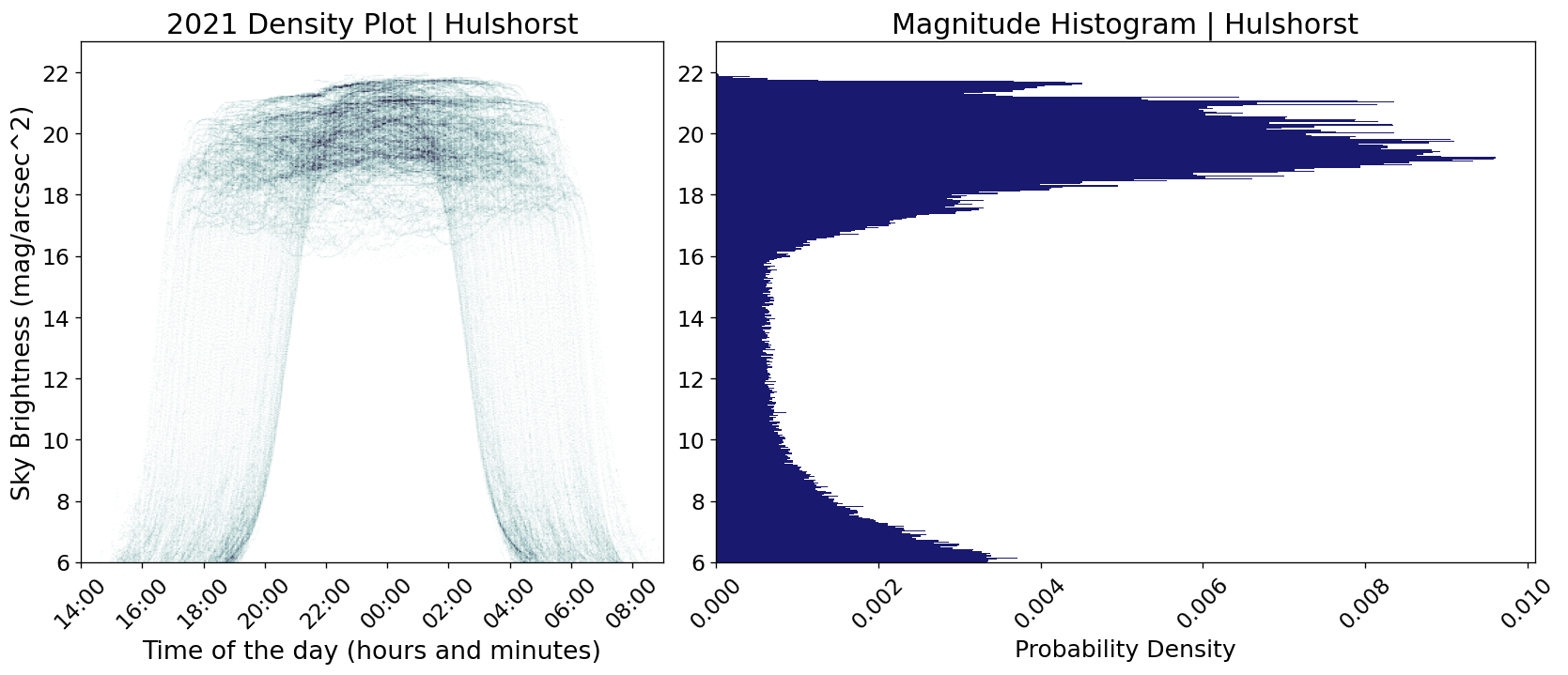}
        \caption{}
        \label{fig:jellyfish16}
    \end{figure}
    \begin{figure}
        \includegraphics[width=\columnwidth]{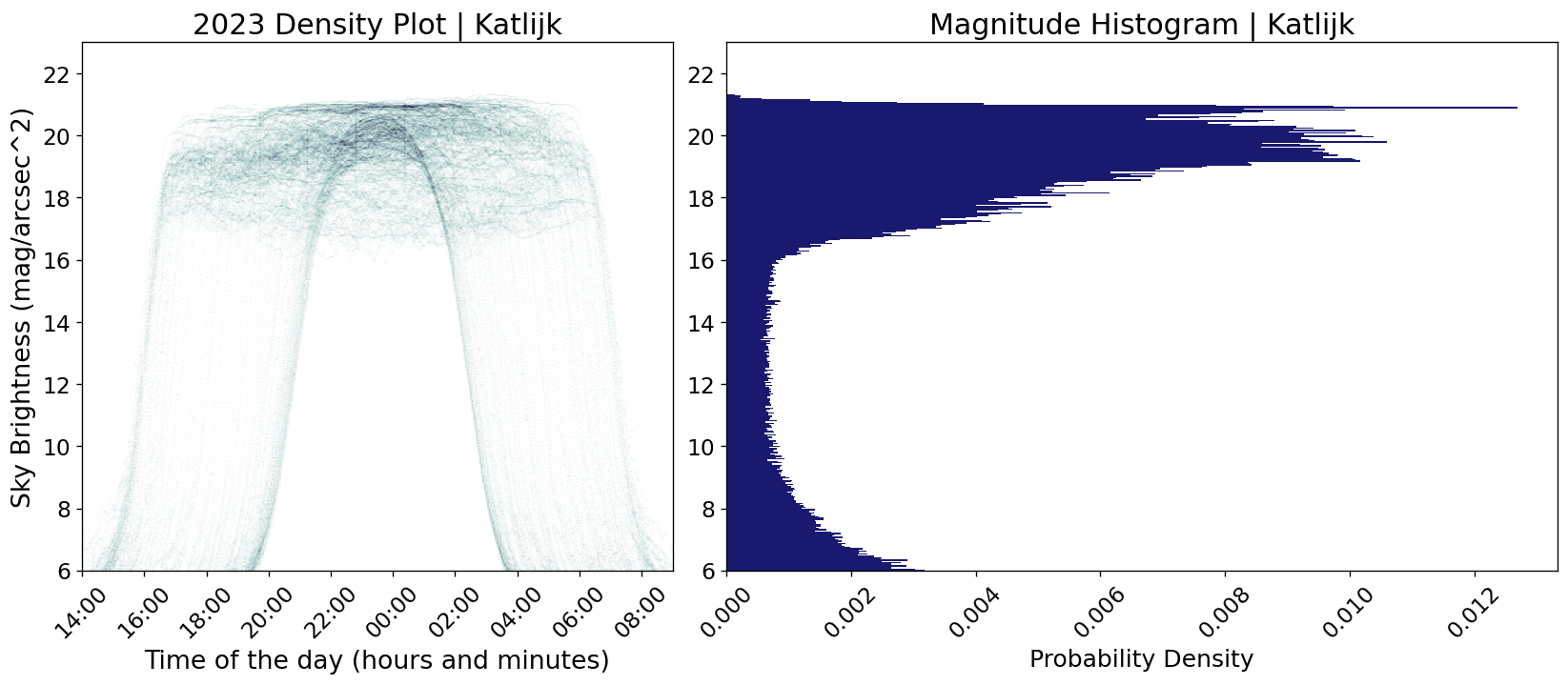}
        \caption{}
        \label{fig:jellyfish17}
    \end{figure}
    \begin{figure}
        \includegraphics[width=\columnwidth]{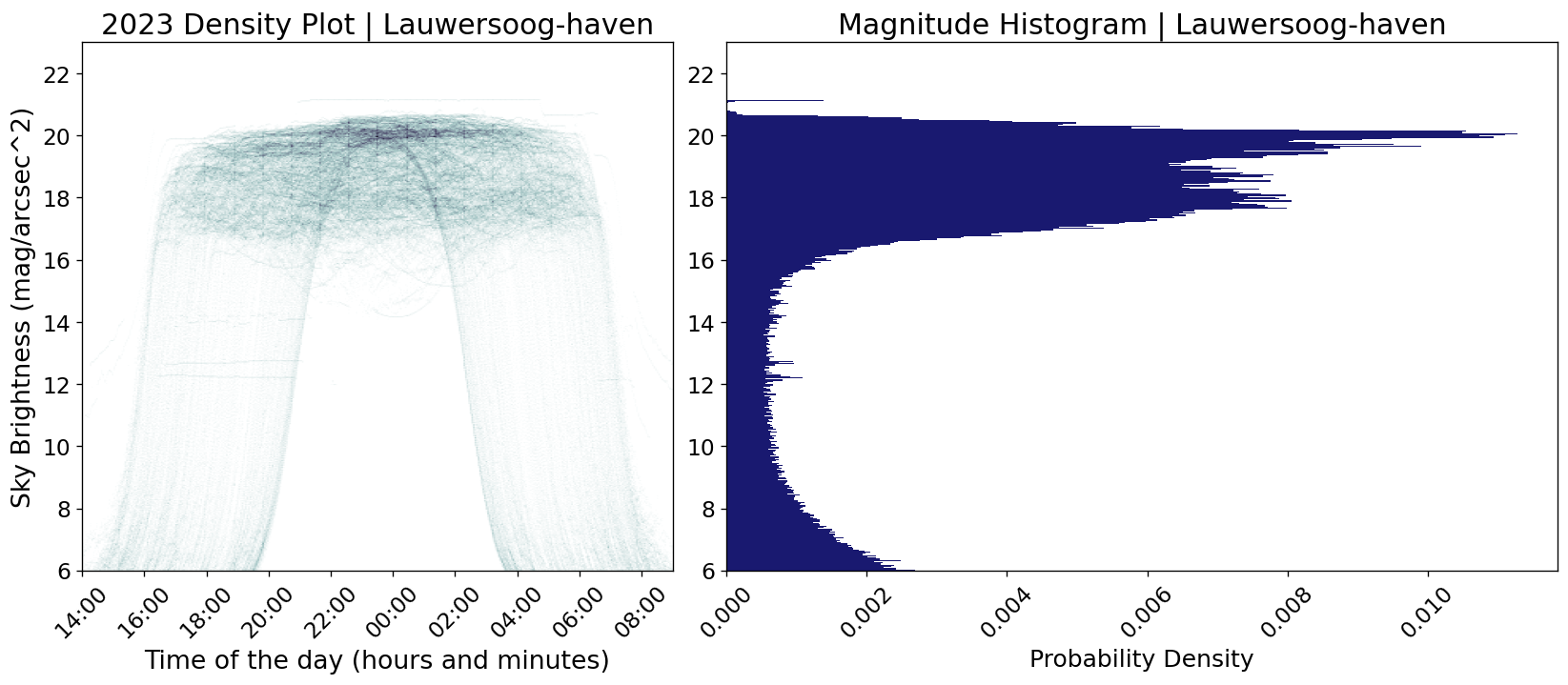}
        \caption{}
        \label{fig:jellyfish18}
    \end{figure}
    \begin{figure}
        \includegraphics[width=\columnwidth]{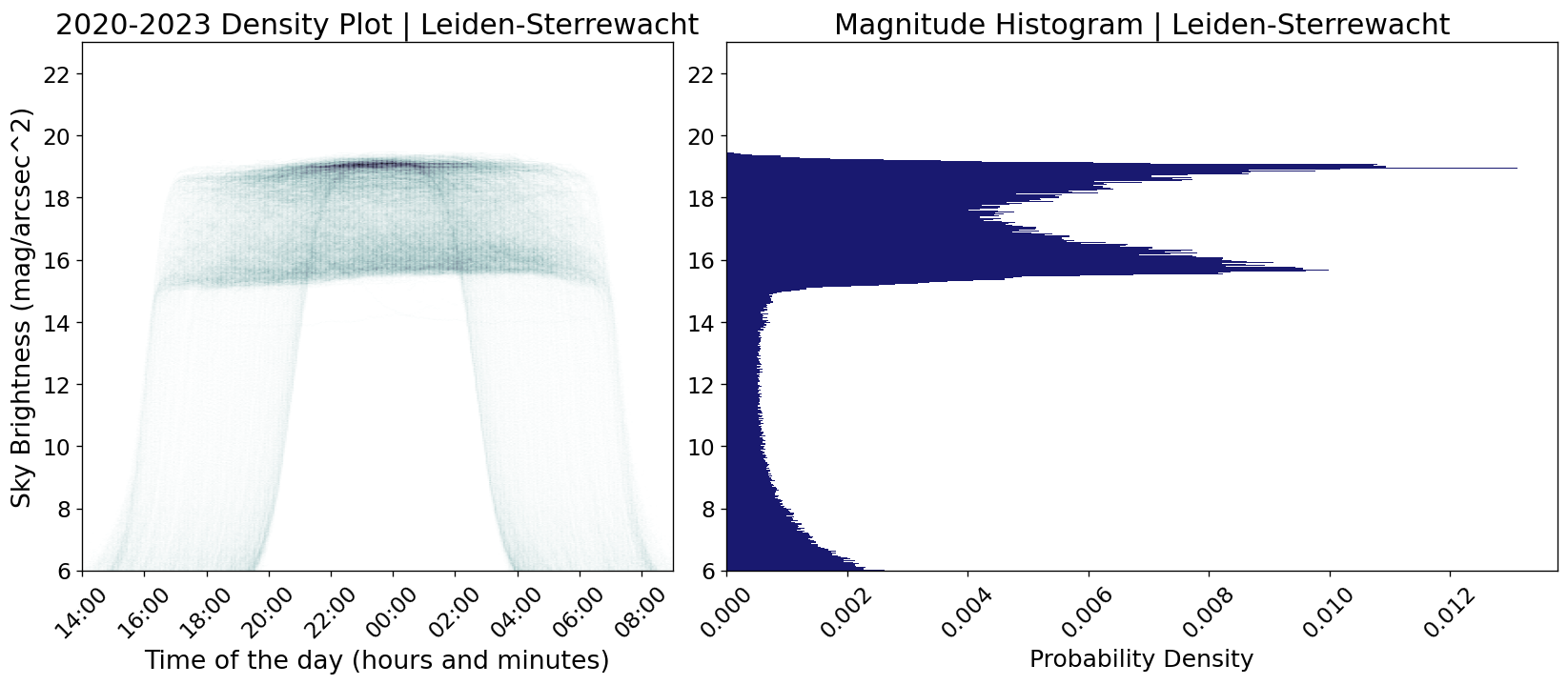}
        \caption{}
        \label{fig:jellyfish19}
    \end{figure}
    \begin{figure}
        \includegraphics[width=\columnwidth]{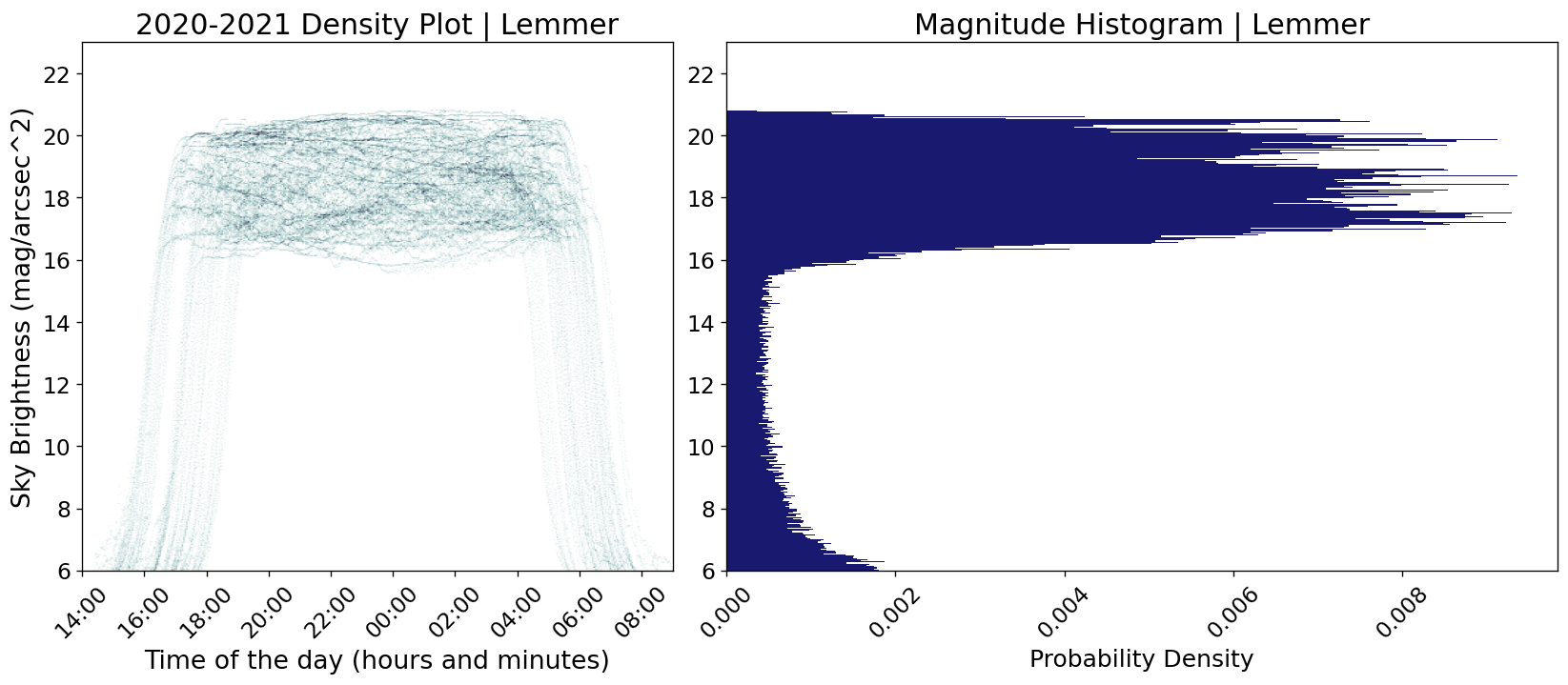}
        \caption{}
        \label{fig:jellyfish20}
    \end{figure}
    \begin{figure}
        \includegraphics[width=\columnwidth]{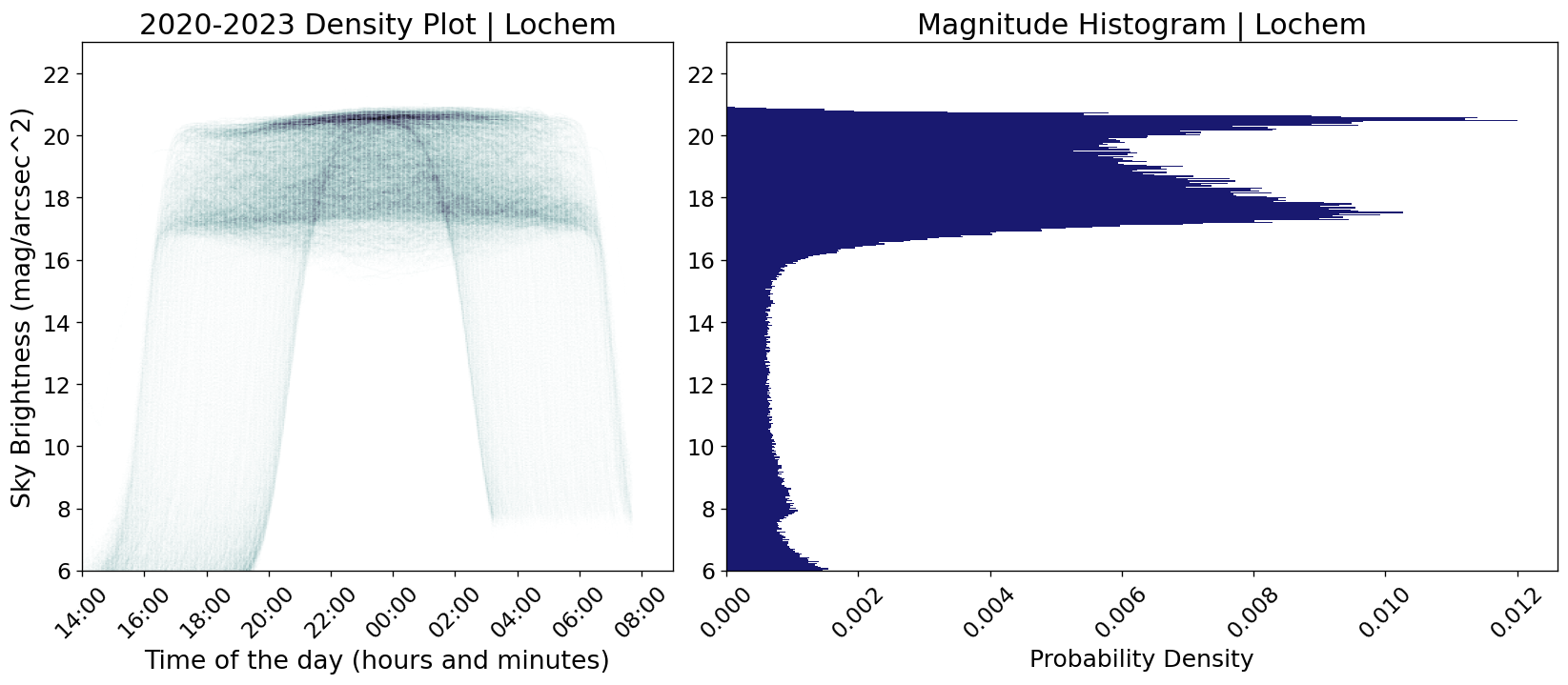}
        \caption{}
        \label{fig:jellyfish21}
    \end{figure}
    \begin{figure}
        \includegraphics[width=\columnwidth]{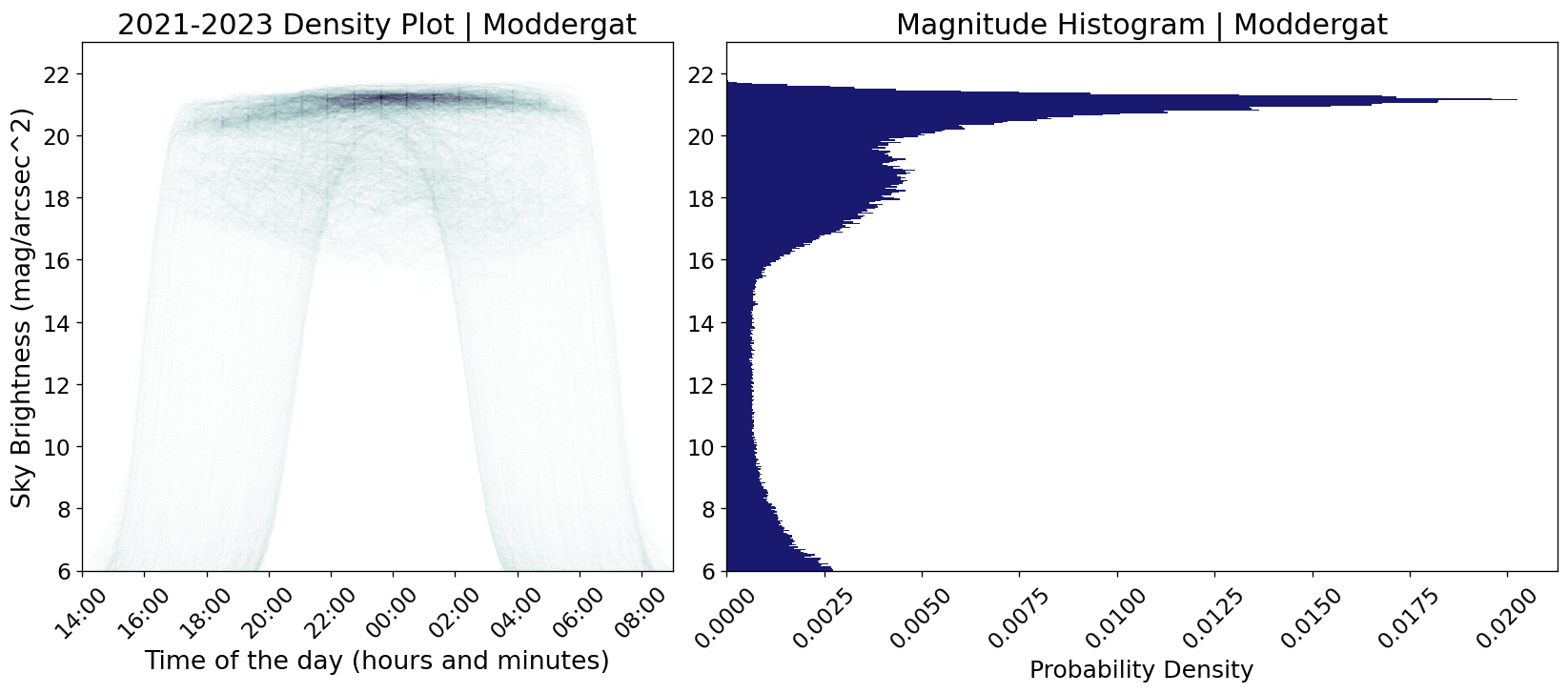}
        \caption{}
        \label{fig:jellyfish22}
    \end{figure}
    \begin{figure}
        \includegraphics[width=\columnwidth]{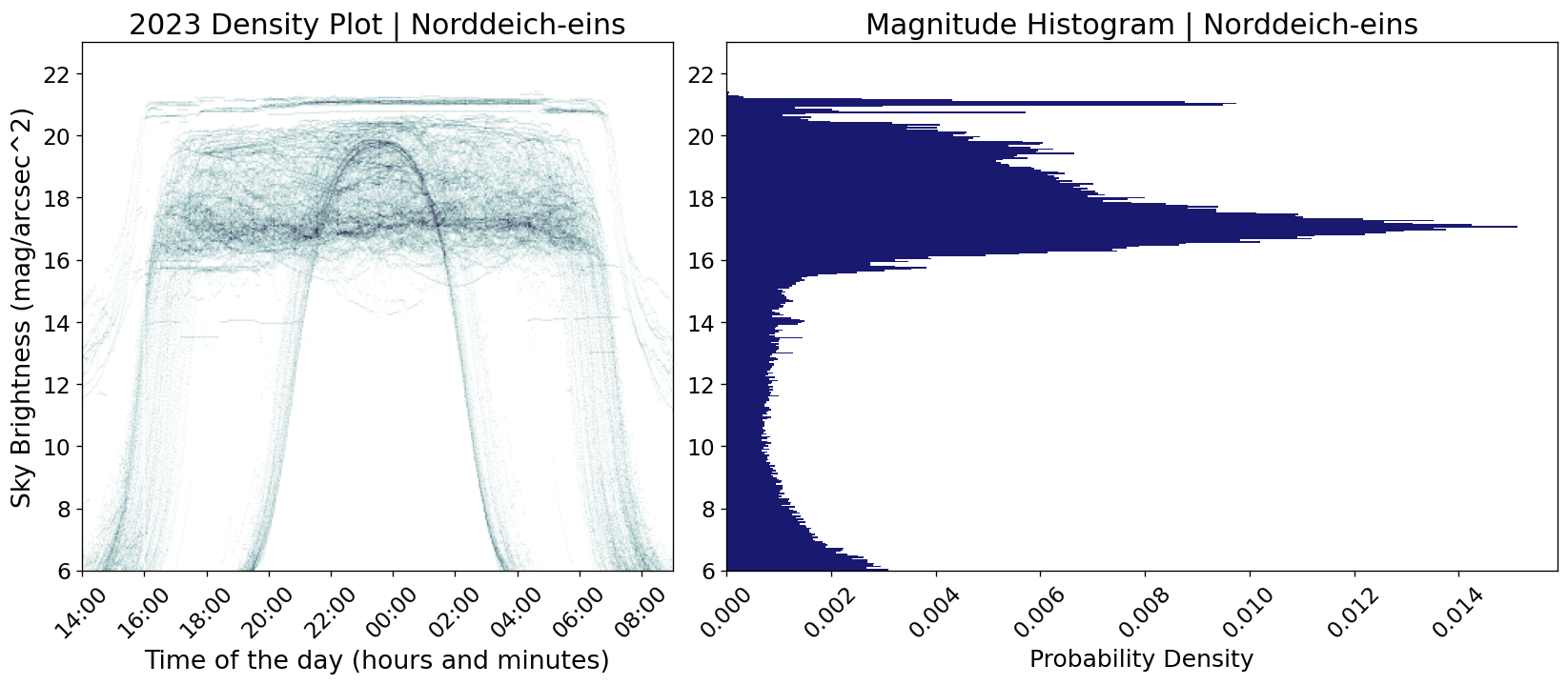}
        \caption{}
        \label{fig:jellyfish23}
    \end{figure}
    \begin{figure}
        \includegraphics[width=\columnwidth]{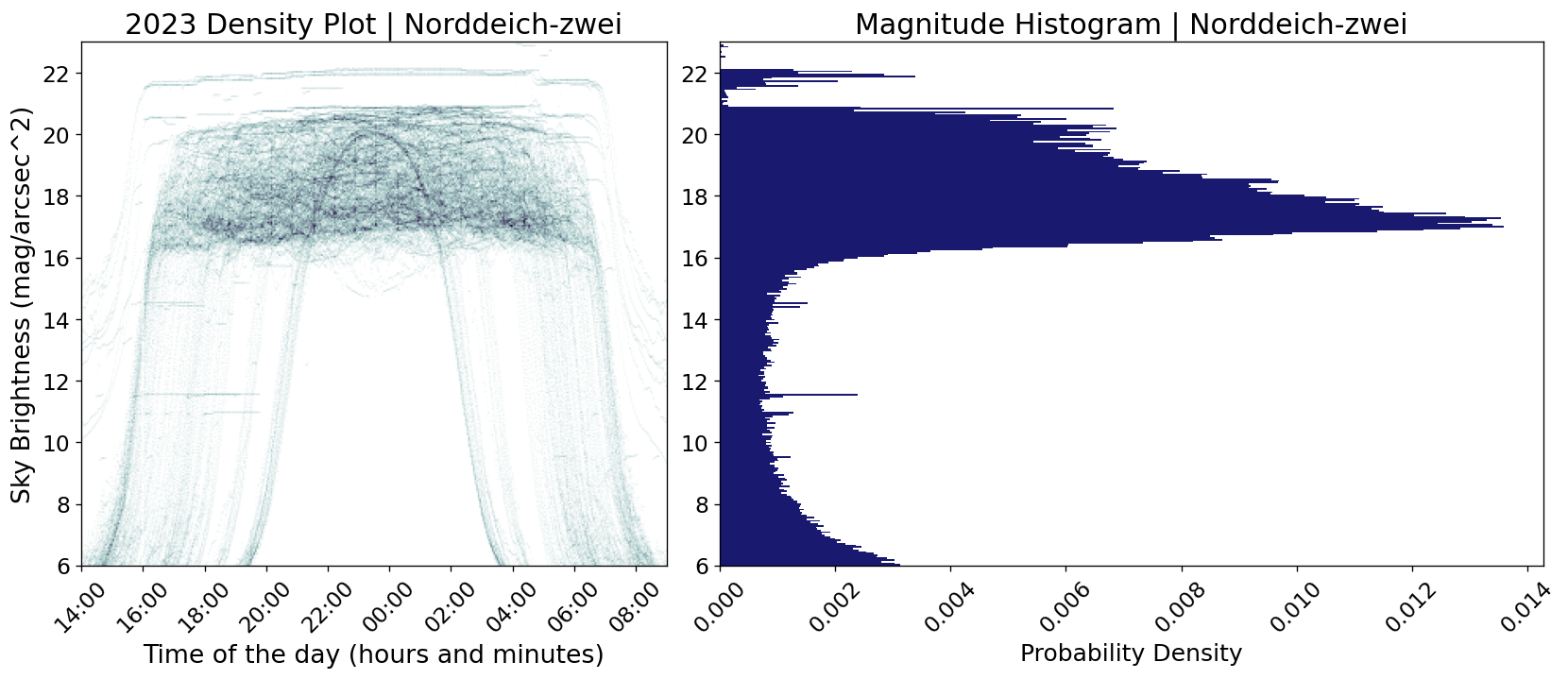}
        \caption{}
        \label{fig:jellyfish24}
    \end{figure}
    \begin{figure}
        \includegraphics[width=\columnwidth]{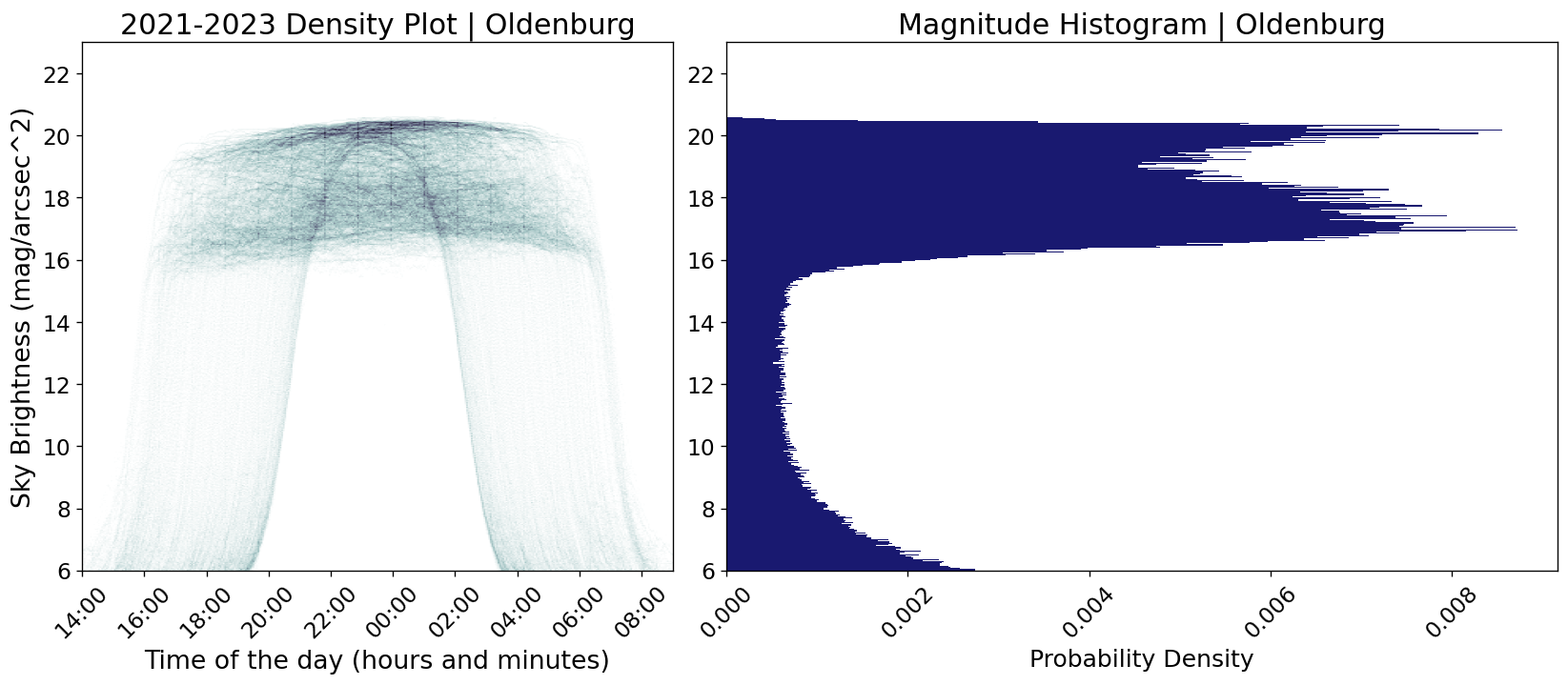}
        \caption{}
        \label{fig:jellyfish25}
    \end{figure}
    \begin{figure}
        \includegraphics[width=\columnwidth]{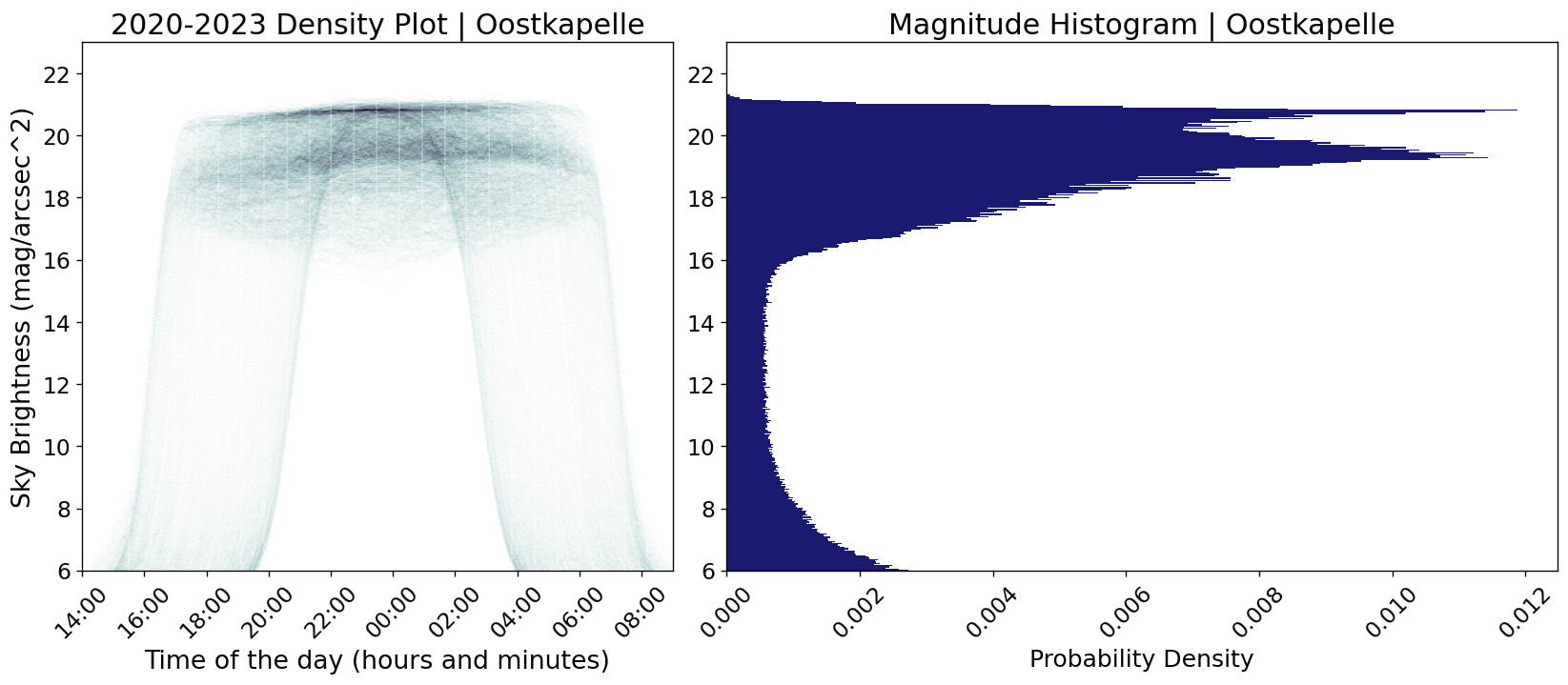}
        \caption{}
        \label{fig:jellyfish26}
    \end{figure}
    \begin{figure}
        \includegraphics[width=\columnwidth]{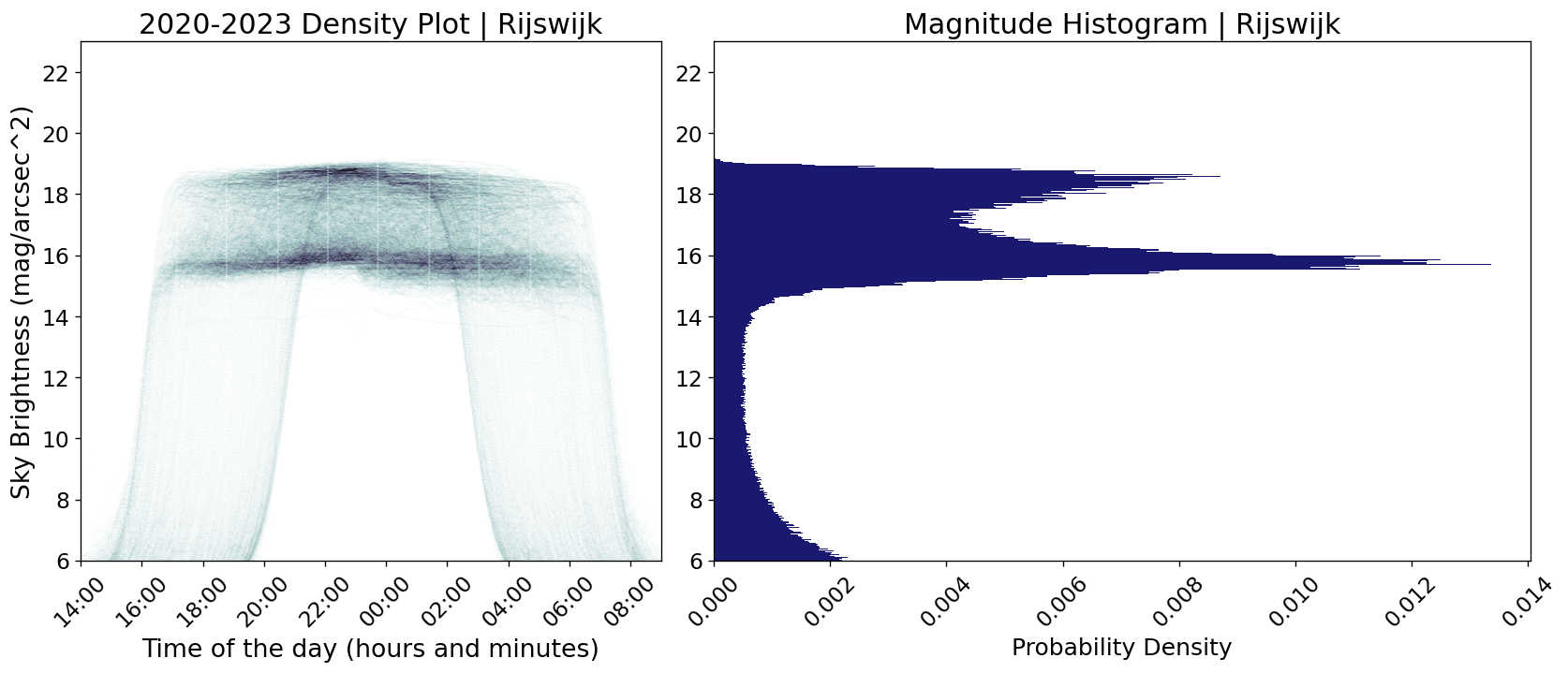}
        \caption{}
        \label{fig:jellyfish27}
    \end{figure}
    \begin{figure}
        \includegraphics[width=\columnwidth]{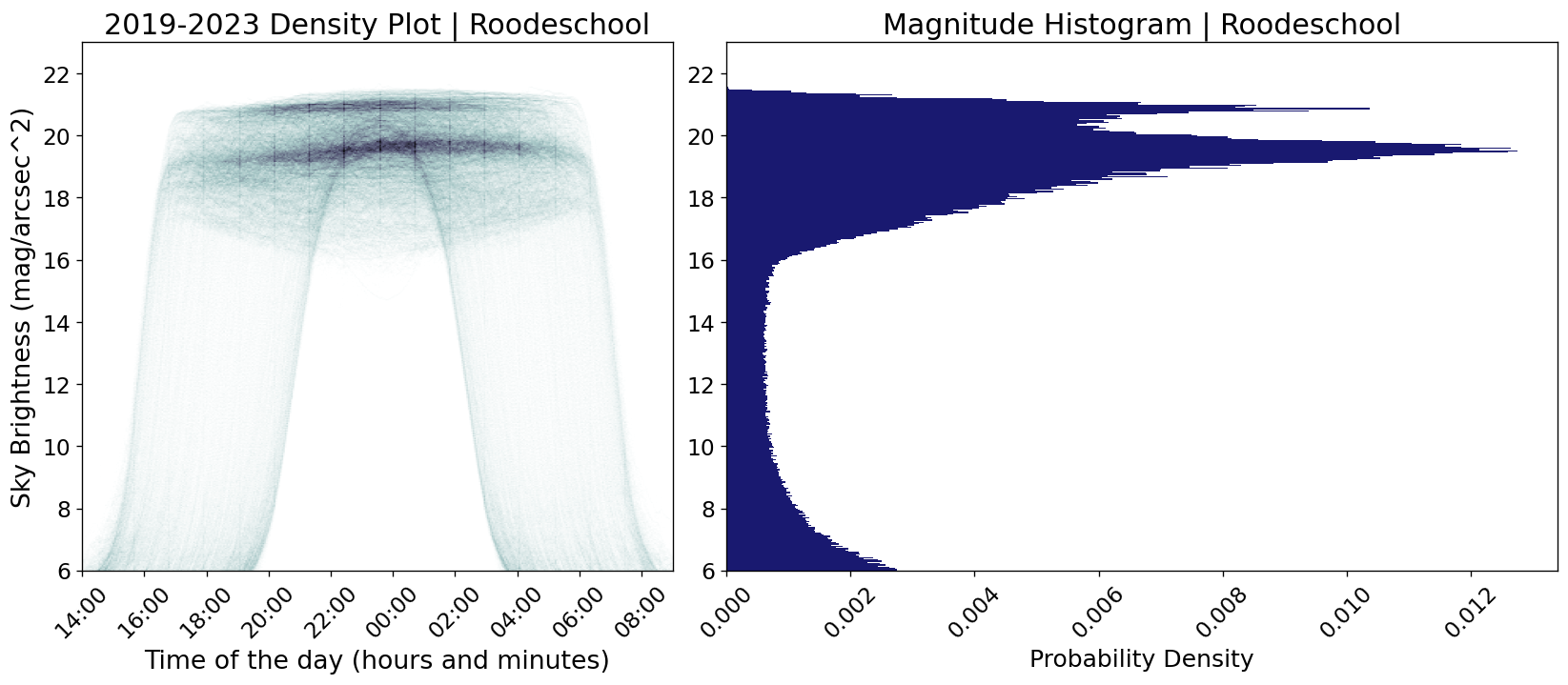}
        \caption{}
        \label{fig:jellyfish28}
    \end{figure}
    \begin{figure}
        \includegraphics[width=\columnwidth]{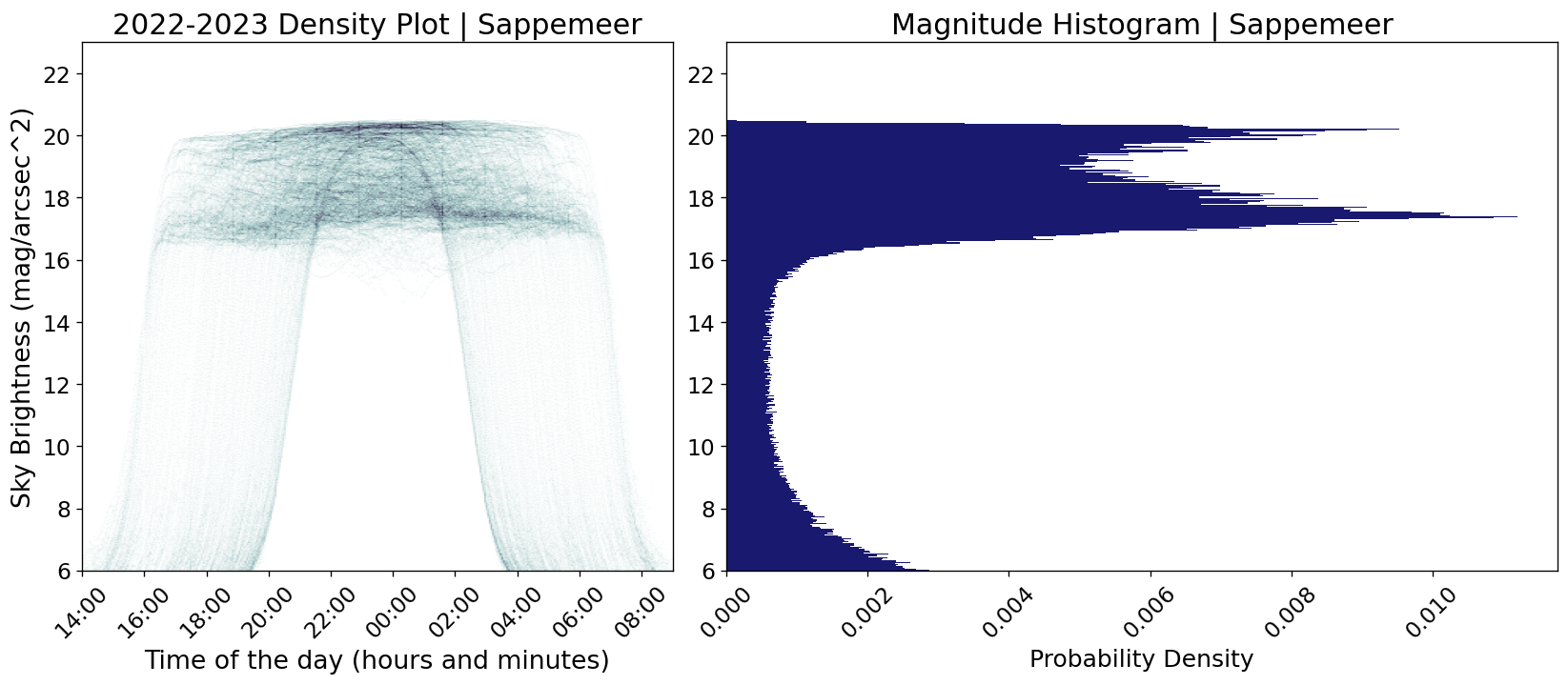}
        \caption{}
        \label{fig:jellyfish29}
    \end{figure}
    \begin{figure}
        \includegraphics[width=\columnwidth]{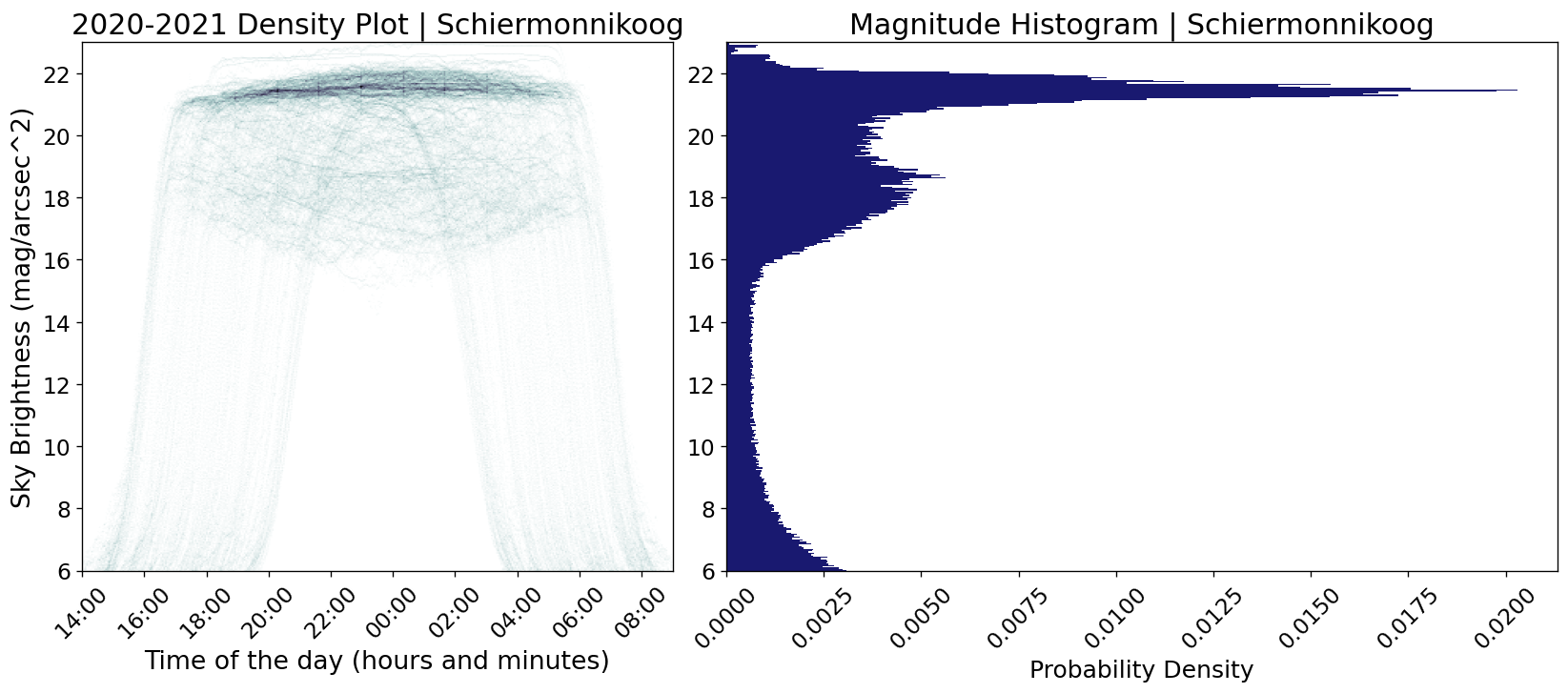}
        \caption{}
        \label{fig:jellyfish30}
    \end{figure}
    \begin{figure}
        \includegraphics[width=\columnwidth]{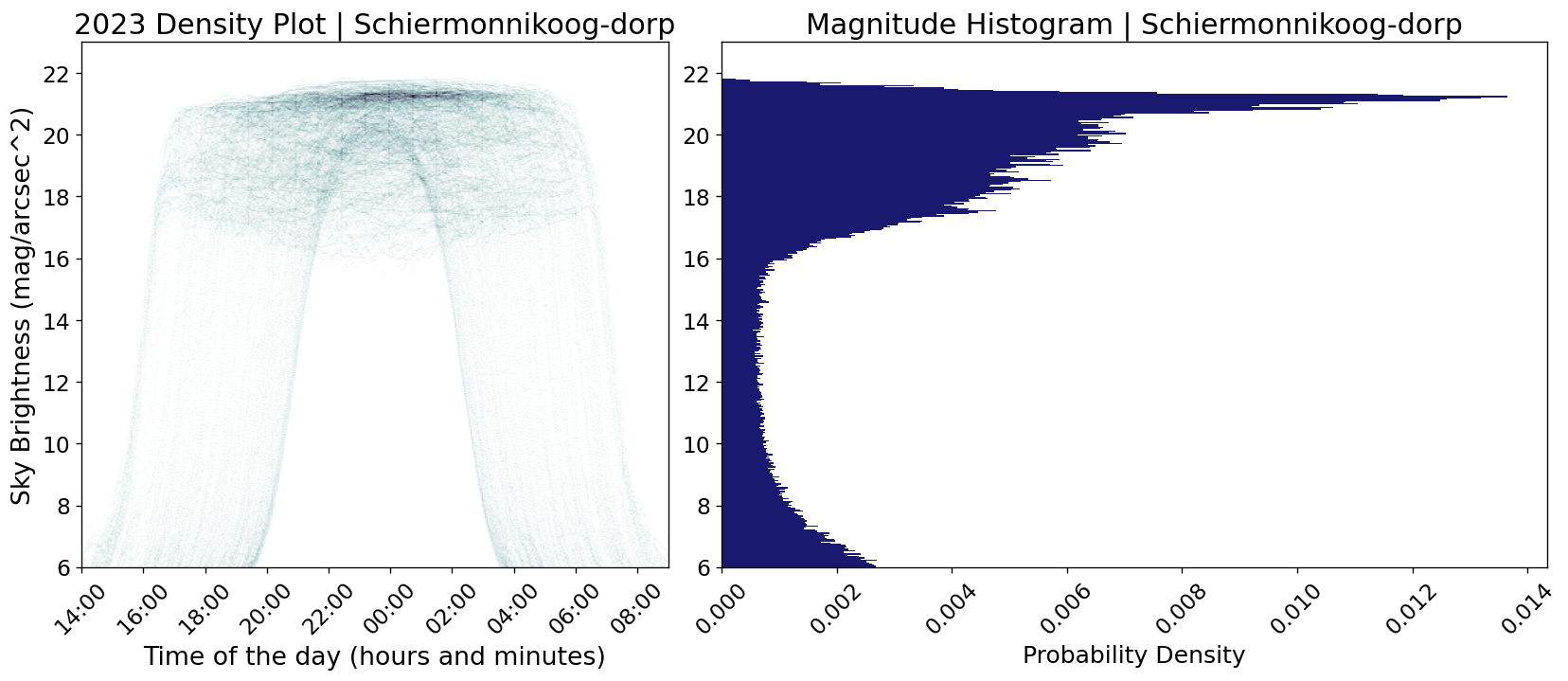}
        \caption{}
        \label{fig:jellyfish31}
    \end{figure}
    \begin{figure}
        \includegraphics[width=\columnwidth]{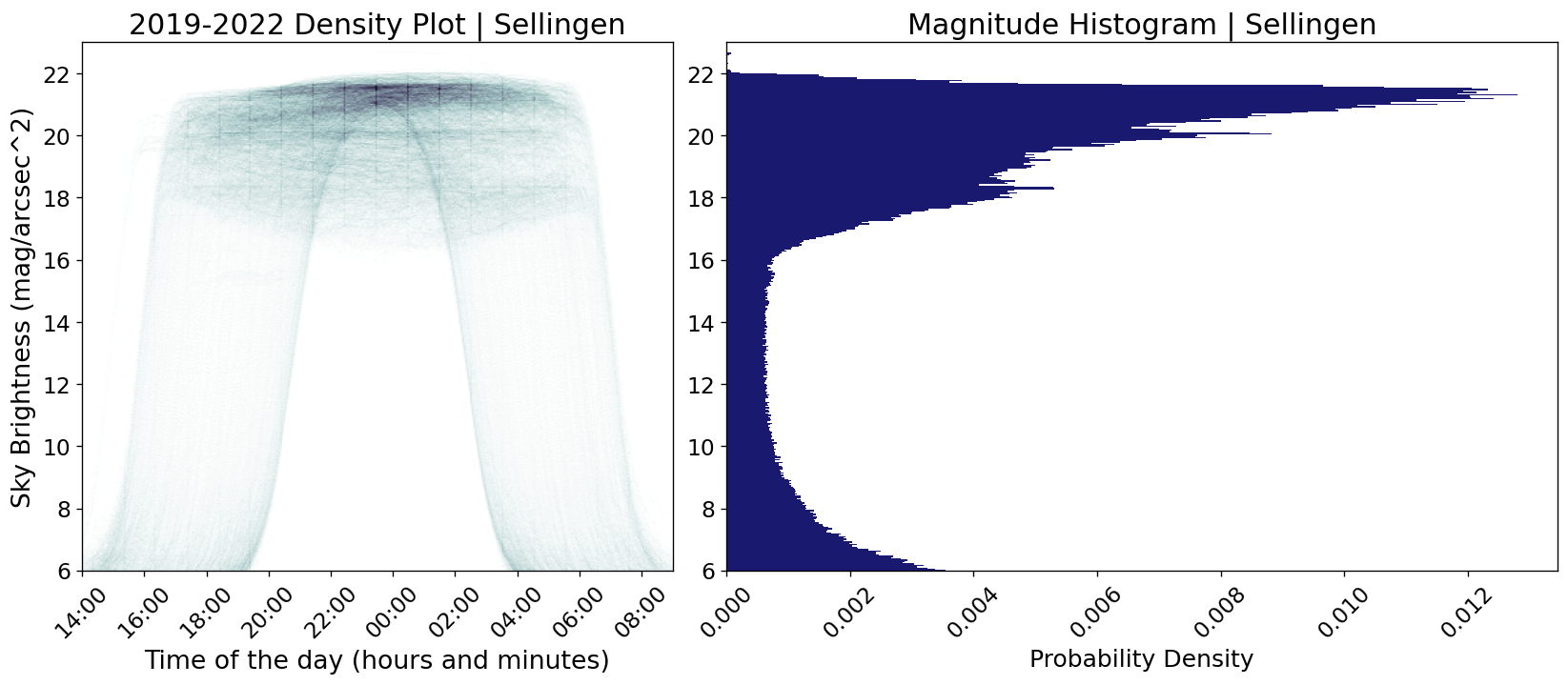}
        \caption{}
        \label{fig:jellyfish32}
    \end{figure}
    \begin{figure}
        \includegraphics[width=\columnwidth]{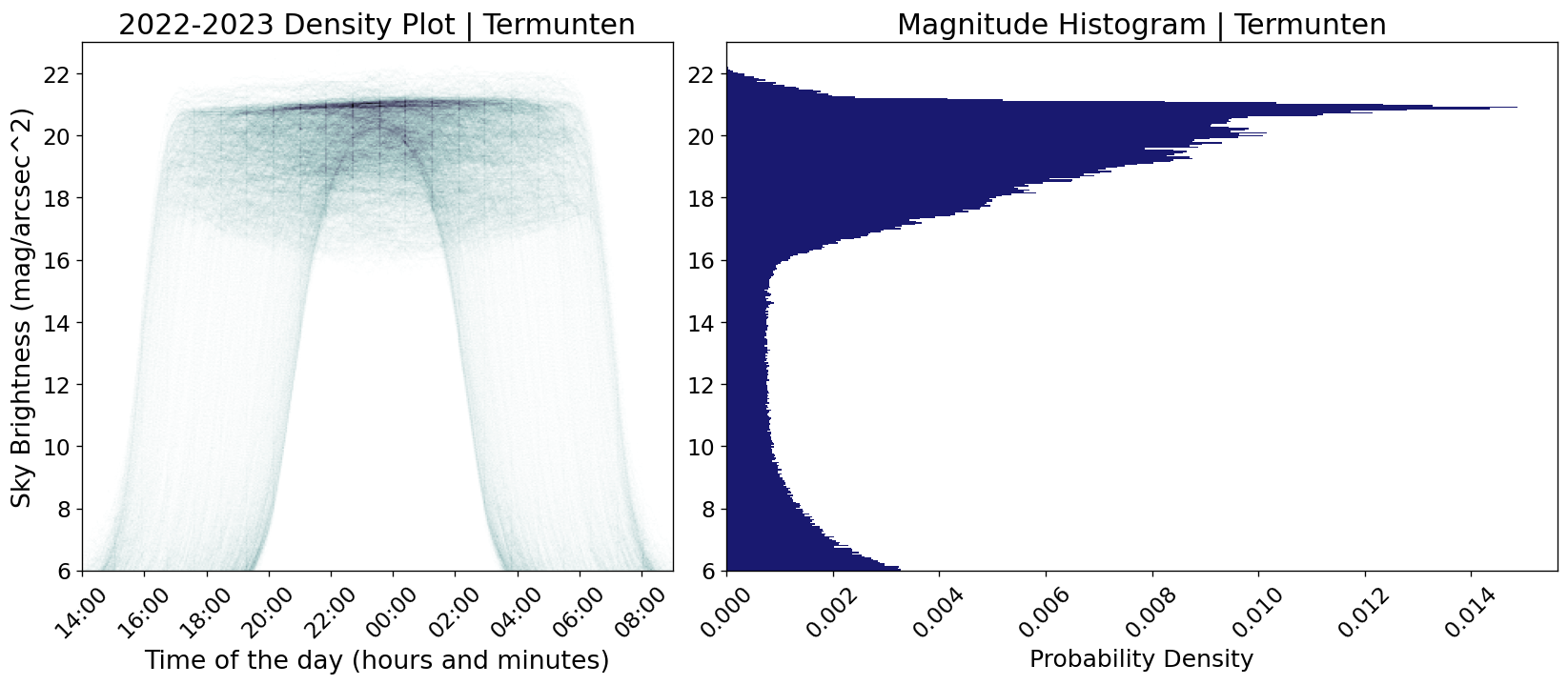}
        \caption{}
        \label{fig:jellyfish33}
    \end{figure}
    \begin{figure}
        \includegraphics[width=\columnwidth]{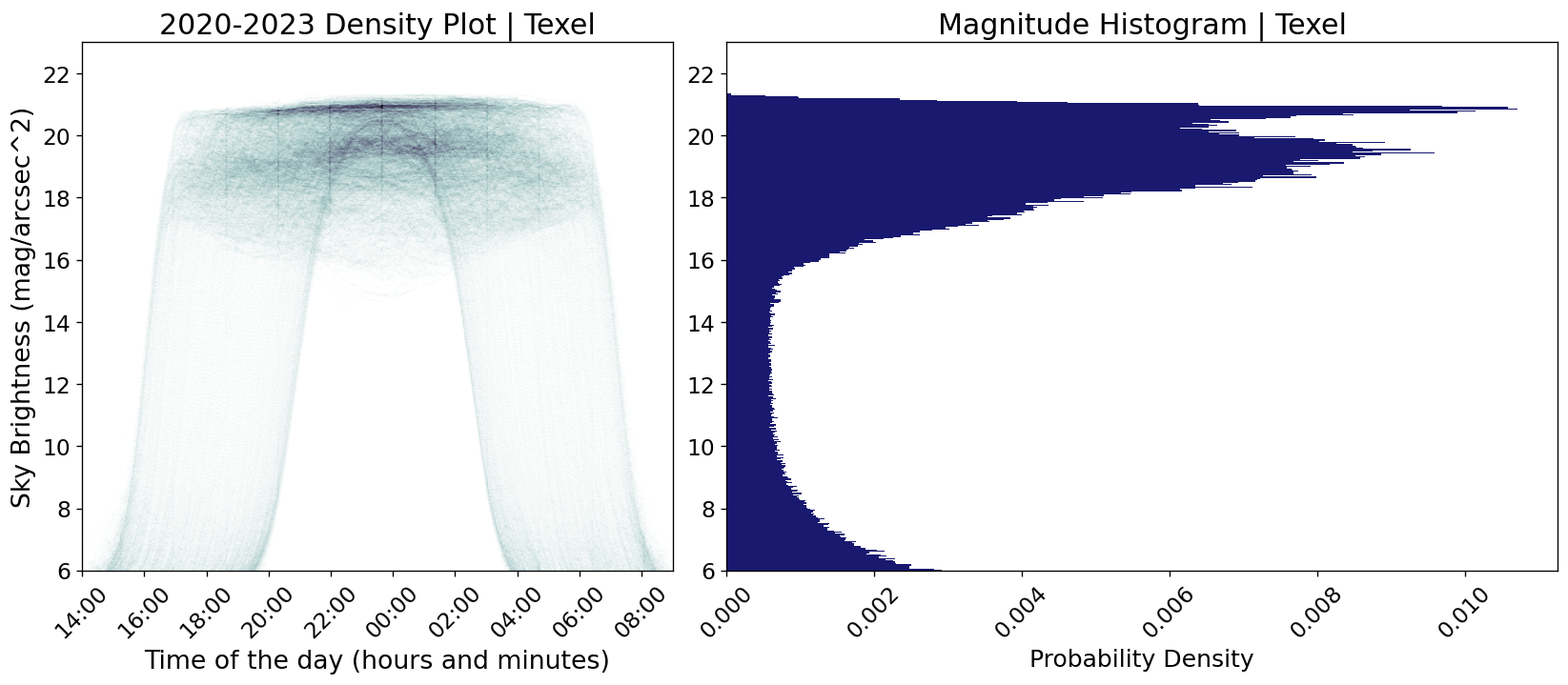}
        \caption{}
        \label{fig:jellyfish34}
    \end{figure}
    \begin{figure}
        \includegraphics[width=\columnwidth]{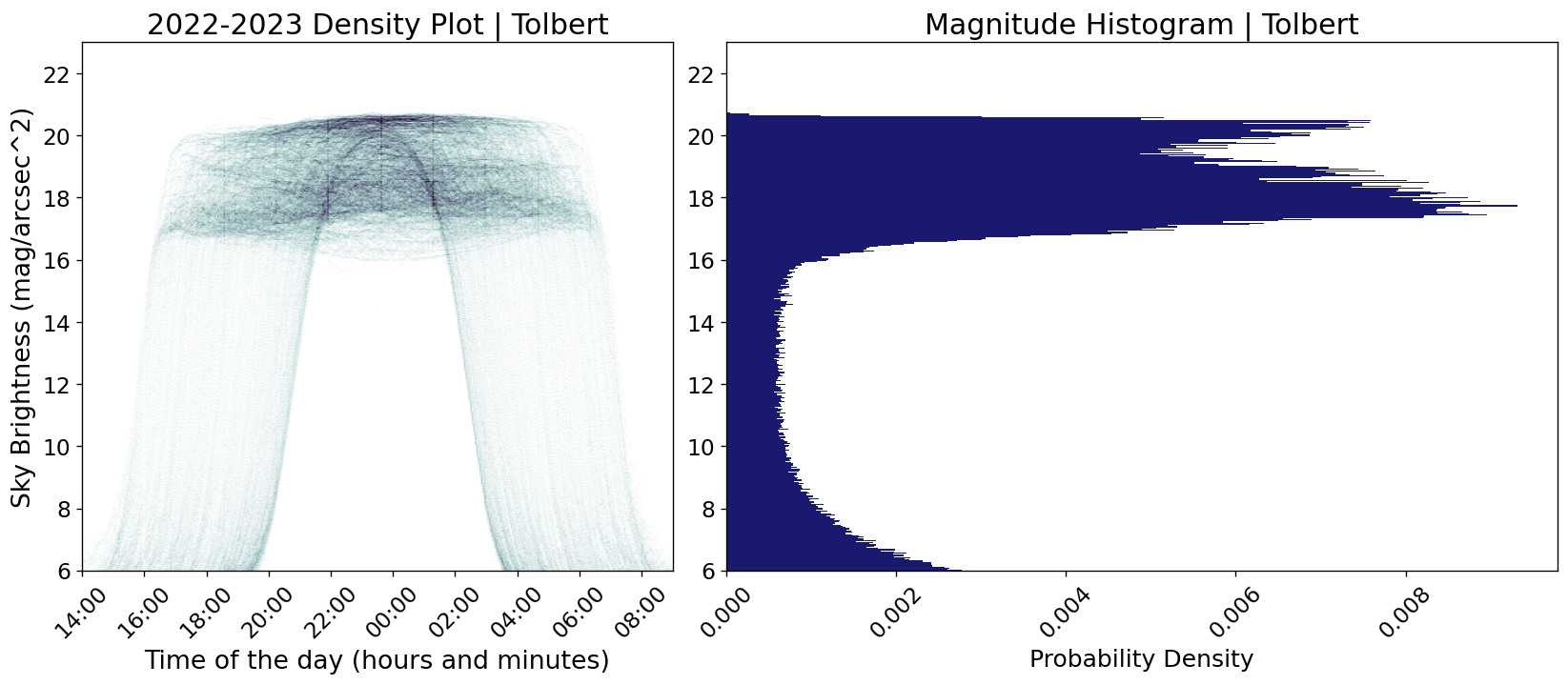}
        \caption{}
        \label{fig:jellyfish35}
    \end{figure}
    \begin{figure}
        \includegraphics[width=\columnwidth]{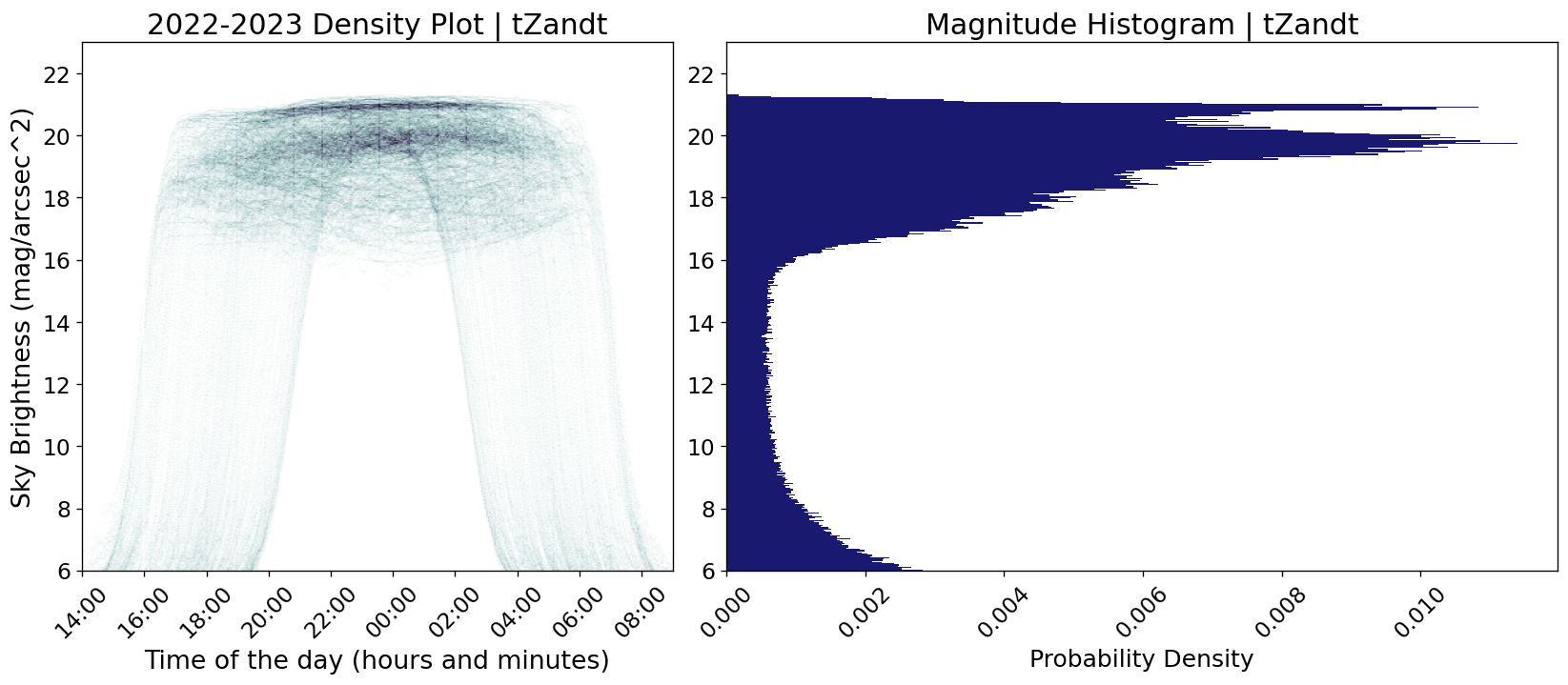}
        \caption{}
        \label{fig:jellyfish36}
    \end{figure}
    \begin{figure}
        \includegraphics[width=\columnwidth]{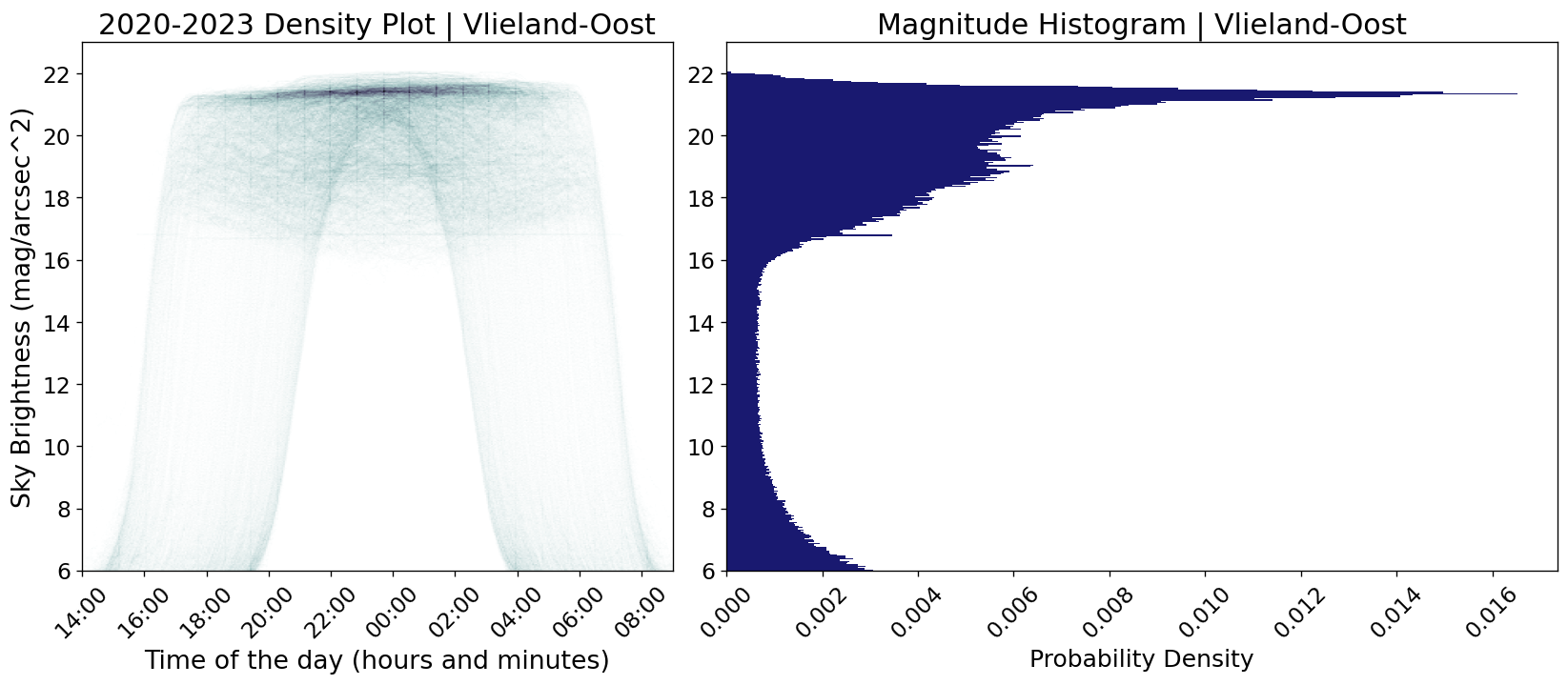}
        \caption{}
        \label{fig:jellyfish37}
    \end{figure}
    \begin{figure}
        \includegraphics[width=\columnwidth]{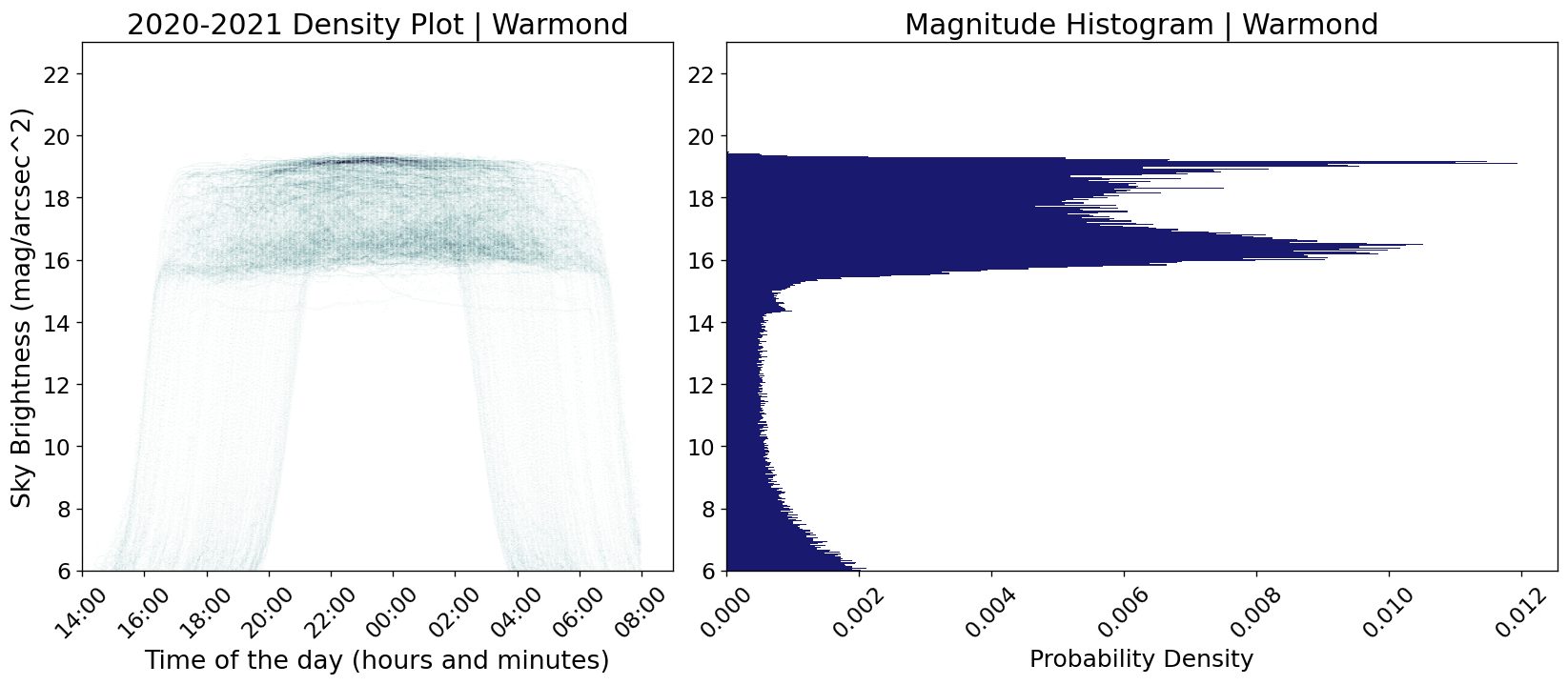}
        \caption{}
        \label{fig:jellyfish38}
    \end{figure}
    \begin{figure}
        \includegraphics[width=\columnwidth]{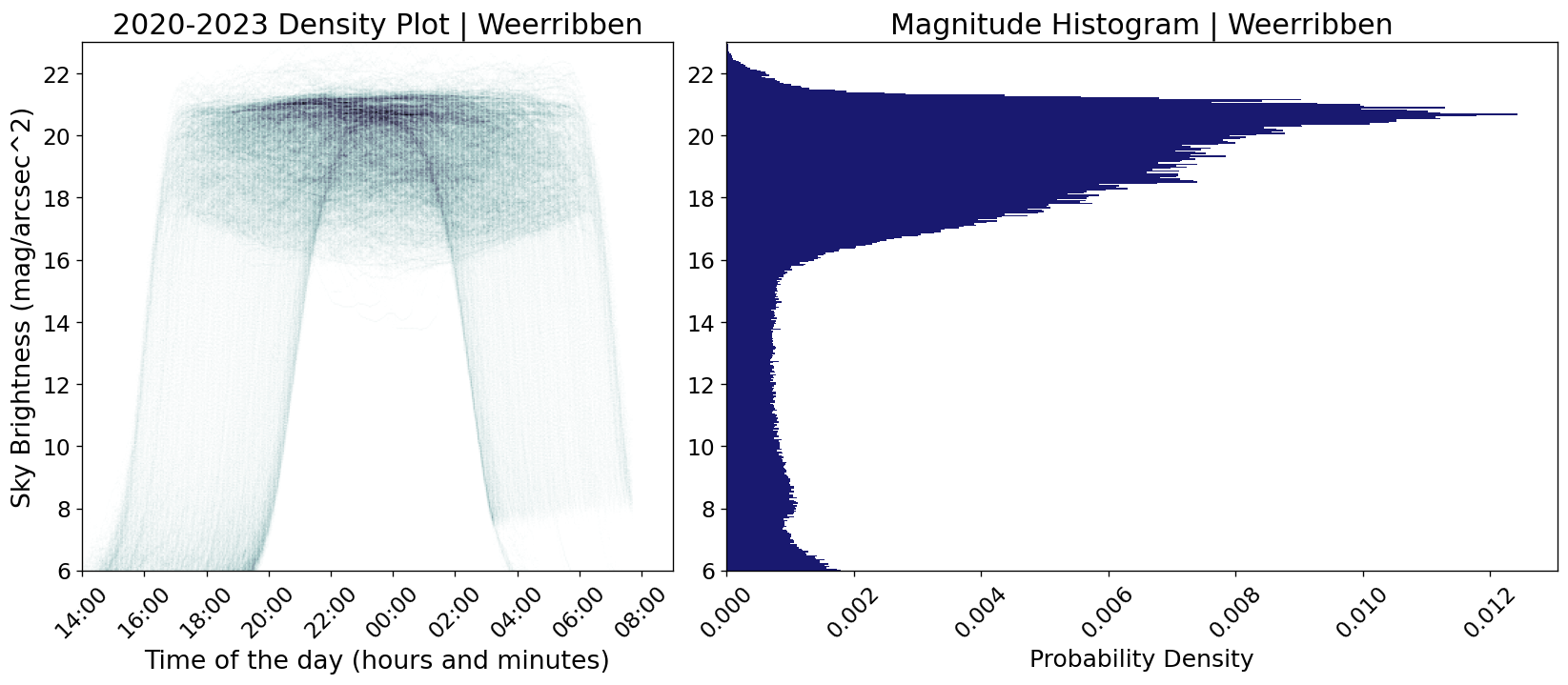}
        \caption{}
        \label{fig:jellyfish39}
    \end{figure}
    \begin{figure}
        \includegraphics[width=\columnwidth]{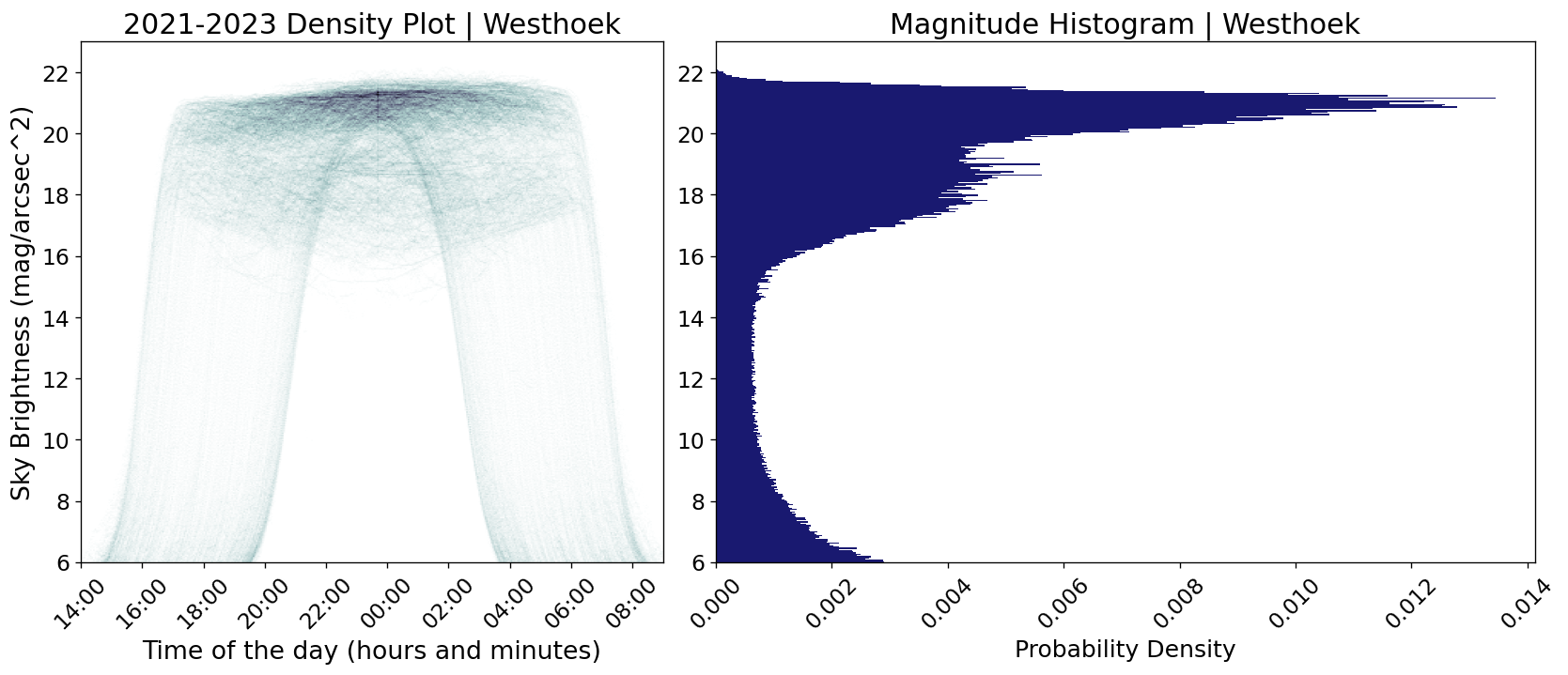}
        \caption{}
        \label{fig:jellyfish40}
    \end{figure}
    \begin{figure}
        \includegraphics[width=\columnwidth]{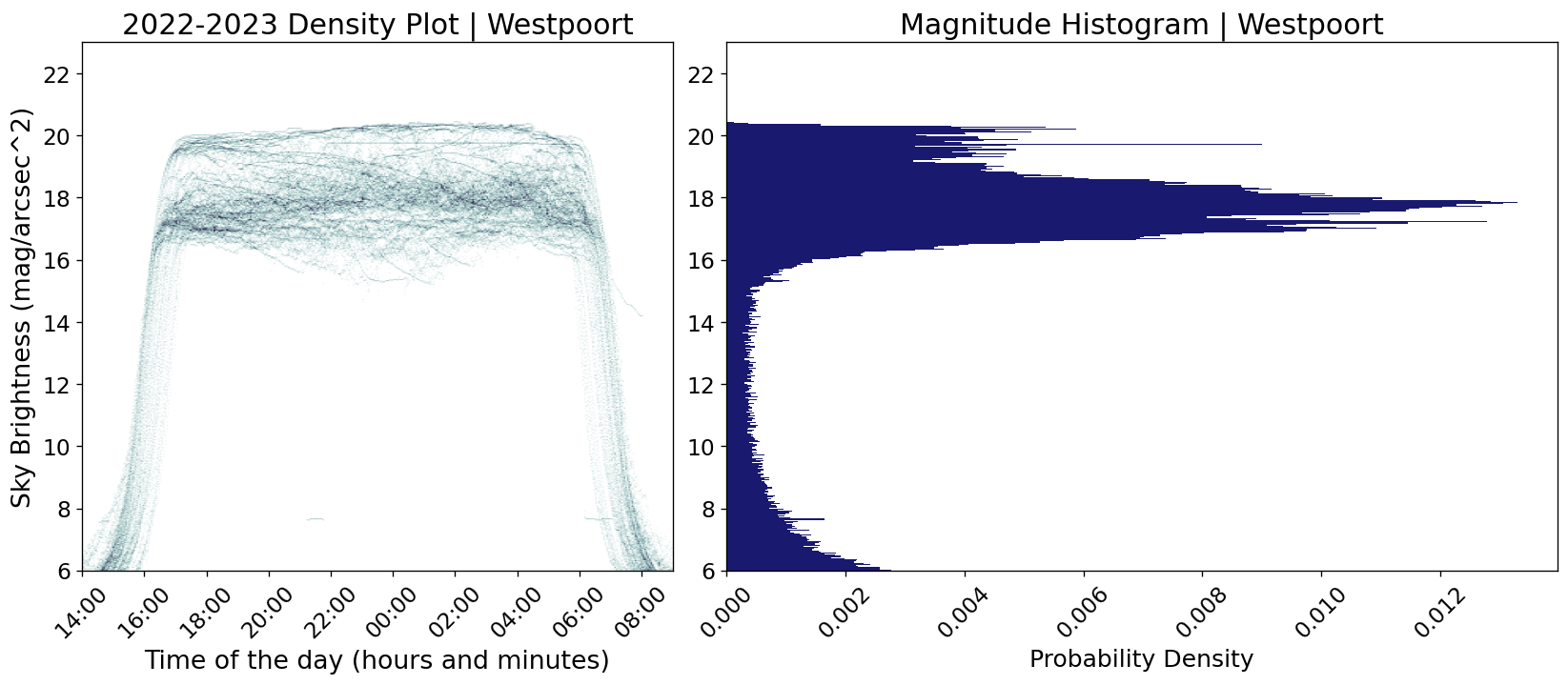}
        \caption{}
        \label{fig:jellyfish41}
    \end{figure}
%}

\FloatBarrier

\section{Time-Magnitude trend plots} \label{appendixtrend}
The trend plots created during this study are included here.

\setcounter{figure}{0}
\renewcommand{\thefigure}{E\arabic{figure}}

% Define the directory for trend images
\newcommand{\trenddir}{Trends/}

% Loop through all 25 trend images
    \begin{figure}
        \includegraphics[width=\columnwidth]{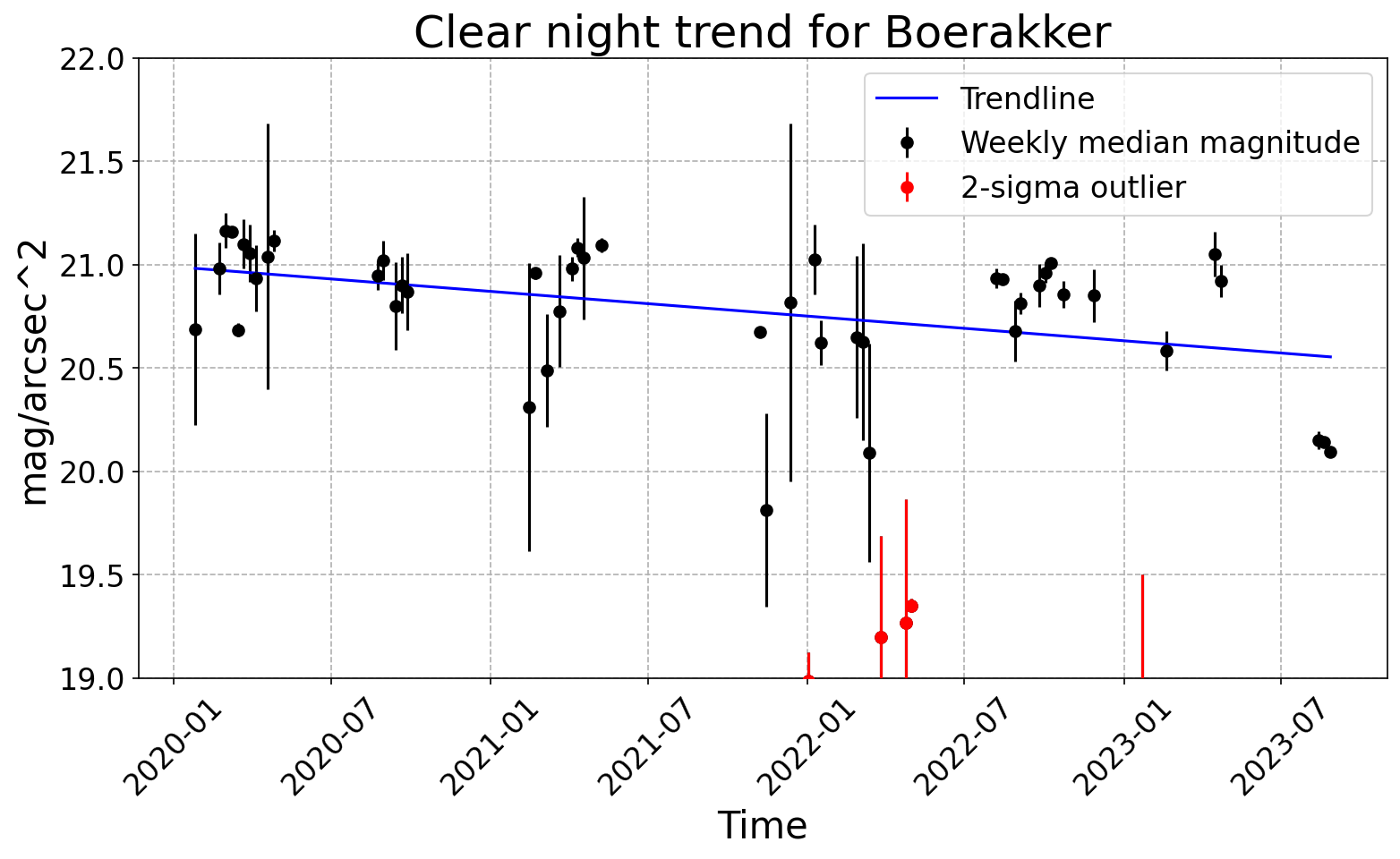}
        \caption{}
        \label{fig:trends1}
    \end{figure}
    \begin{figure}
        \includegraphics[width=\columnwidth]{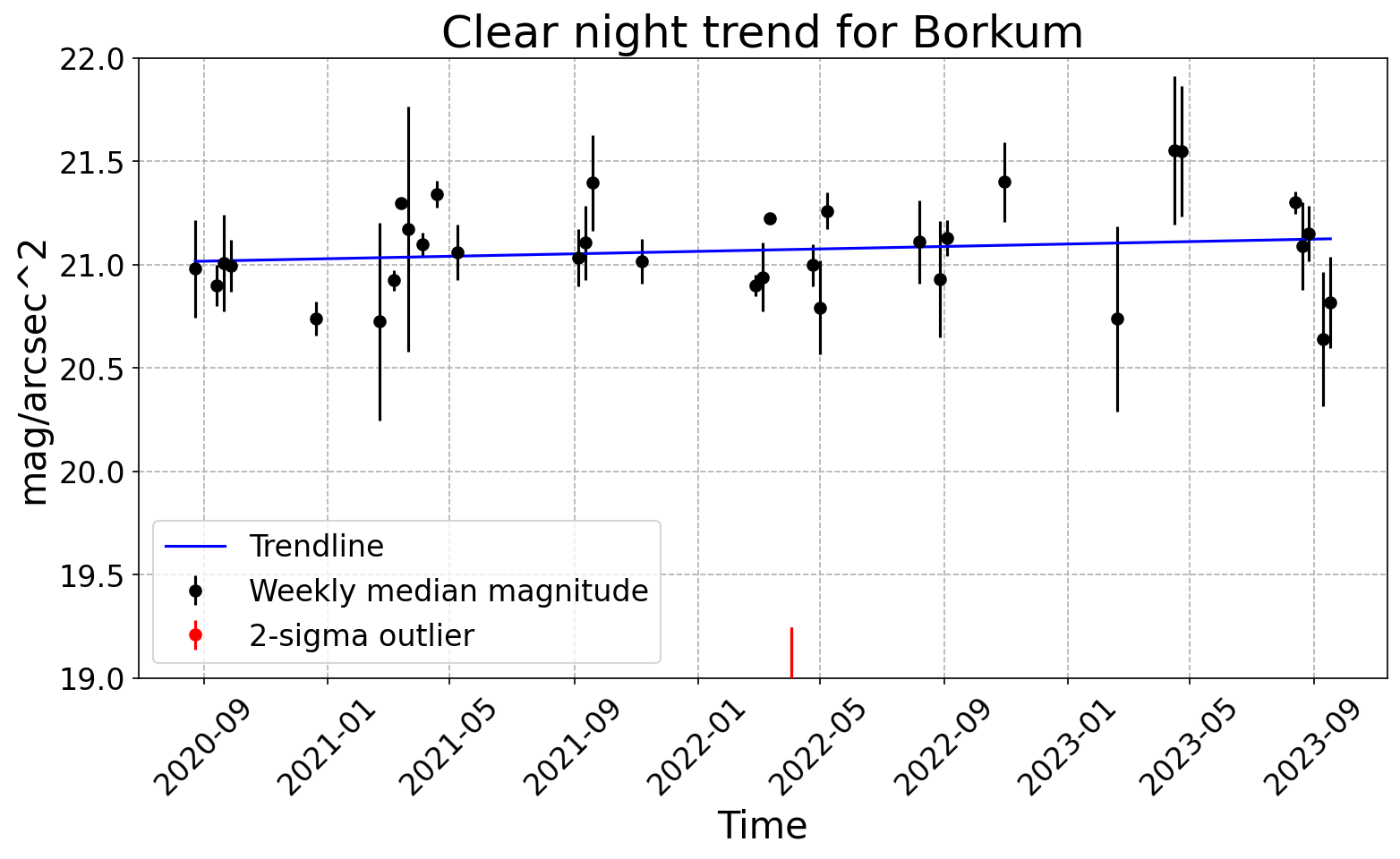}
        \caption{}
        \label{fig:trends2}
    \end{figure}
    \begin{figure}
        \includegraphics[width=\columnwidth]{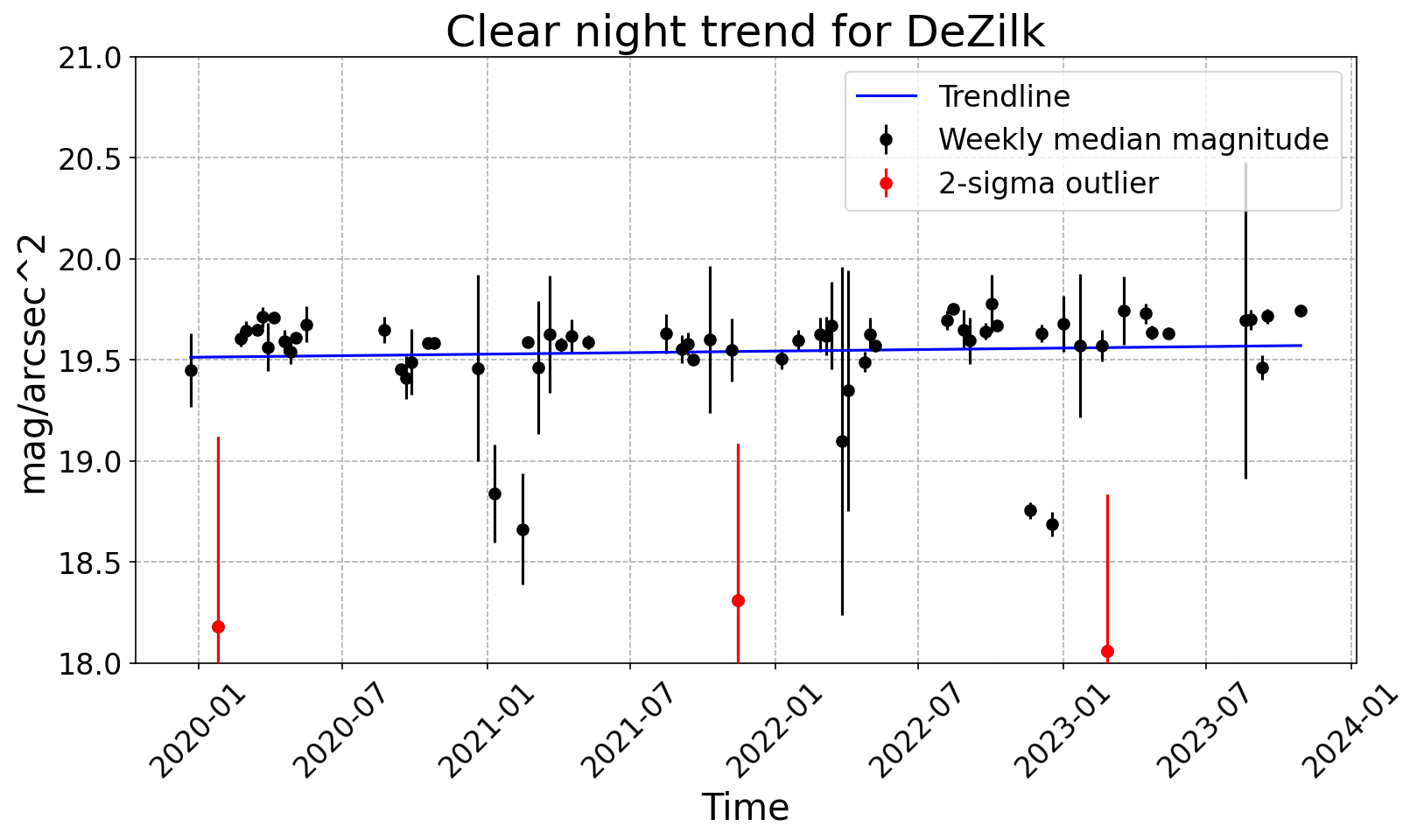}
        \caption{}
        \label{fig:trends3}
    \end{figure}
    \begin{figure}
        \includegraphics[width=\columnwidth]{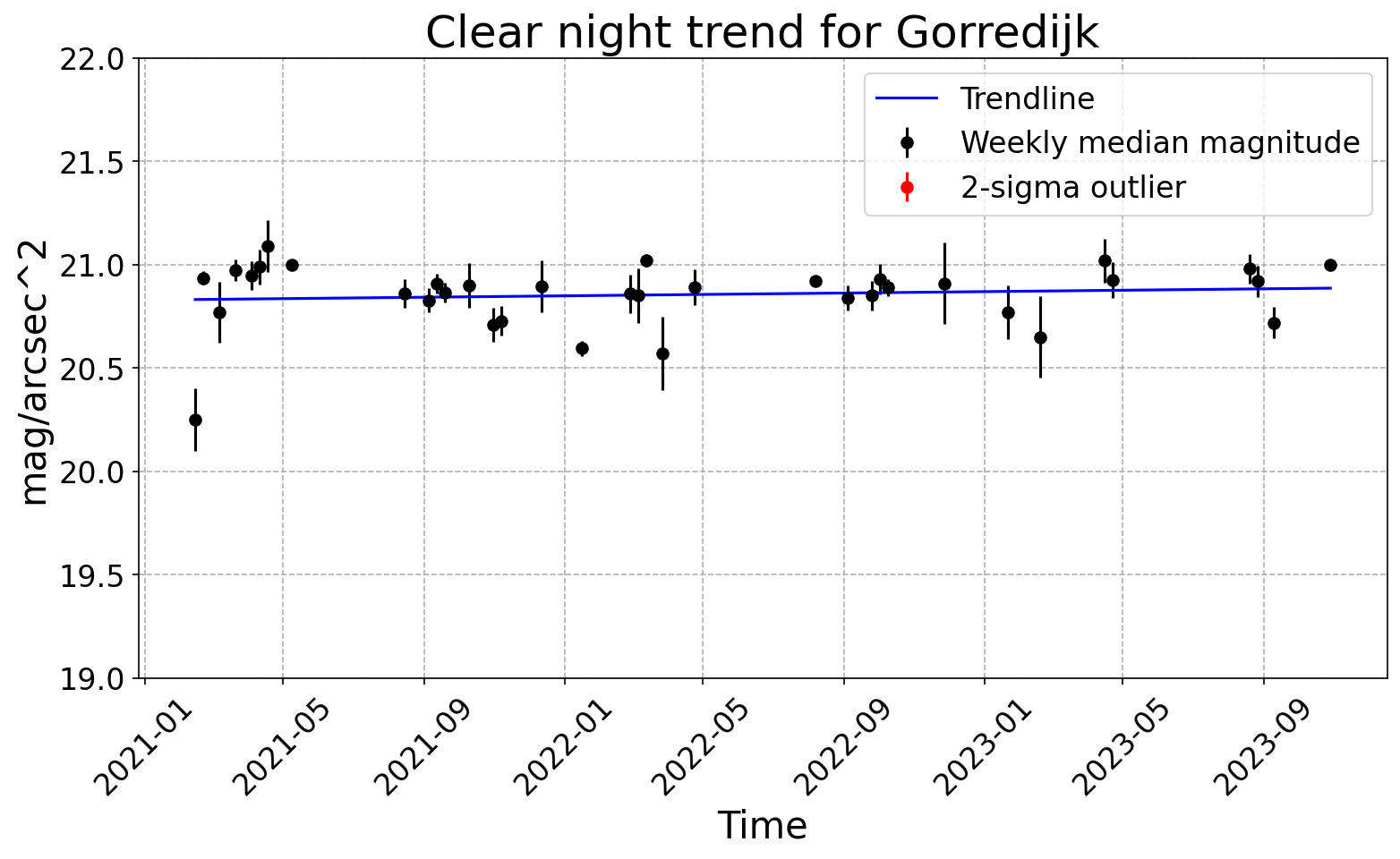}
        \caption{}
        \label{fig:trends4}
    \end{figure}
    \begin{figure}
        \includegraphics[width=\columnwidth]{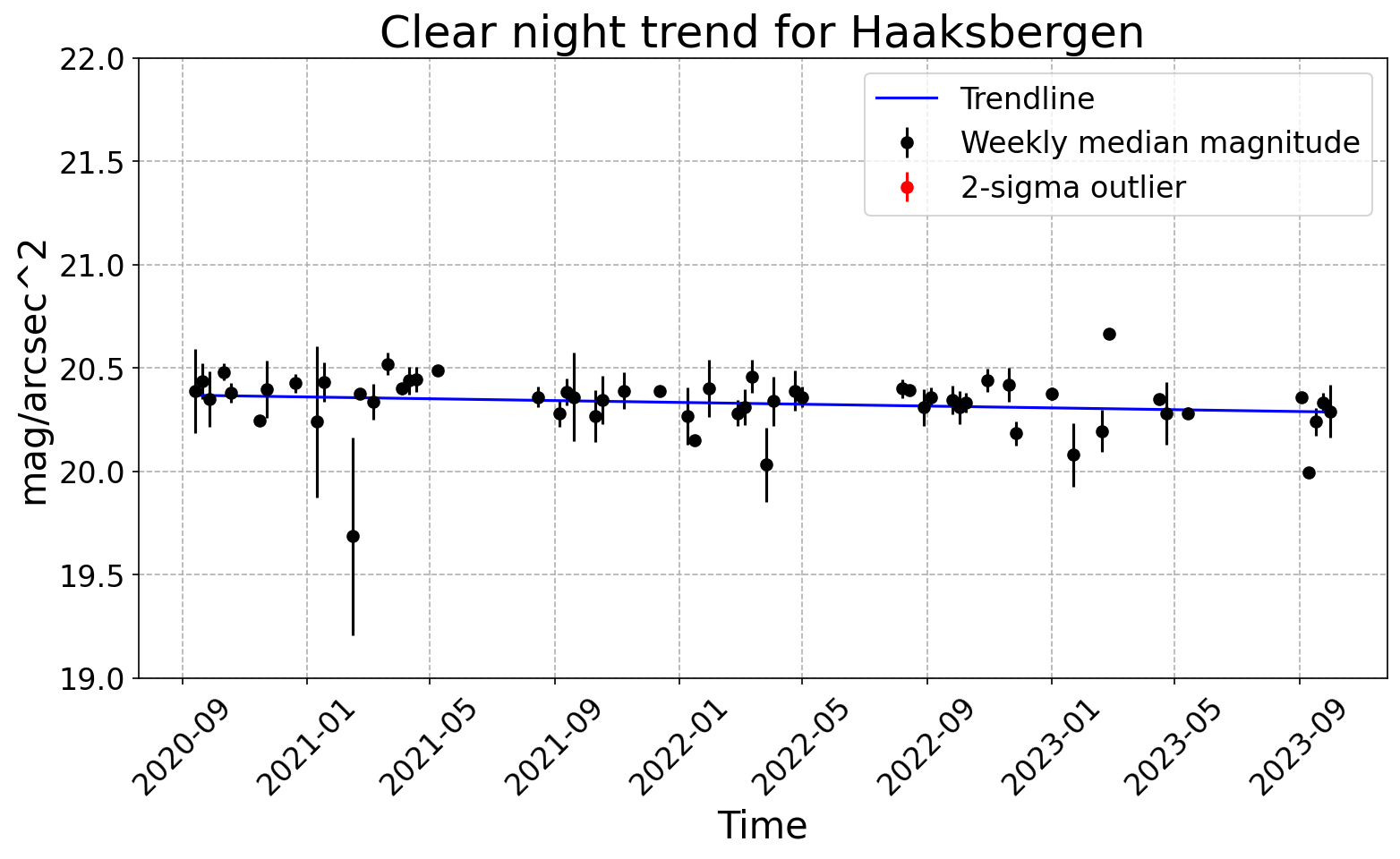}
        \caption{}
        \label{fig:trends5}
    \end{figure}
    \begin{figure}
        \includegraphics[width=\columnwidth]{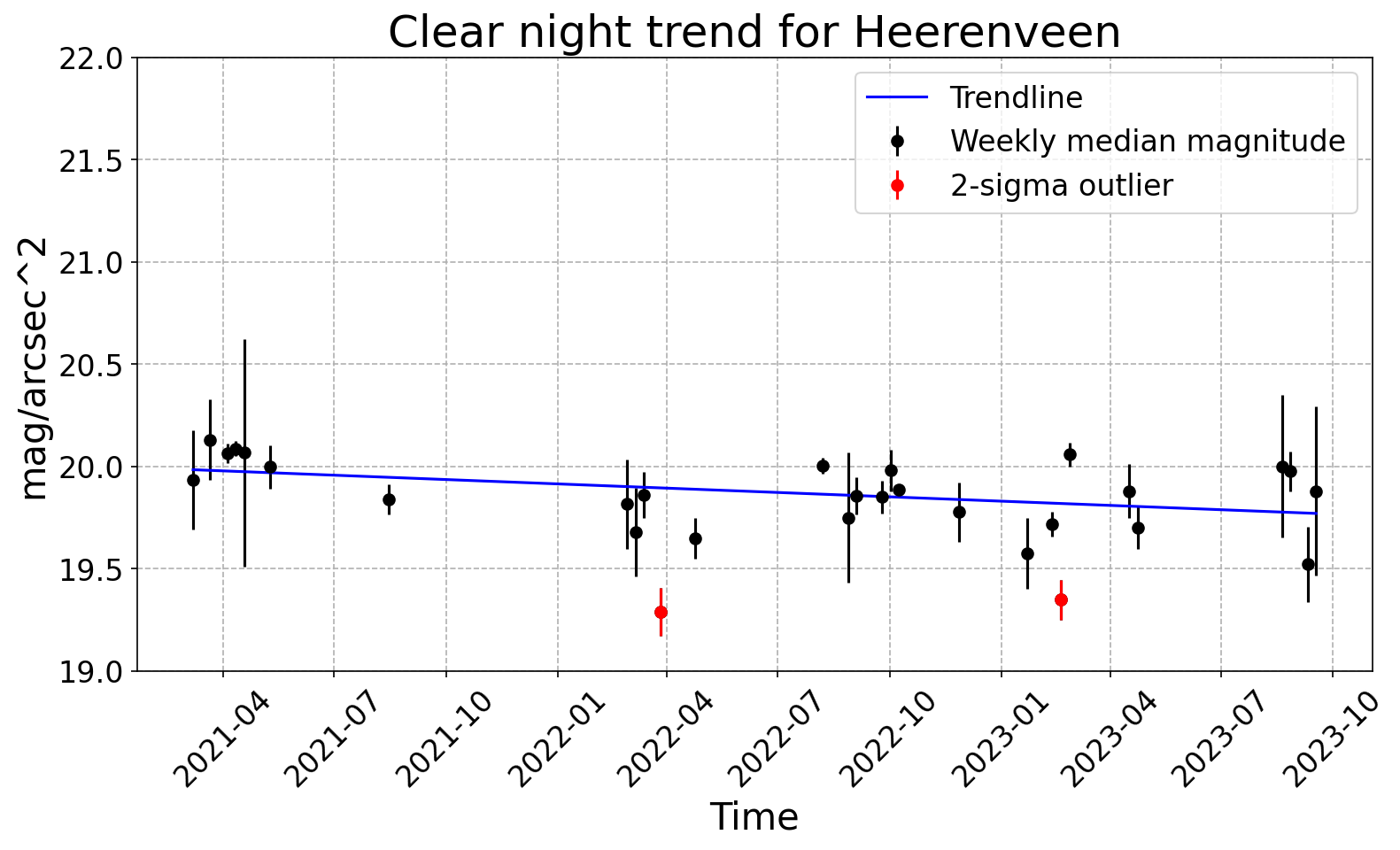}
        \caption{}
        \label{fig:trends6}
    \end{figure}
    \begin{figure}
        \includegraphics[width=\columnwidth]{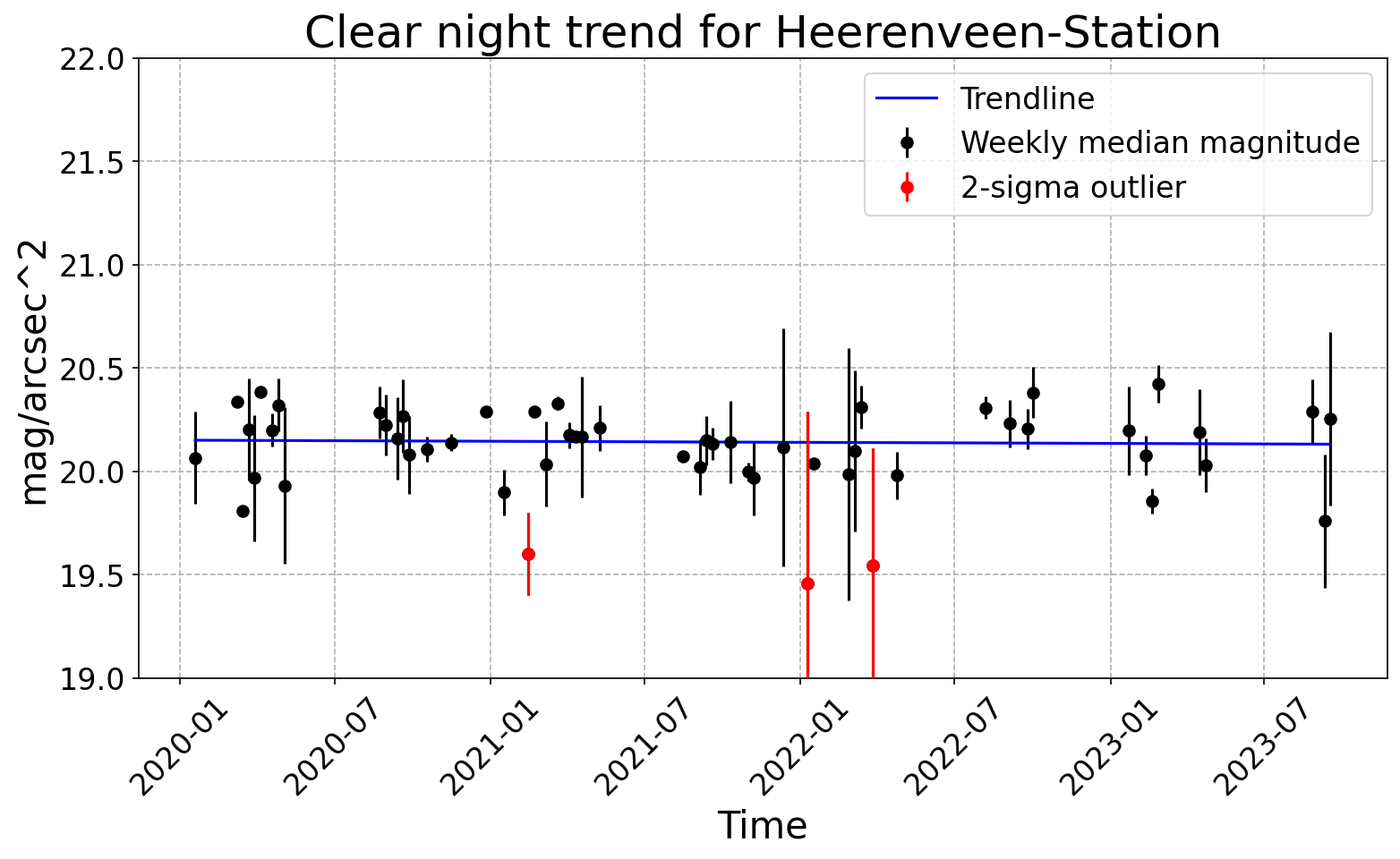}
        \caption{}
        \label{fig:trends7}
    \end{figure}
    \begin{figure}
        \includegraphics[width=\columnwidth]{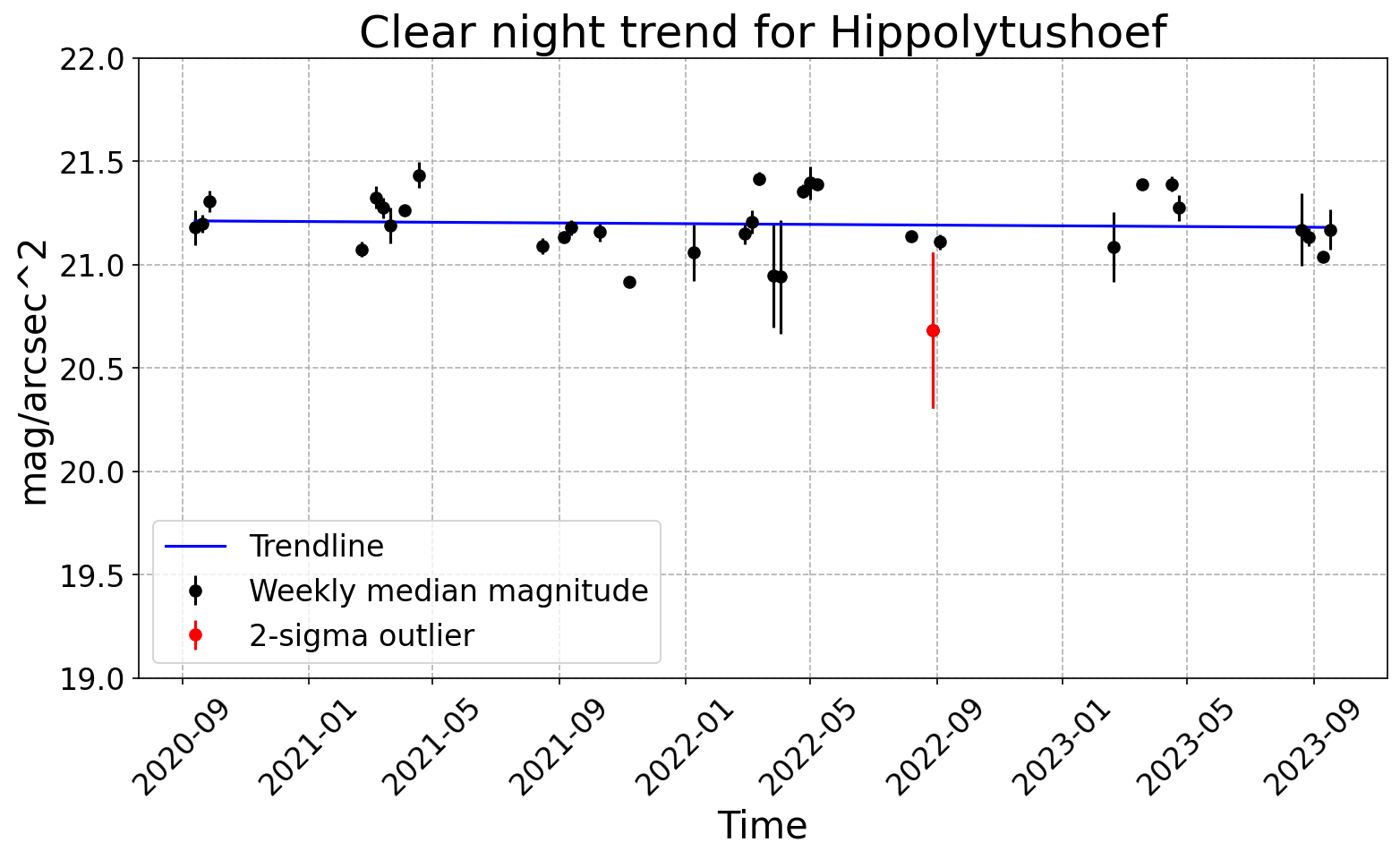}
        \caption{}
        \label{fig:trends8}
    \end{figure}
    \begin{figure}
        \includegraphics[width=\columnwidth]{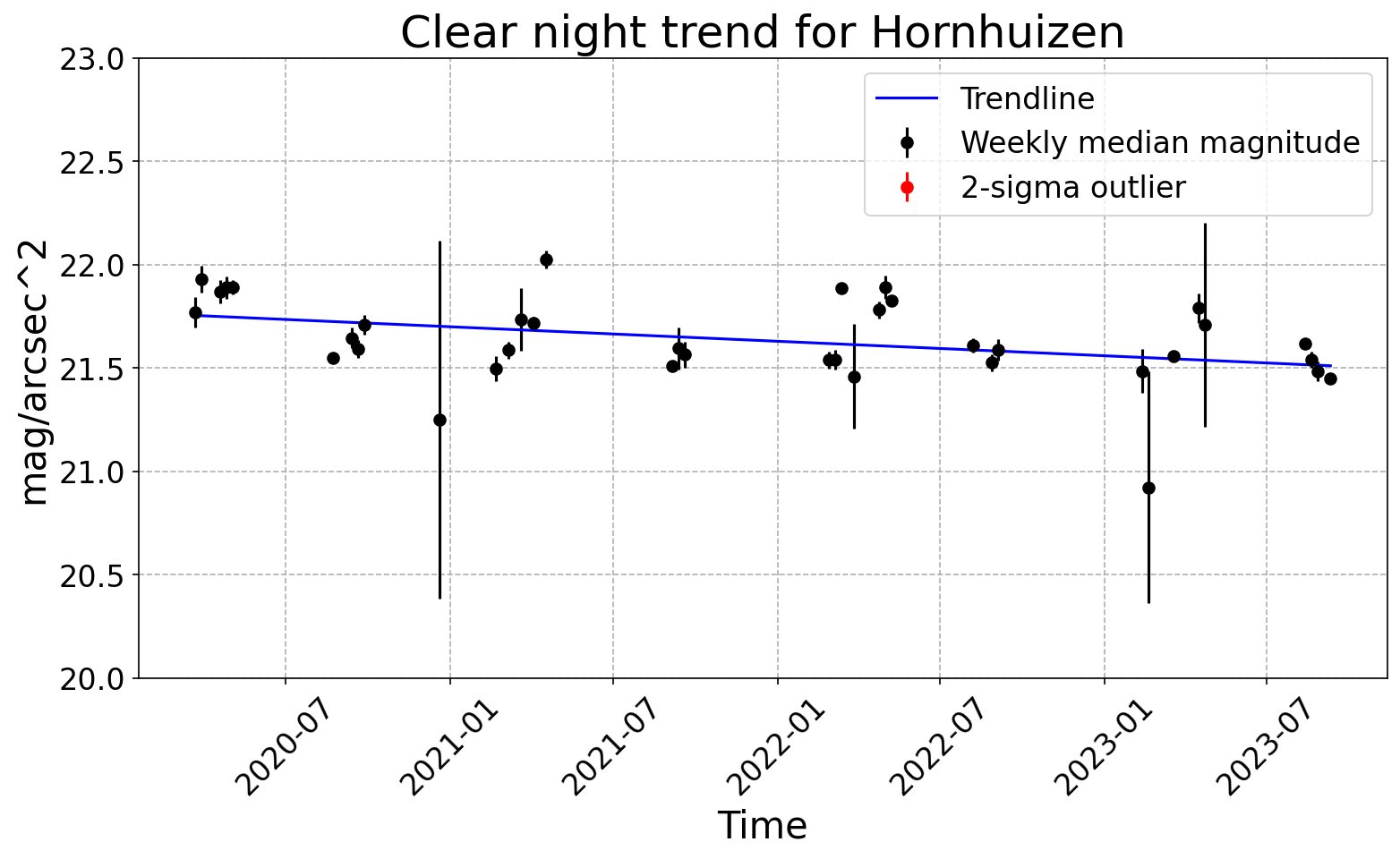}
        \caption{}
        \label{fig:trends9}
    \end{figure}
    \begin{figure}
        \includegraphics[width=\columnwidth]{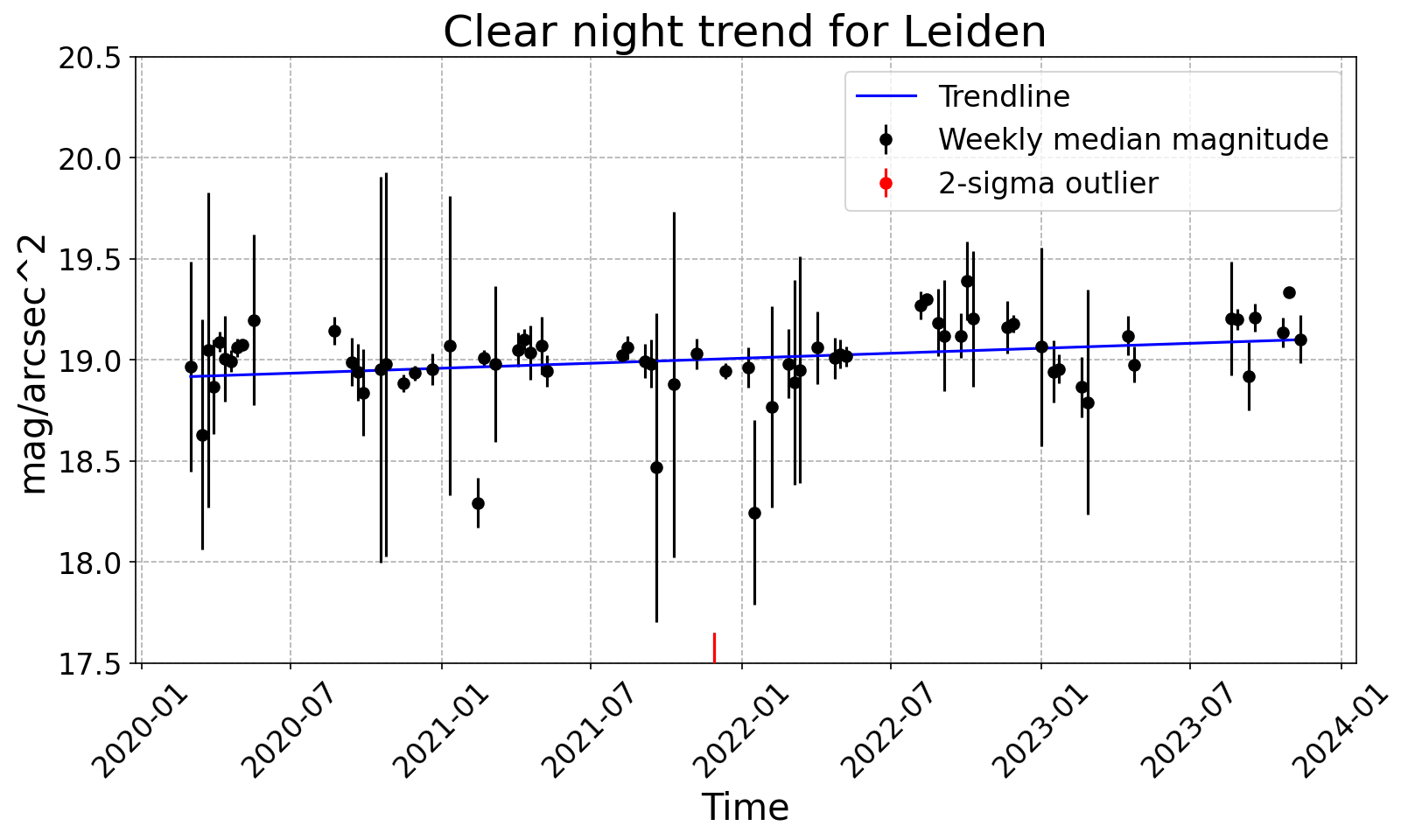}
        \caption{}
        \label{fig:trends10}
    \end{figure}
    \begin{figure}
        \includegraphics[width=\columnwidth]{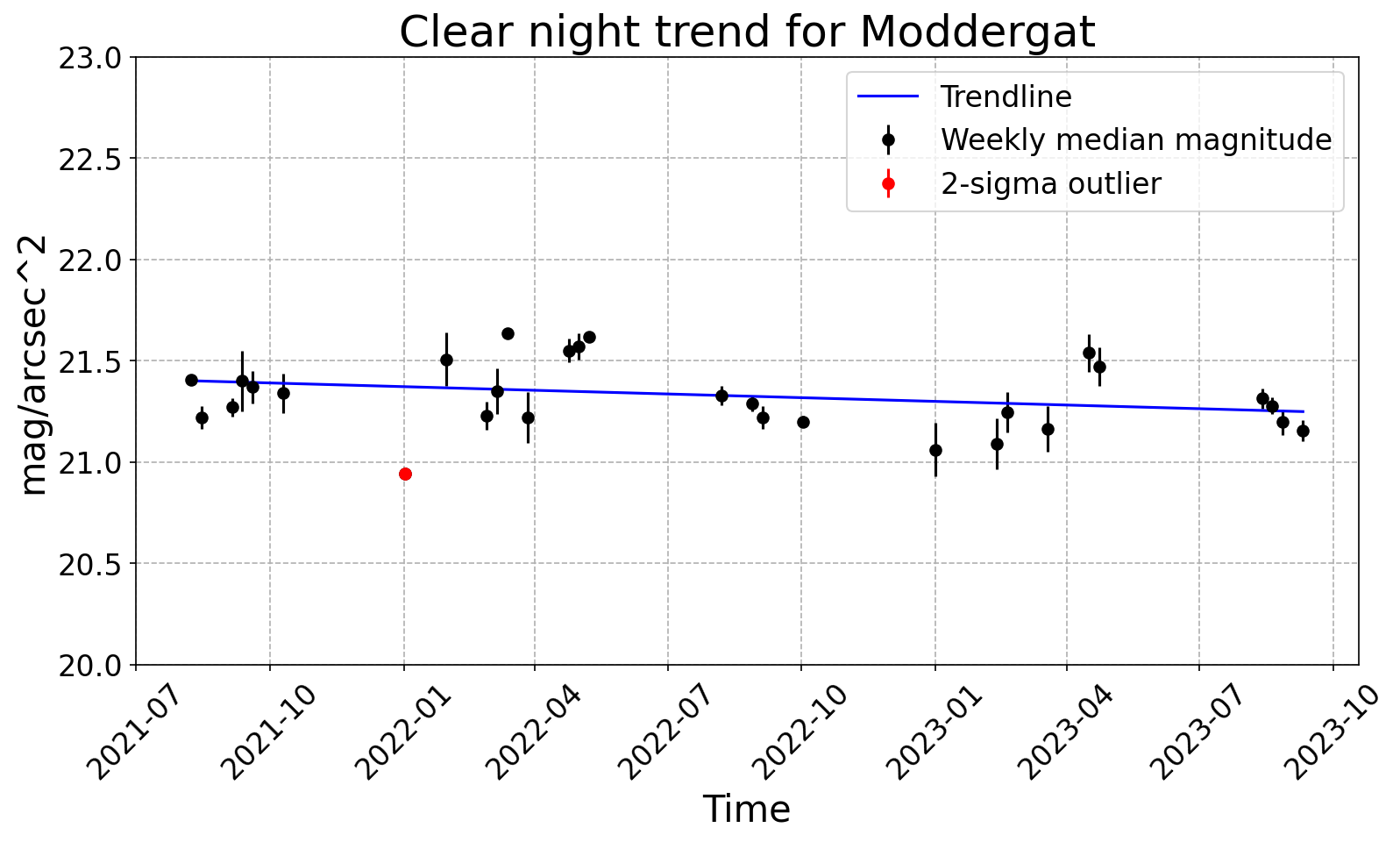}
        \caption{}
        \label{fig:trends11}
    \end{figure}
    \begin{figure}
        \includegraphics[width=\columnwidth]{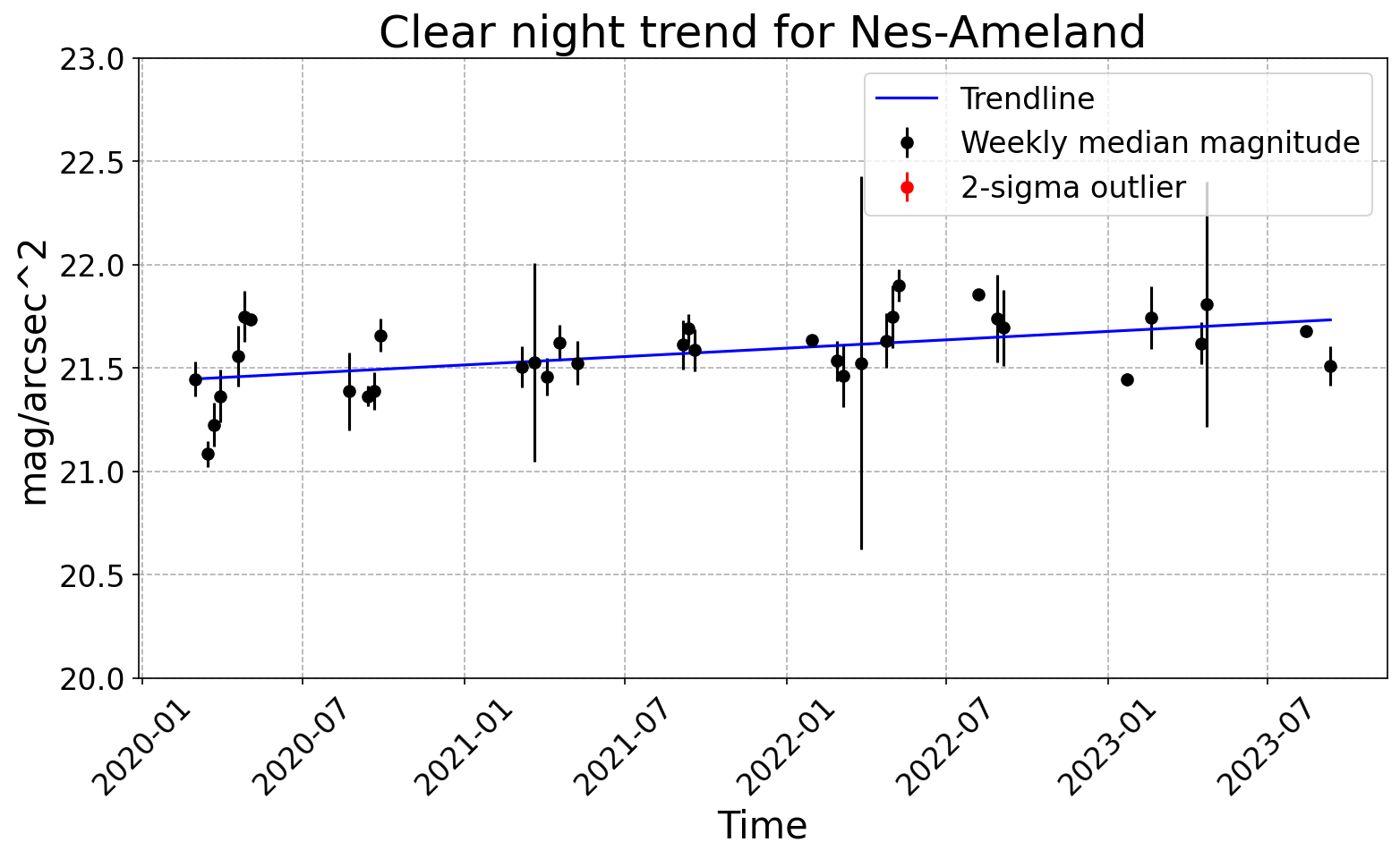}
        \caption{}
        \label{fig:trends12}
    \end{figure}
    \begin{figure}
        \includegraphics[width=\columnwidth]{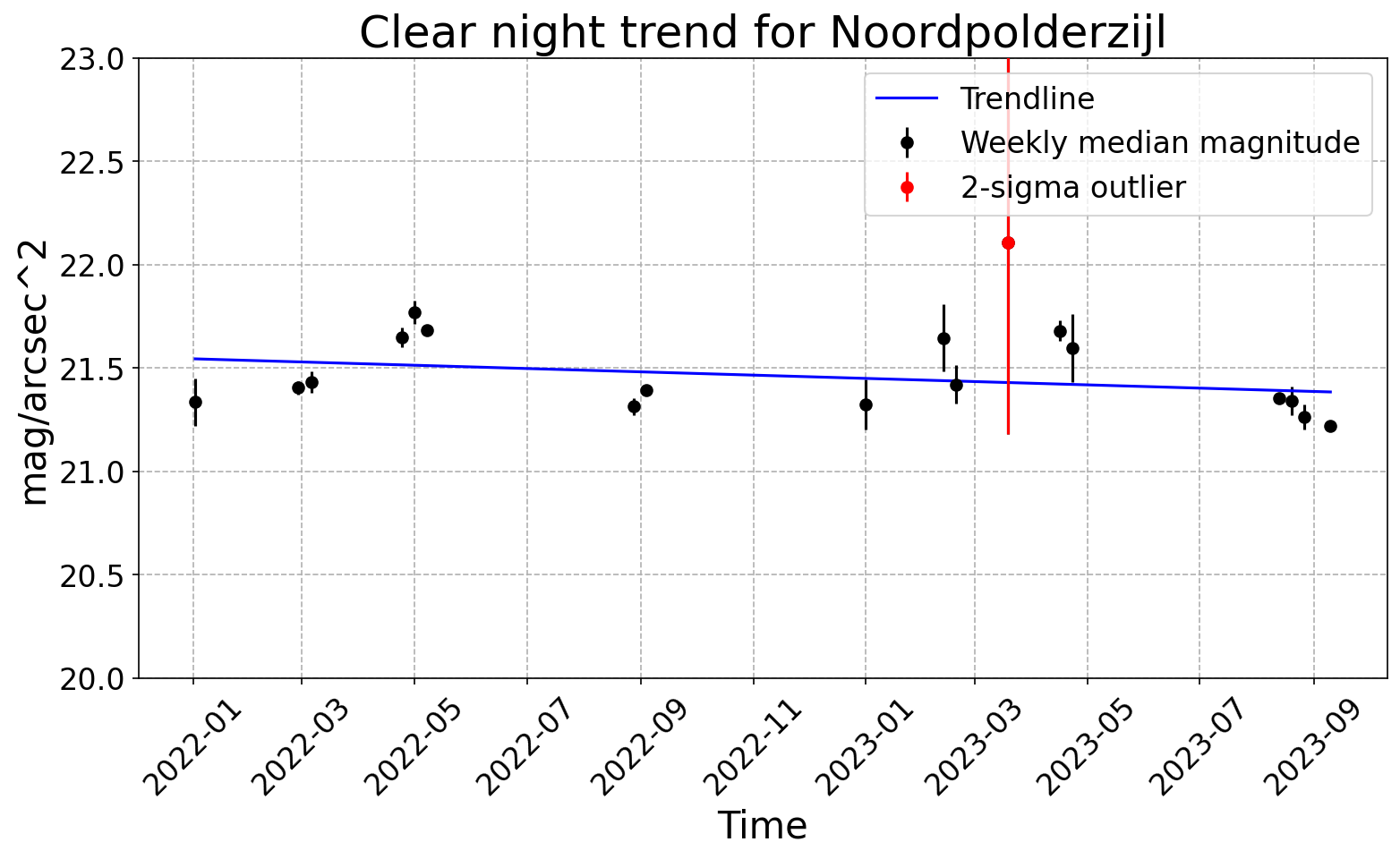}
        \caption{}
        \label{fig:trends13}
    \end{figure}
    \begin{figure}
        \includegraphics[width=\columnwidth]{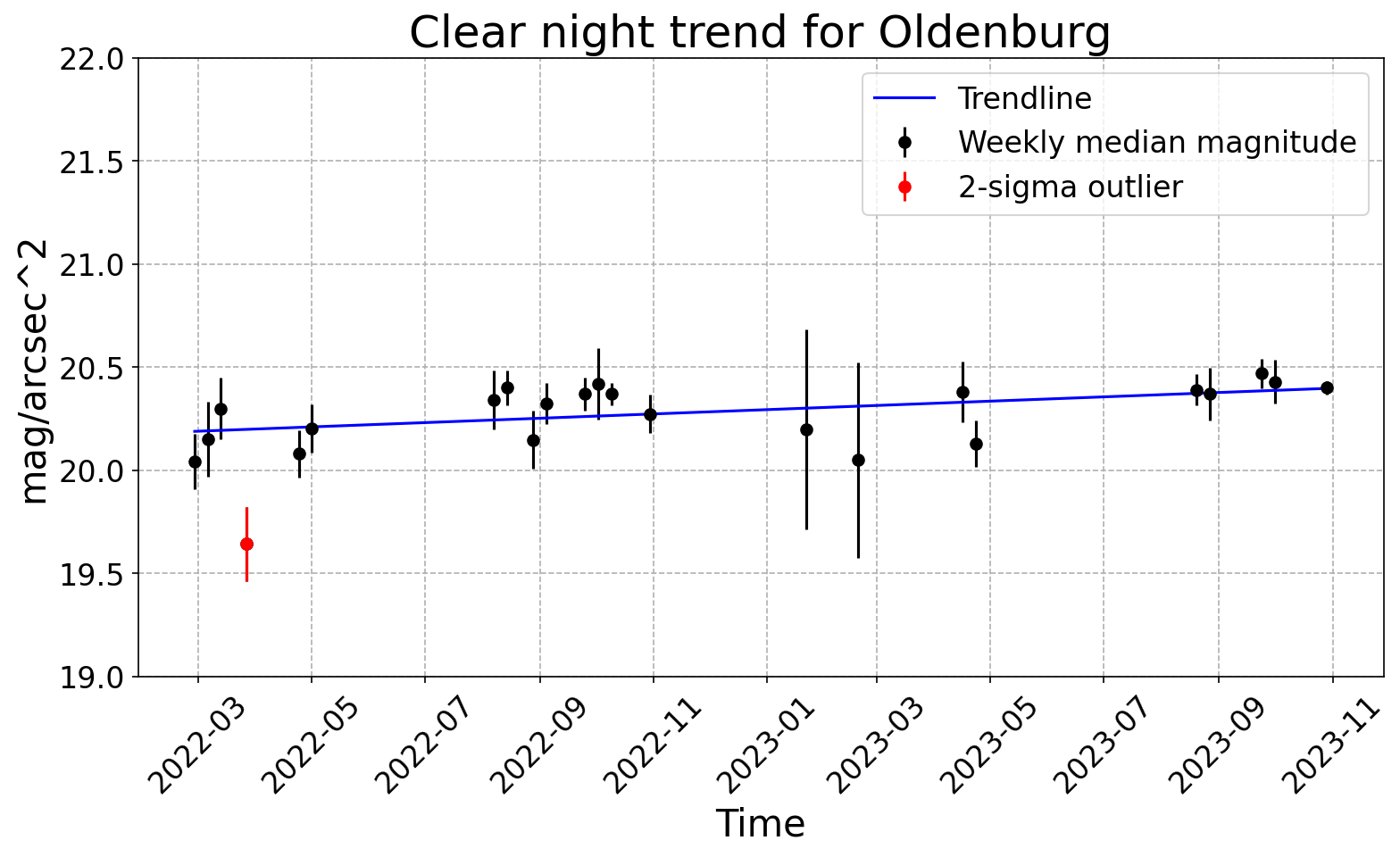}
        \caption{}
        \label{fig:trends14}
    \end{figure}
    \begin{figure}
        \includegraphics[width=\columnwidth]{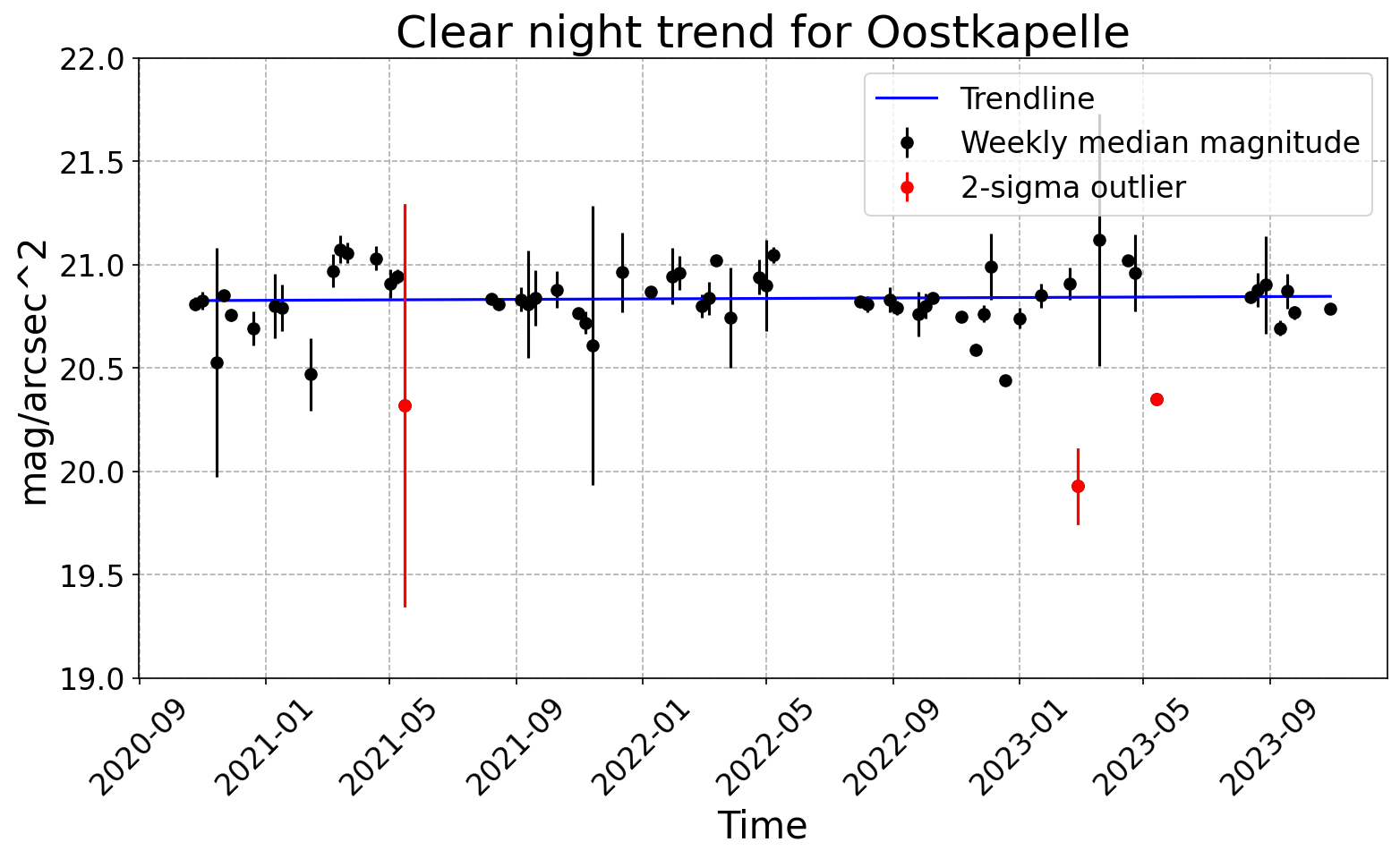}
        \caption{}
        \label{fig:trends15}
    \end{figure}
    \begin{figure}
        \includegraphics[width=\columnwidth]{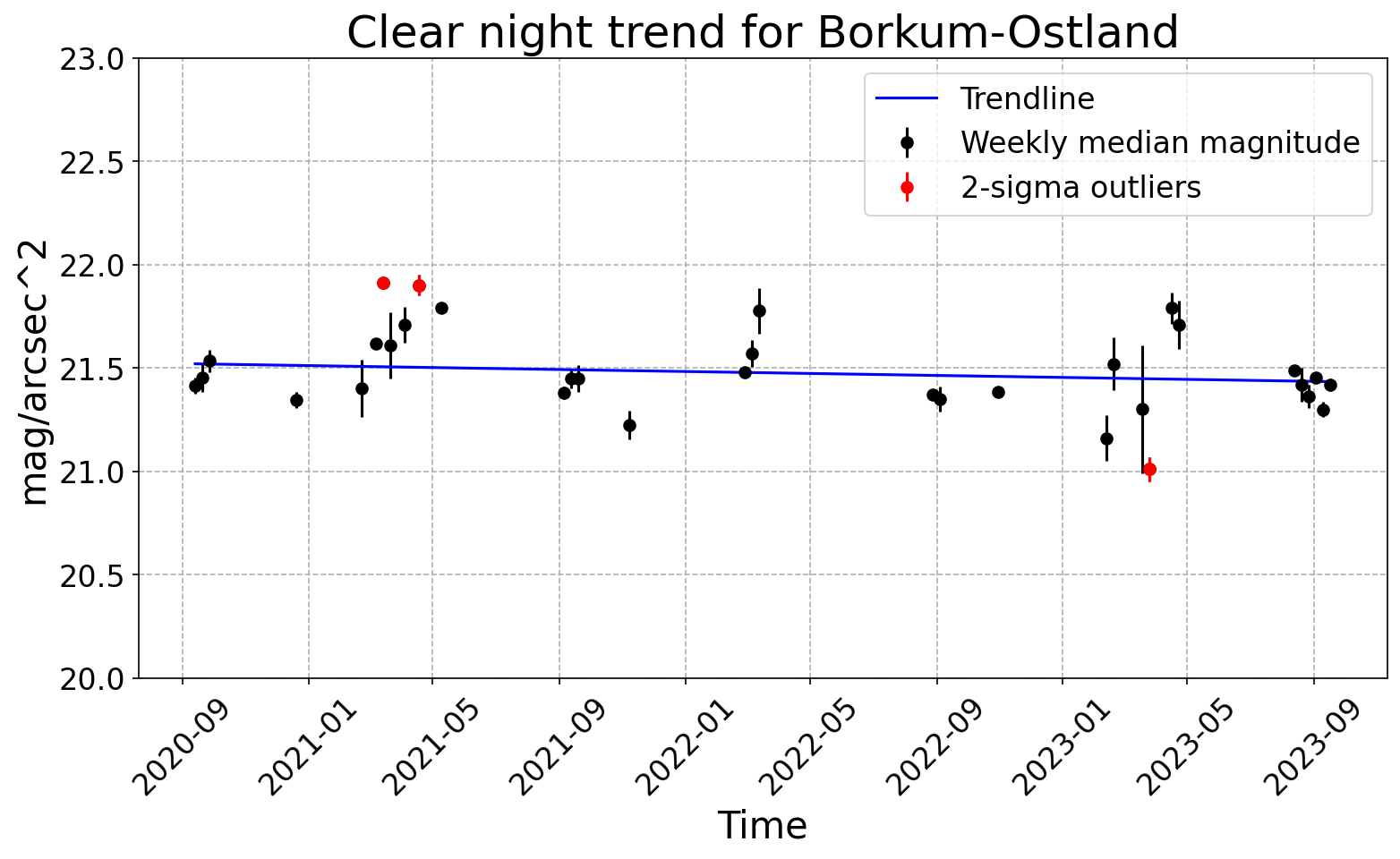}
        \caption{}
        \label{fig:trends16}
    \end{figure}
    \begin{figure}
        \includegraphics[width=\columnwidth]{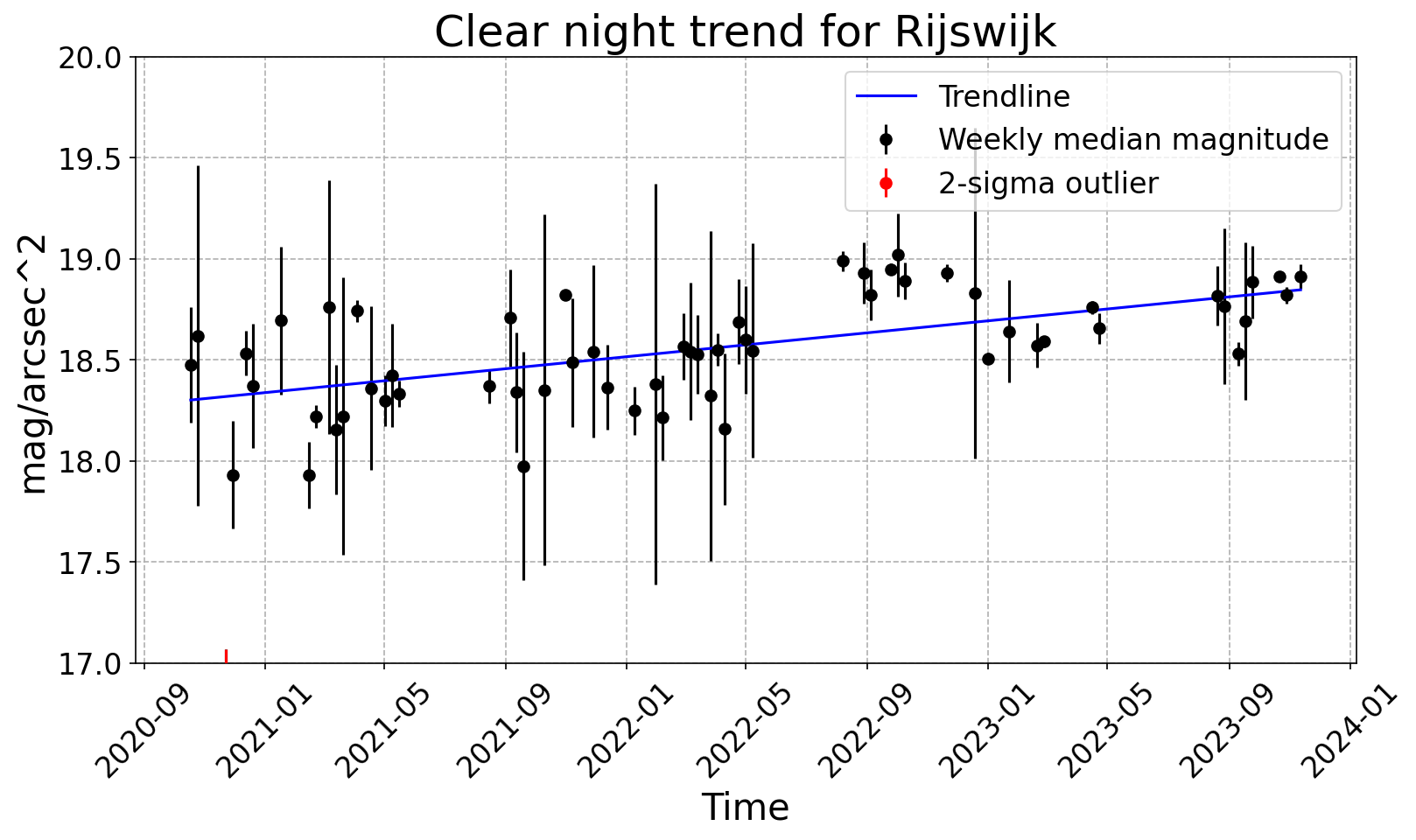}
        \caption{}
        \label{fig:trends17}
    \end{figure}
    \begin{figure}
        \includegraphics[width=\columnwidth]{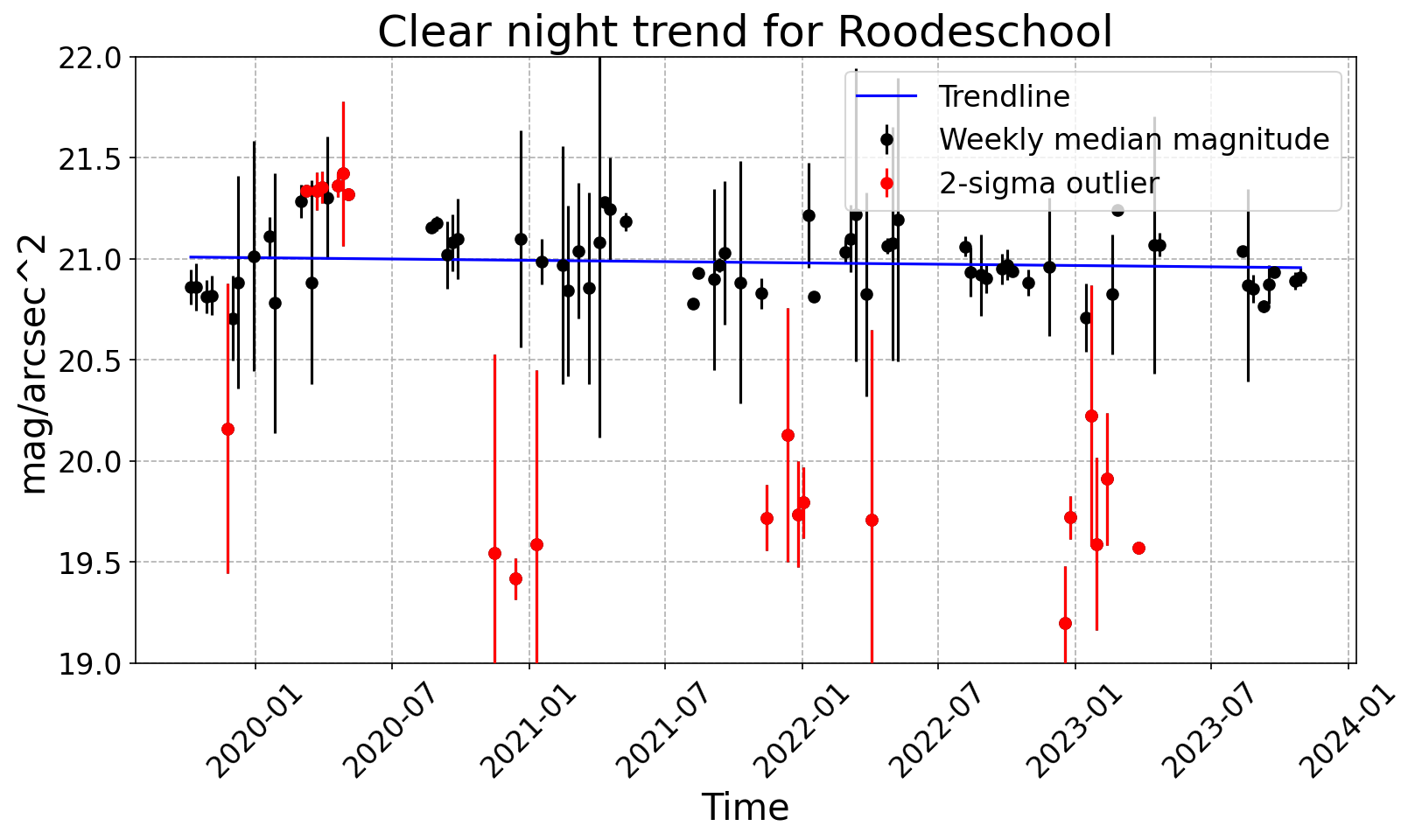}
        \caption{}
        \label{fig:trends18}
    \end{figure}
    \begin{figure}
        \includegraphics[width=\columnwidth]{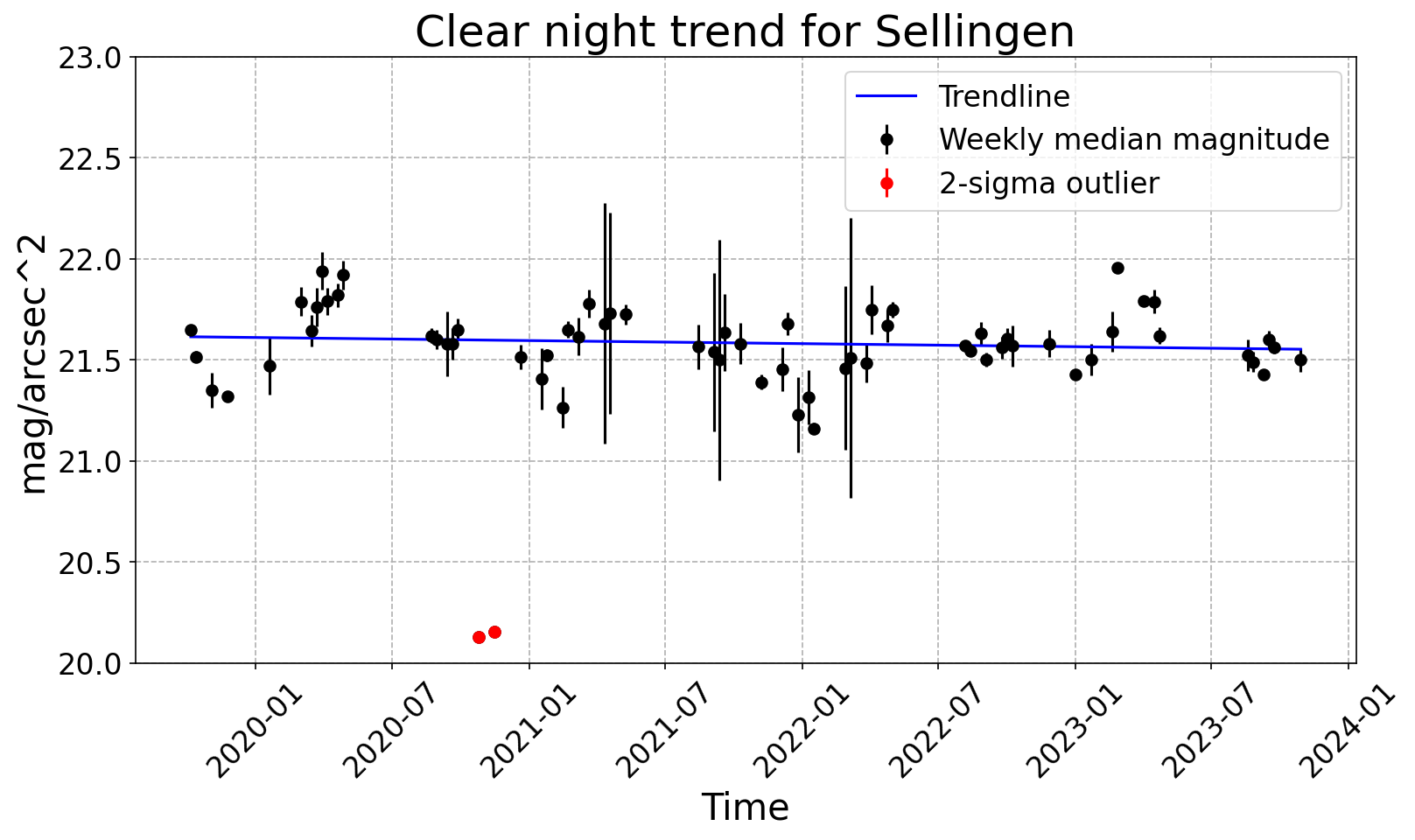}
        \caption{}
        \label{fig:trends19}
    \end{figure}
    \begin{figure}
        \includegraphics[width=\columnwidth]{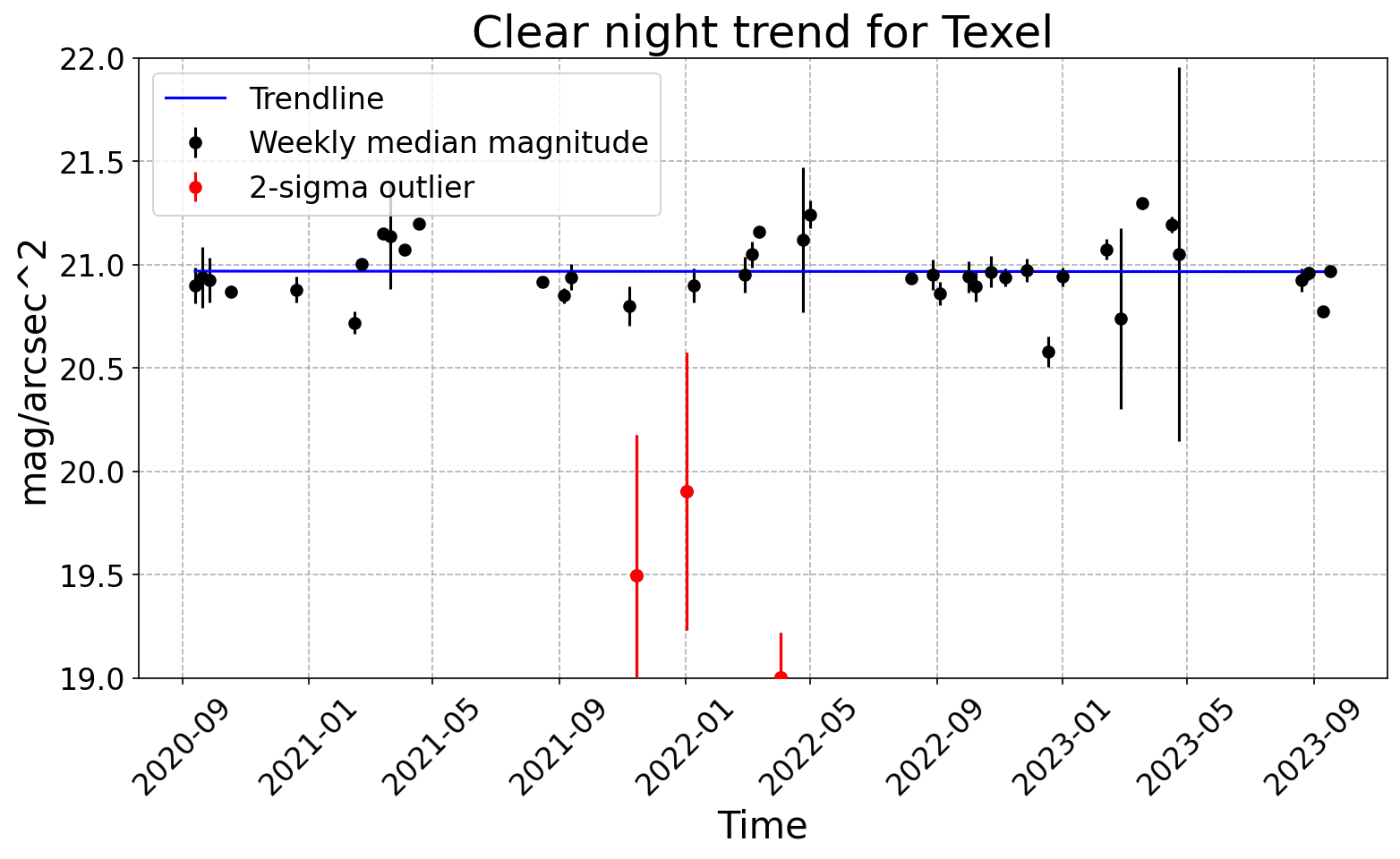}
        \caption{}
        \label{fig:trends20}
    \end{figure}
    \begin{figure}
        \includegraphics[width=\columnwidth]{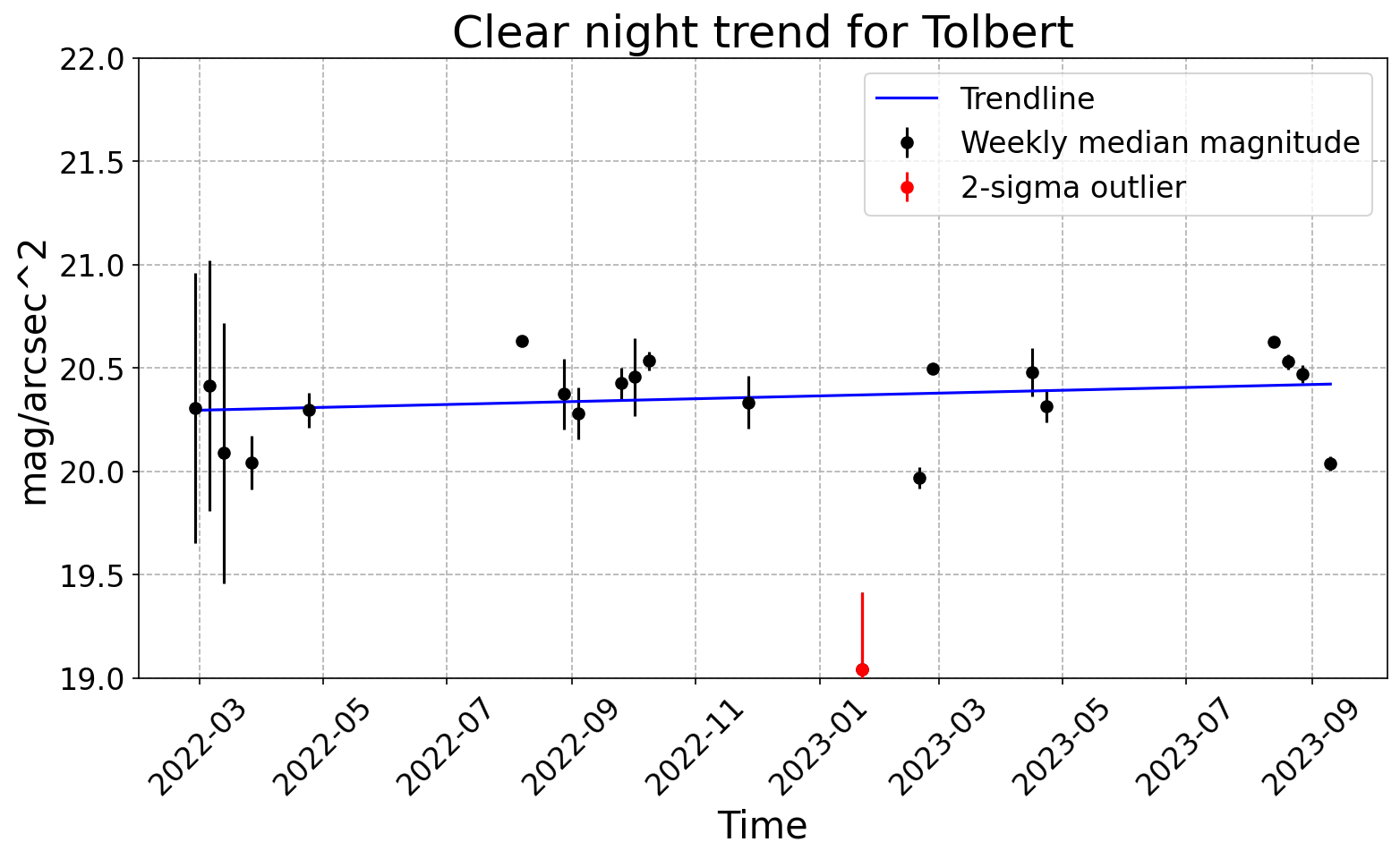}
        \caption{}
        \label{fig:trends21}
    \end{figure}
    \begin{figure}
        \includegraphics[width=\columnwidth]{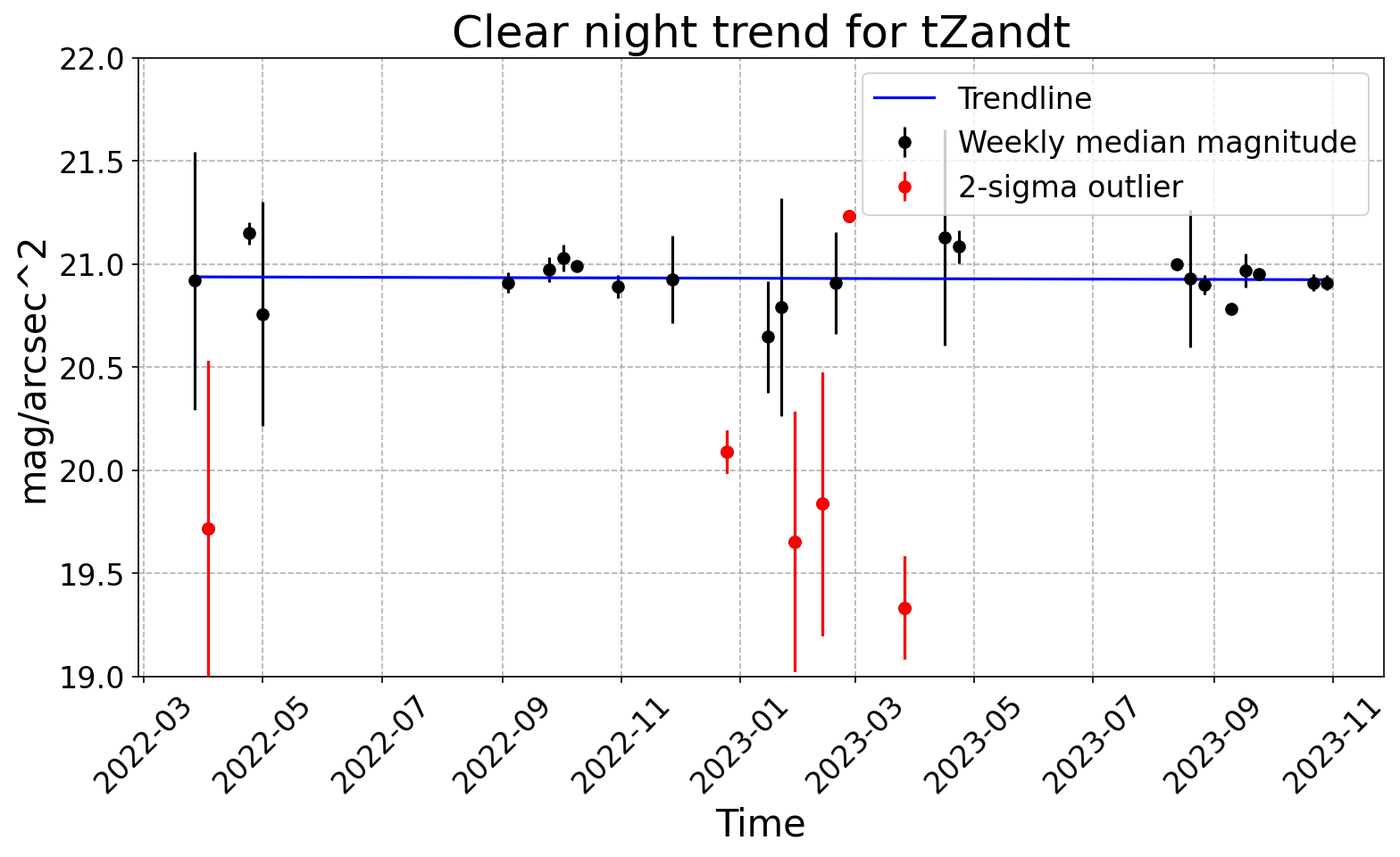}
        \caption{}
        \label{fig:trends22}
    \end{figure}
    \begin{figure}
        \includegraphics[width=\columnwidth]{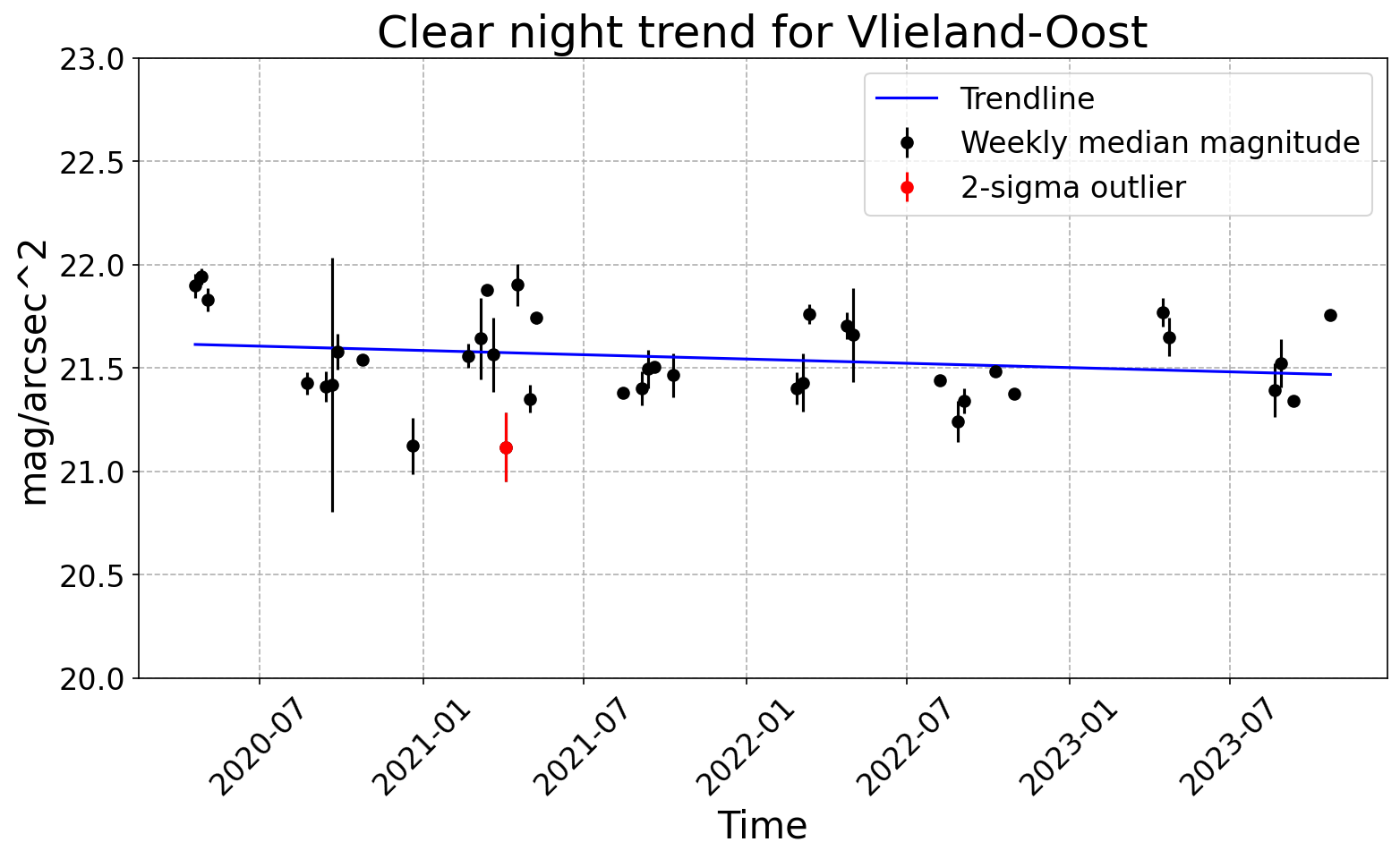}
        \caption{}
        \label{fig:trends23}
    \end{figure}
    \begin{figure}
        \includegraphics[width=\columnwidth]{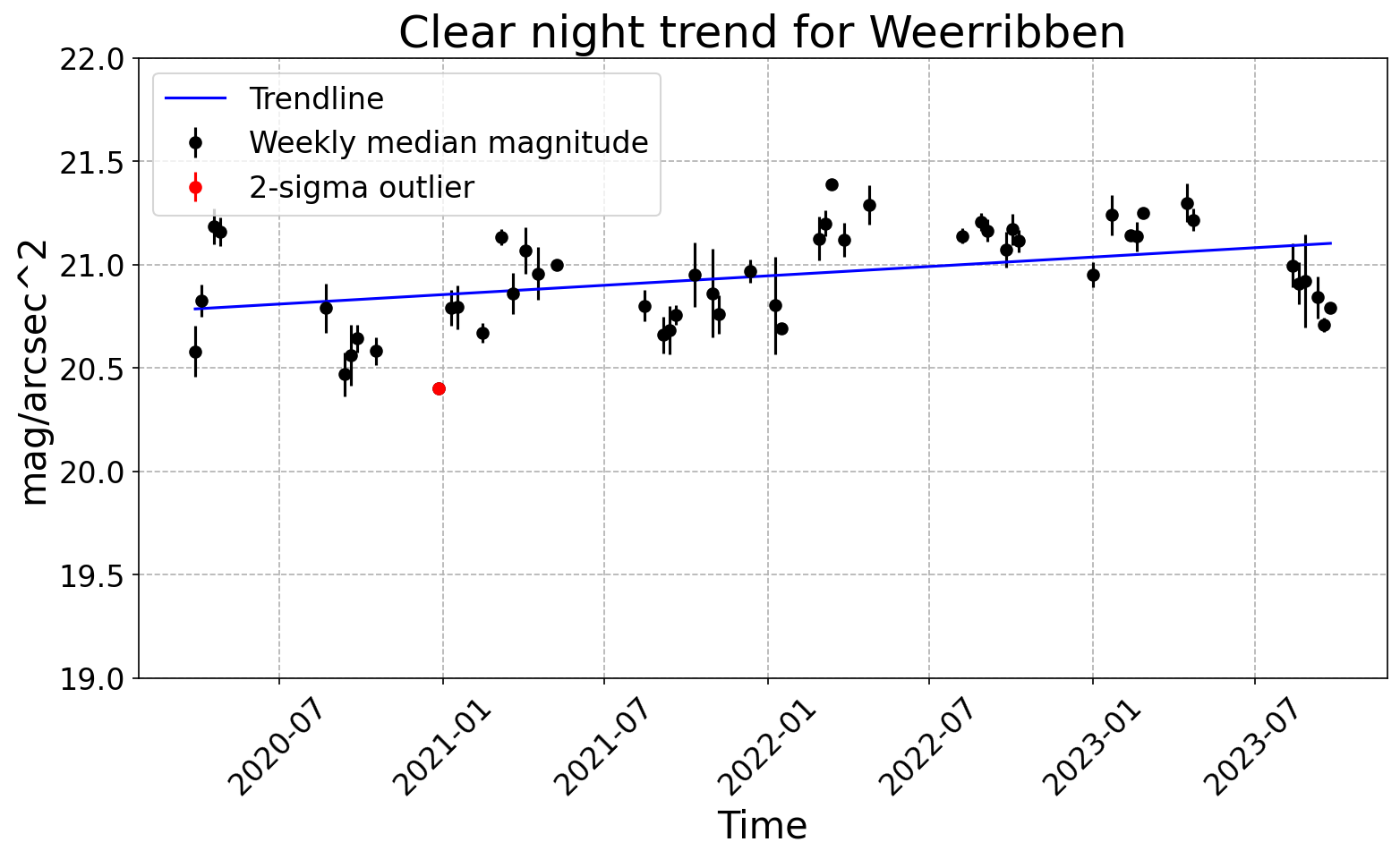}
        \caption{}
        \label{fig:trends24}
    \end{figure}
    \begin{figure}
        \includegraphics[width=\columnwidth]{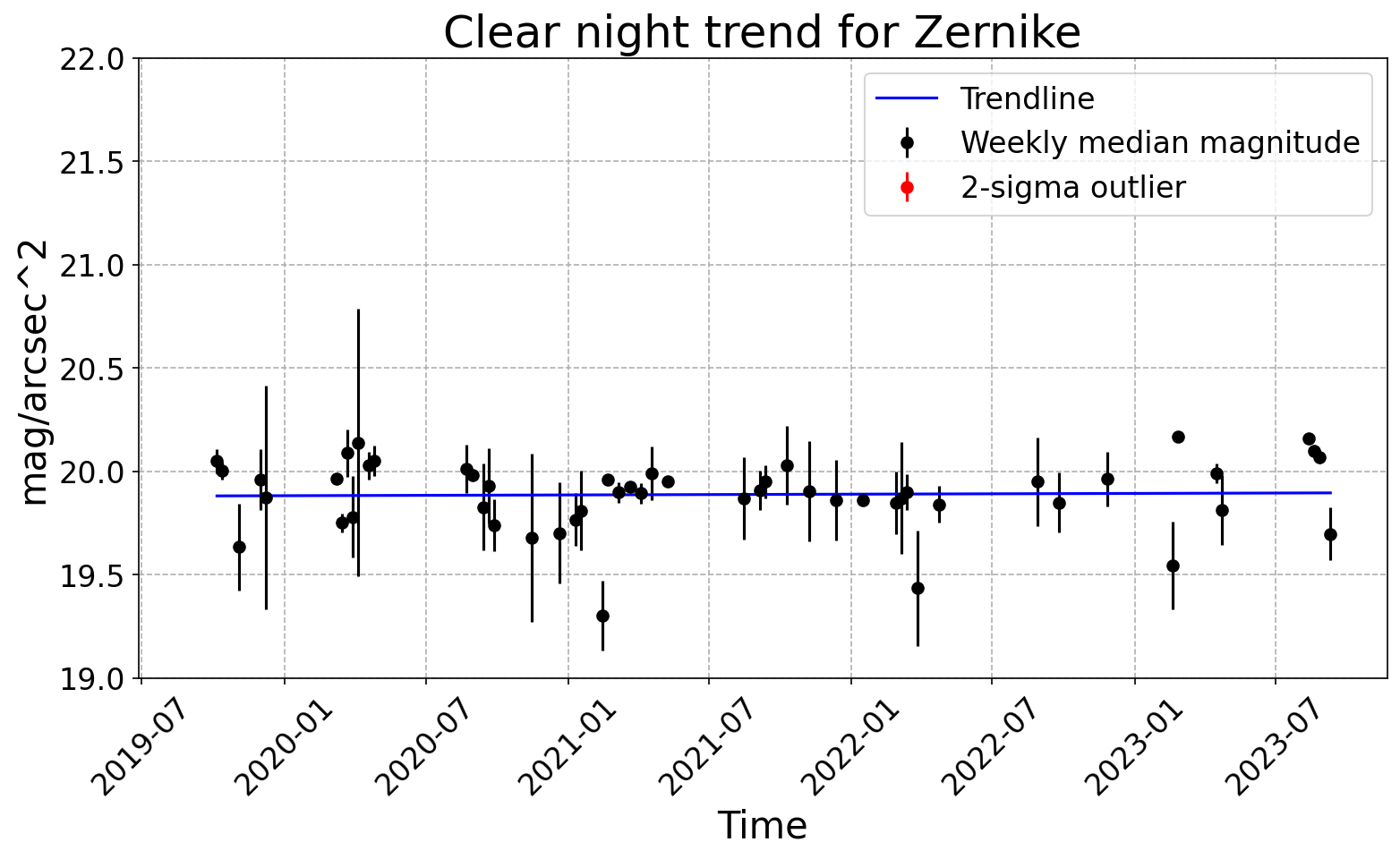}
        \caption{}
        \label{fig:trends}
    \end{figure}

\FloatBarrier
\section{Seasonal fits and sampling diagnostics}
\label{seasonal_appendix}

Table~\ref{tab:sine_overview} summarises, for nine representative
stations, (i) the amount of usable data, and (ii) the parameters of the
annual sine+linear model  

\[
m(t)\;=\;m_{0}+A\sin\bigl[2\pi(t-t_{0})\bigr]+mt ,
\]

where $t$ is time in years since 2020\,Jan 1.  Uncertainties on the
weekly‑only slopes are the total errors of Eq.~(\ref{eq:sigma_tot}). These 9 sites are the ones with the most seasonal trend and that the others are negligible. 

\begin{table*}
\centering
\caption{Sampling and annual‑sine diagnostics}
\label{tab:sine_overview}
\begin{tabular}{lrrrrrr}
\hline
Location & Weeks obs. & Weeks missed & $m_{\rm week}$\,(mag yr$^{-1}$) &
Amplitude $A$ (mag) & $m_{\rm sine}$\,(mag yr$^{-1}$) & Phase (rad)\\
\hline
Hippolytushoef & 49 & 109 & $-0.010\pm0.026$ & $0.153$ & $-0.025$ & $0.28$\\
Hornhuizen      & 57 & 148 & $-0.070\pm0.029$ & $0.271$ & $-0.056$ & $0.09$\\
Oostkapelle     & 76 &  82 & $-0.007\pm0.020$ & $-0.103$& $-0.002$ & $1.33$\\
Ostland         & 44 & 114 & $-0.029\pm0.035$ & $-0.234$& $-0.050$ & $0.41$\\
Roodeschool     &109 & 107 & $-0.013\pm0.039$ & $-0.363$& $-0.094$ & $0.41$\\
Sellingen       & 91 & 122 & $-0.015\pm0.028$ & $-0.182$& $-0.009$ & $0.72$\\
Lauwersoog      &101 & 117 & $-0.027\pm0.017$ & $-0.211$& $-0.005$ & $1.18$\\
Texel           & 69 &  95 & $-0.001\pm0.032$& $-0.141$& $\;\;0.002$& $0.65$\\
Weerribben      & 72 & 112 & $\;\;0.091\pm0.030$& $0.170$ & $\;\;0.125$& $1.11$\\
\hline
\end{tabular}
\end{table*}

% Reset figure counter for appendix and set format to F1, F2, etc.
\setcounter{figure}{0}
\renewcommand{\thefigure}{F\arabic{figure}}

% Define the directory containing your images
\newcommand{\sinedir}{Sine-Fits/}

% Loop through all 6 images
%\foreach \i in {1,...,6} {%
    \begin{figure}
        \includegraphics[width=\columnwidth]{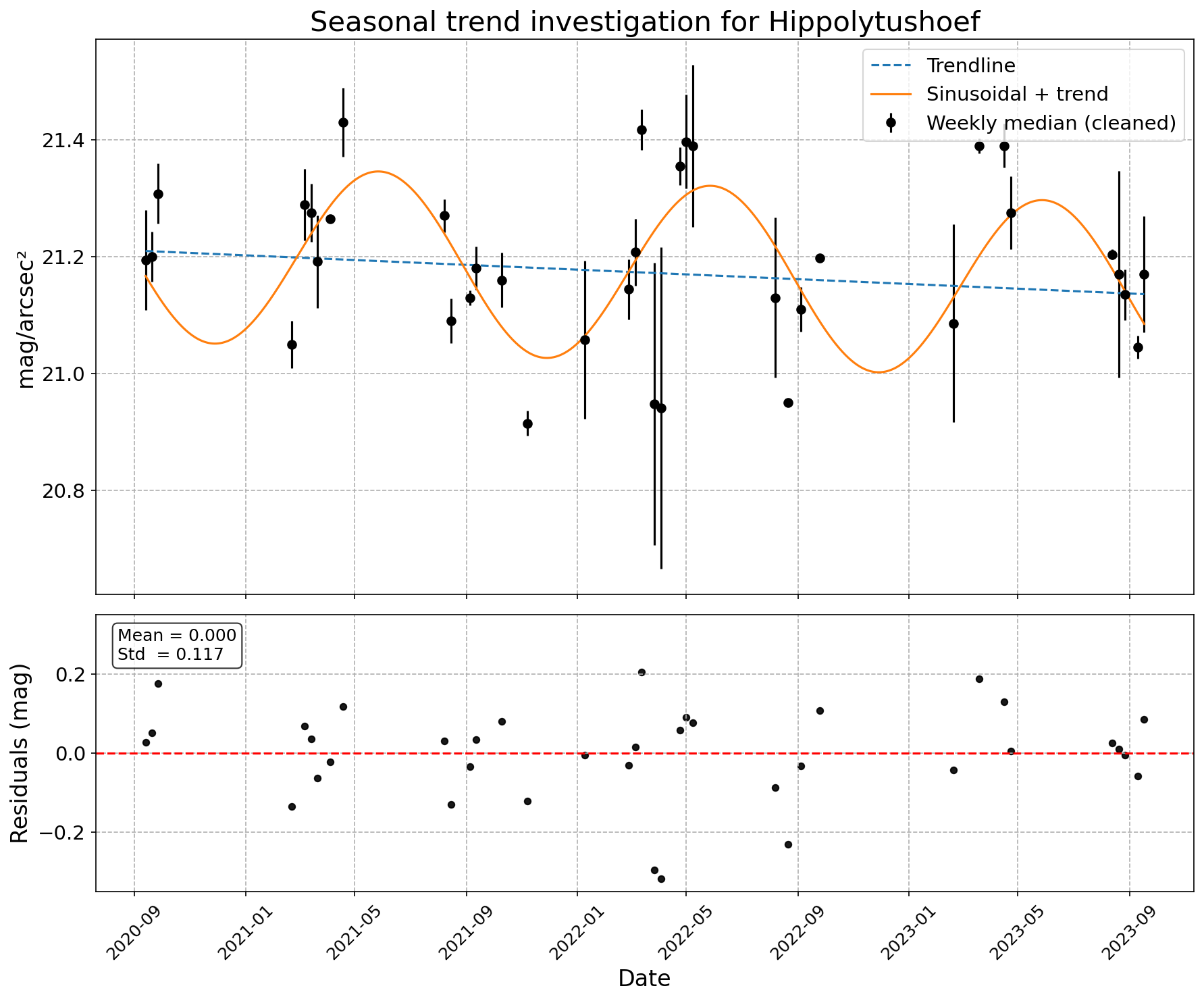}
       \caption{}
        \label{fig:Sine1}
    \end{figure}
    \begin{figure}
        \includegraphics[width=\columnwidth]{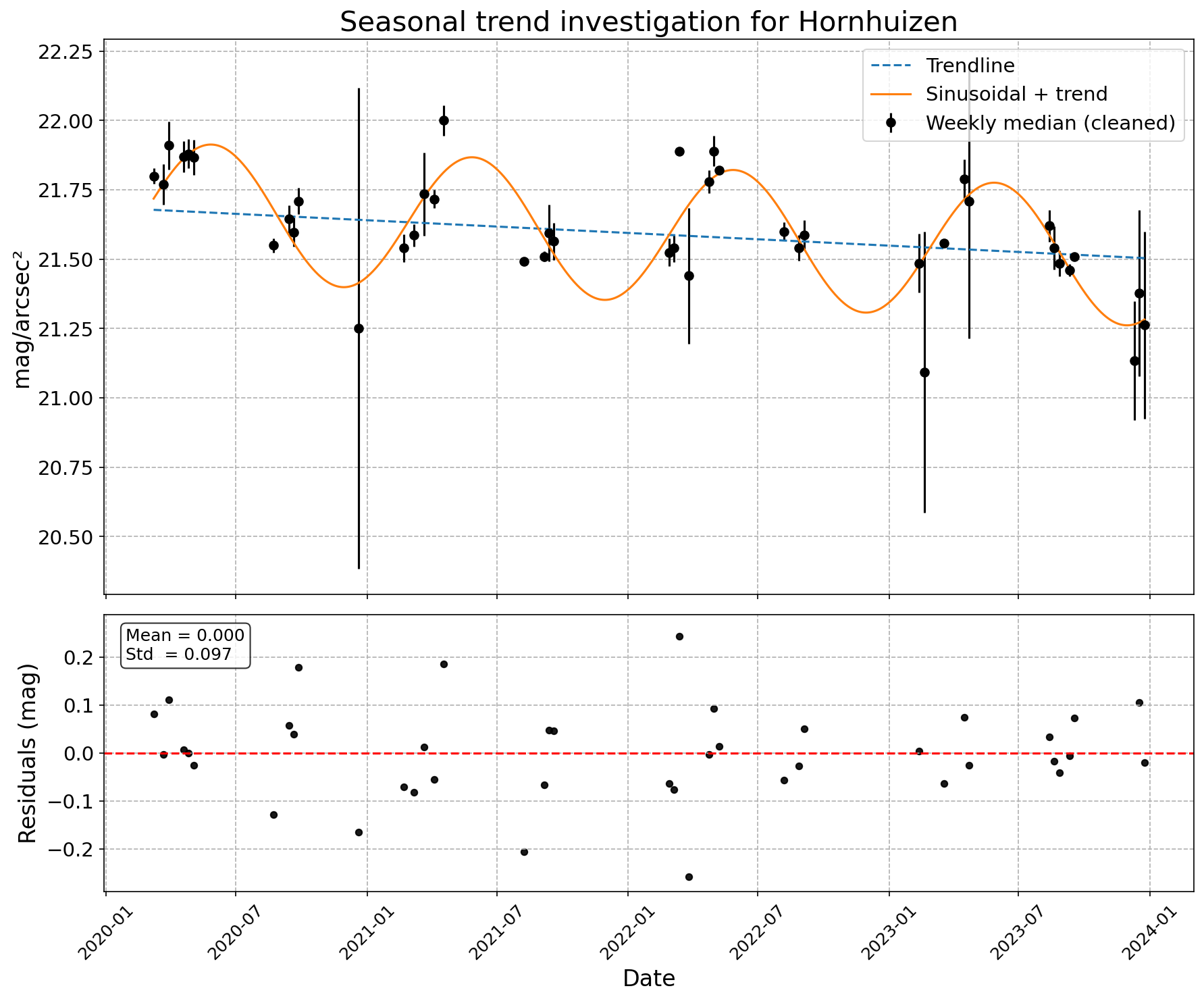}
       \caption{}
        \label{fig:Sine2}
    \end{figure}
    \begin{figure}
        \includegraphics[width=\columnwidth]{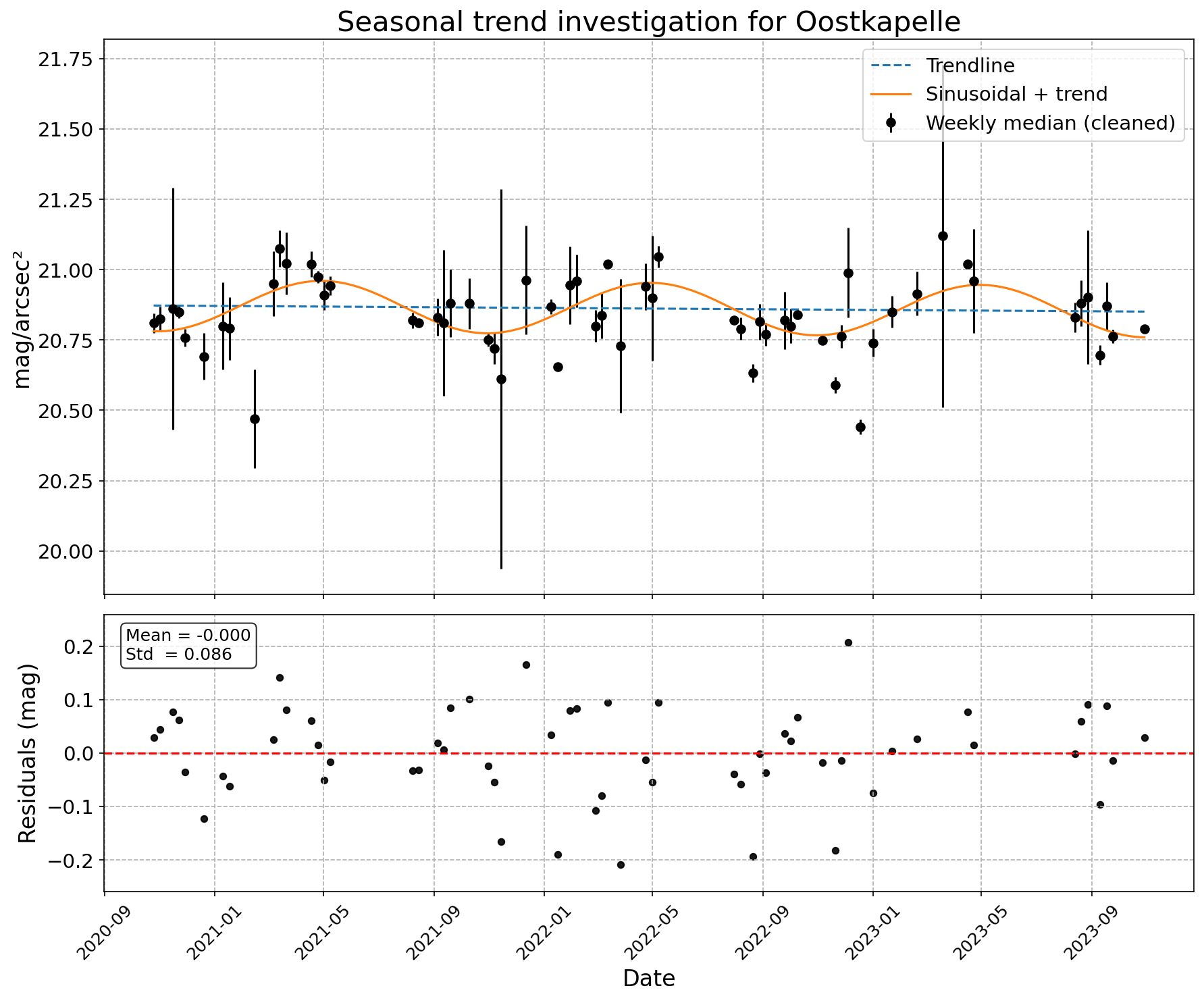}
       \caption{}
        \label{fig:Sine3}
    \end{figure}
    \begin{figure}
        \includegraphics[width=\columnwidth]{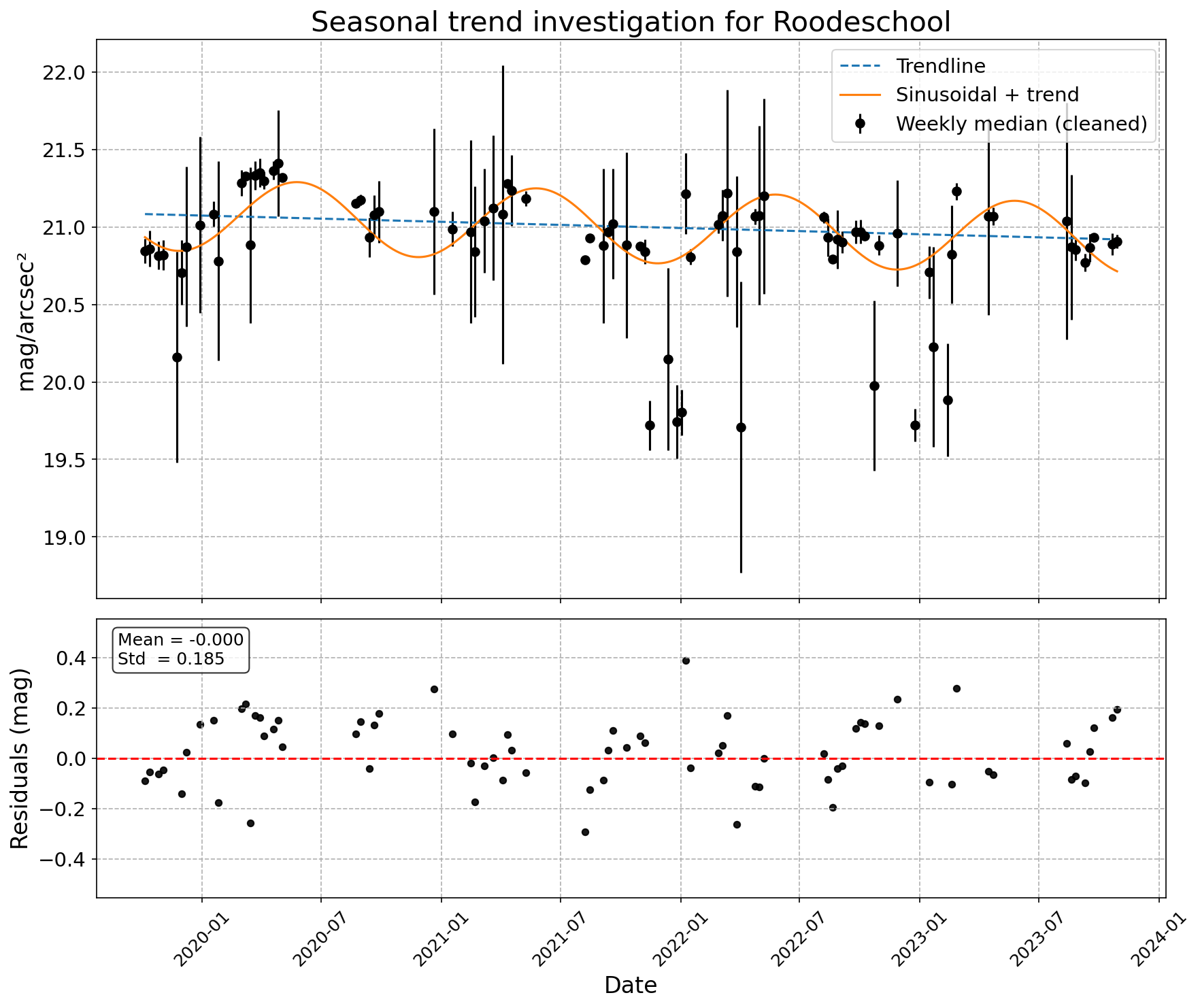}
       \caption{}
        \label{fig:Sine4}
    \end{figure}
    \begin{figure}
        \includegraphics[width=\columnwidth]{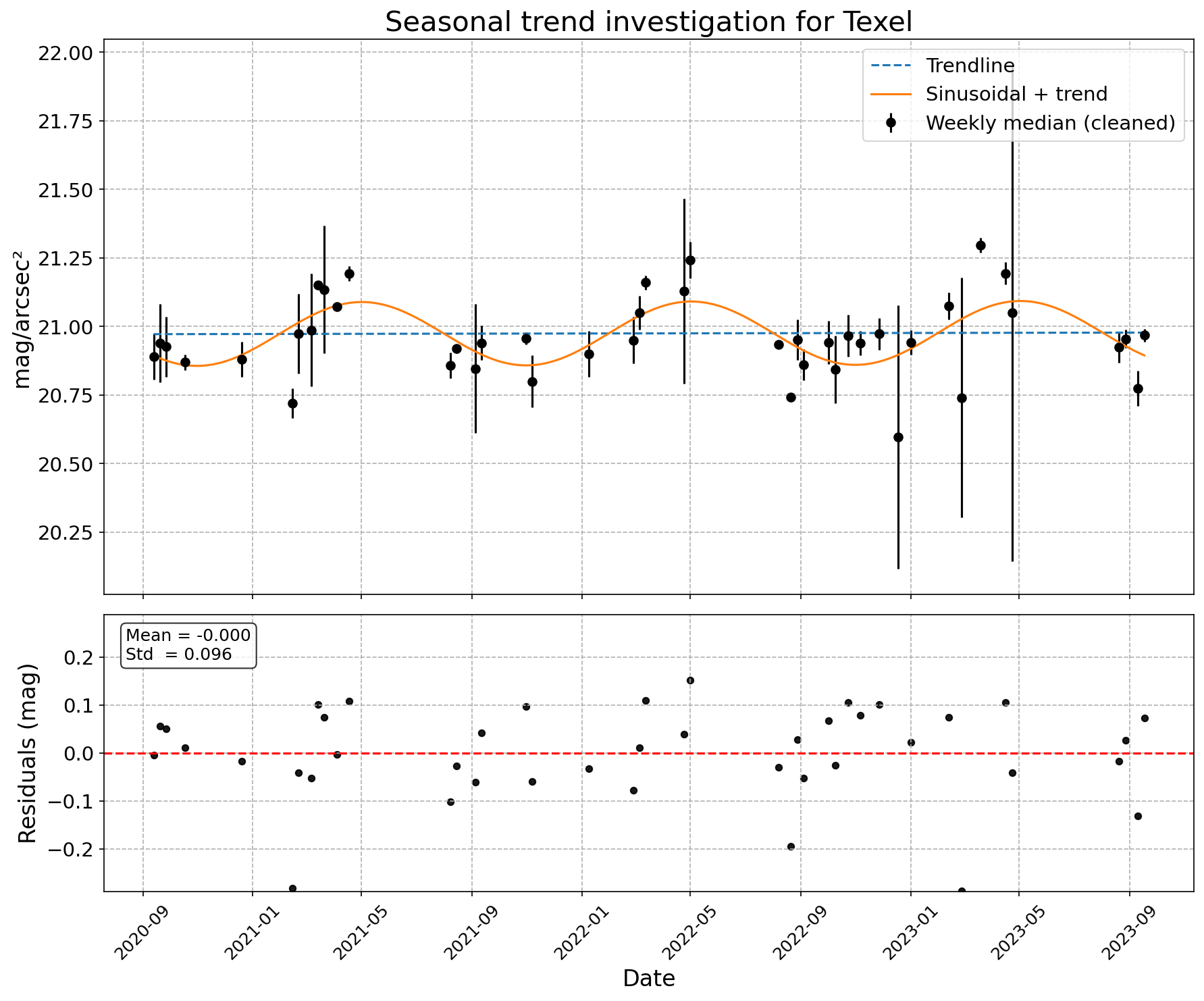}
       \caption{}
        \label{fig:Sine5}
    \end{figure}
    \begin{figure}
        \includegraphics[width=\columnwidth]{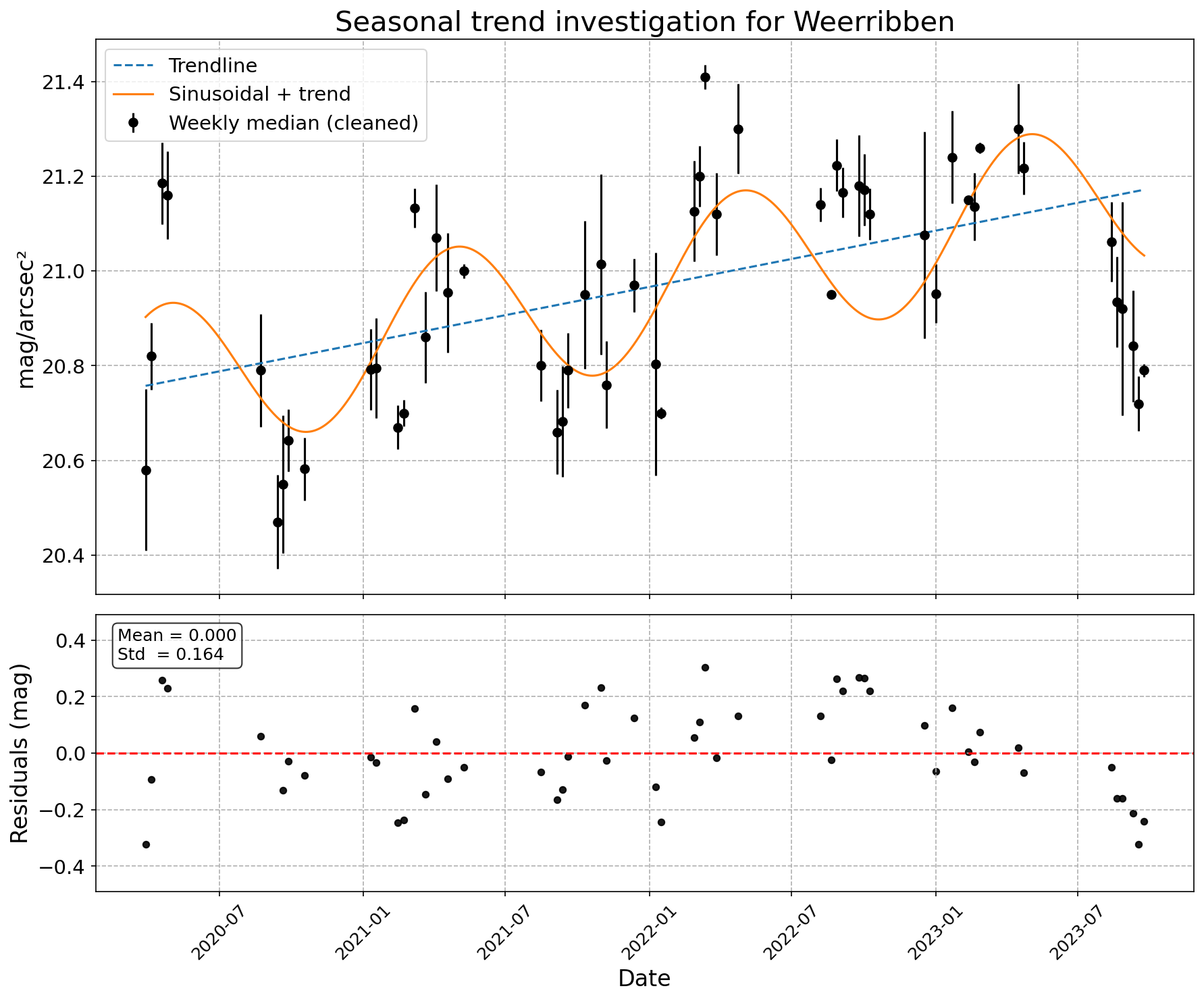}
       \caption{}
        \label{fig:Sine6}
    \end{figure}
%}

%%%%%%%%%%%%%%%%%%%%%%%%%%%%%%%%%%%%%%%%%%%%%%%%%%

% Don't change these lines
\label{lastpage}
\end{document}